\documentclass[aps,prd,twocolumn,showpacs,notitlepage,eqsecnum,
superscriptaddress]{revtex4-1}

\usepackage[centertags]{amsmath}
\usepackage{amssymb}
\usepackage{latexsym}
\usepackage{enumerate}
\usepackage{enumitem}
\usepackage{graphicx}
\usepackage{mathrsfs}
\usepackage{stmaryrd}
\usepackage{import}
\usepackage{tensor}
\usepackage{color}
\usepackage{bm}
\usepackage{multirow}
\usepackage{breakurl}
\usepackage{float}
\usepackage{placeins}
\usepackage{autobreak}
\usepackage{array}
\usepackage{mathtools}
\usepackage[table,dvipsnames]{xcolor}
\usepackage{pbox}
\usepackage{hyperref}

\newcommand{\bi}{\begin{itemize}}
\newcommand{\ei}{\end{itemize}}
\newcommand{\be}{\begin{equation}}
\newcommand{\ee}{\end{equation}}

\renewcommand{\arraystretch}{2}

\newcommand{\e}{\epsilon}
\newcommand{\ve}{\varepsilon}

\renewcommand{\O}{\Omega}

\newcommand{\q}{\quad}
\newcommand{\qq}{\qquad}

\newcommand{\vp}{\varphi}

\allowdisplaybreaks

\begin{document}

\title{Eccentric-orbit extreme-mass-ratio-inspiral radiation II: 1PN correction to leading-logarithm and 
subleading-logarithm flux sequences and the entire perturbative 4PN flux}

\author{Christopher Munna}
\author{Charles R. Evans}
\affiliation{Department of Physics and Astronomy, University of North 
Carolina, Chapel Hill, North Carolina 27599, USA}

\begin{abstract}
In a recent paper we showed that for eccentric-orbit extreme-mass-ratio inspirals the analytic forms of the 
leading-logarithm energy and angular momentum post-Newtonian (PN) flux terms (radiated to infinity) can, 
to arbitrary PN order, be determined by sums over the Fourier spectrum of the Newtonian quadrupole moment.  We 
further showed that an essential part of the eccentricity dependence of the related subleading-logarithm PN sequences,  
at lowest order in the symmetric mass ratio $\nu$, stems as well from the Newtonian quadrupole moment.  Once that 
part is factored out, the remaining eccentricity dependence is more easily determined by black hole perturbation 
theory.  In this paper we show how the sequences that are the 1PN corrections to the entire leading-logarithm series, 
namely terms that appear at PN orders $x^{3k+1} \log^k(x)$ and $x^{3k+5/2} \log^k(x)$ (for PN compactness 
parameter $x$ and integers $k\ge 0$), at lowest order in $\nu$, are determined by the Fourier spectra of the 
Newtonian mass octupole, Newtonian current quadrupole, and 1PN part of the mass quadrupole moments.  We also 
develop a conjectured (but plausible) form for 1PN correction to the leading logs at second order in $\nu$.  
Further, in analogy to the first paper, we show that these same source multipole moments also yield nontrivial parts 
of the 1PN correction to the subleading-logarithm series, and that the remaining eccentricity dependence (at lowest 
order in $\nu$) can then more easily be determined using black hole perturbation theory.  We use this method to 
determine the entire analytic eccentricity dependence of the perturbative (i.e., lowest order in $\nu$) 4PN non-log 
terms, $\mathcal{R}_4(e_t)$ and $\mathcal{Z}_4(e_t)$, for energy and angular momentum respectively.
\end{abstract}


\maketitle

\section{Introduction}
\label{sec:intro}

With development proceeding on the Laser Interferometer Space Antenna (LISA) gravitational wave mission 
\cite{eLISA,LISA}, the need for accurate theoretical models of eccentric extreme-mass-ratio inspirals (EMRIs) has 
continued to grow \cite{BerrETC19,AmarETC17,BabaETC17,BaraETC18}.  In previous work \cite{ForsEvanHopp16, 
MunnEvan19a,MunnETC20} complementary approaches from post-Newtonian (PN) theory and black hole 
perturbation theory (BHPT) were combined to generate new information on the orbit-averaged energy and angular 
momentum fluxes radiated to infinity in (non-spinning) eccentric-orbit systems.  In a recent one of these papers 
\cite{MunnEvan19a} (hereafter Paper I), we showed that the Fourier amplitudes of the Newtonian mass quadrupole 
moment, and the function $g(n,e_t)$ \cite{PeteMath63,Blan14} proportional to their complex square, determine the 
functional dependence in the quasi-Keplerian (time) eccentricity $e_t$ \cite{DamoDeru85} of the entire 
leading-logarithm sequence (i.e., to arbitrary PN order) of these fluxes.  The functional dependence in eccentricity of 
each such flux term, relative to the circular orbit limit, is commonly referred to as an enhancement function.  

We then went further in Paper I to show that additional sums over the quadrupole spectrum determine essential parts 
of the eccentricity dependence of the subleading-logarithm series, which are terms associated with leading logs at 
the same PN order but with one less power of $\log(x)$, where $x$ is a PN compactness parameter.  Specifically, we 
define $x = ((m_1 + m_2) \O_\vp)^{2/3}$ \cite{Blan14}, where $m_1$ and $m_2$ are the primary and secondary 
masses and $\O_\vp$ is the mean frequency of azimuthal motion.  A subleading-logarithm term can be thought of 
alternatively as the 3PN correction to the leading-logarithm term of the same power of $\log(x)$ (or henceforth 
referred to as the corresponding term in the \emph{3PN log series}).  

At lowest order in $\nu$, these quadrupole-dependent parts can be re-expressed in terms of the Darwin 
\cite{Darw59,Darw61} definition of eccentricity $e$.  Each entire subleading-log term is then taken to have an 
assumed form for its expansion in powers of $e^2$, with the quadrupole-dependent part being built in.  This 
quadrupole portion subsumes all of the transcendental number coefficients.  The remaining unknown structure in 
each flux term is found to be either a closed form expression (at integer PN orders) or an infinite series (at 
half-integer PN orders) with rational number coefficients that can then in principle be determined by BHPT 
calculations.  At lowest order in $\nu$, fluxes can then be transformed back to $e_t$.  We showed this procedure 
in action by extracting the entire analytic dependence of the 6PN log energy and angular momentum terms, 
$\mathcal{R}_{6L}(e_t)$ and $\mathcal{Z}_{6L}(e_t)$, to arbitrary powers of $e_t^2$.  

\begin{figure}[ht!]
\includegraphics[width=0.95\columnwidth]{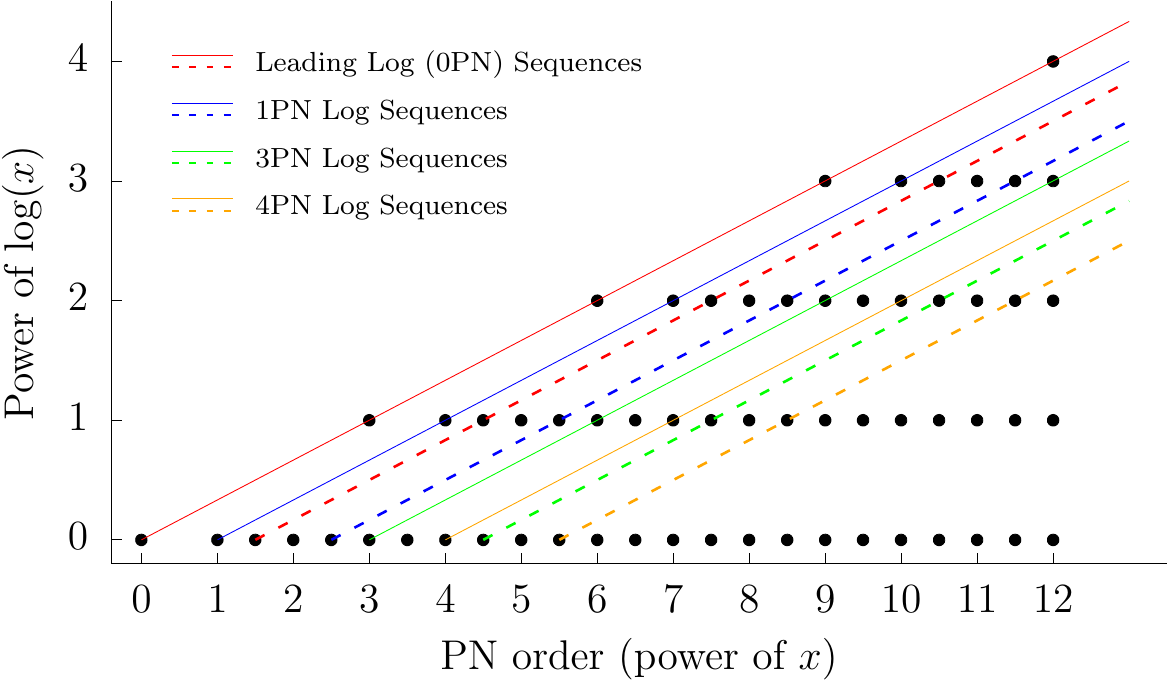}
\caption{Schematic depiction of the presence of terms (as black filled circles) in the high PN order relative fluxes 
for successively higher powers of compactness $x$ (horizontal axis) and higher powers of $\log(x)$ (vertical axis).  
The Peters-Mathews \cite{PeteMath63} flux is symbolized by the left-most point at the origin of the plot.  This 
representation of the PN structure of the fluxes allows a graphical explanation of the various ``log'' sequences 
that are the focus of this paper and Paper I.  The red lines show the leading-log sequences, both integer-order 
(solid) and half-integer-order (dashed) detailed previously in Paper I.  The 3PN log sequences (previously called 
sub-leading logs), also the subject of Paper I, are shown as green lines, both integer-order and half-integer-order.  
The blue lines represent the 1PN log sequences and the orange lines denote the 4PN log sequences, all of which are 
the focus of this paper.}
\label{fig:logSequences}
\end{figure}

Thus, one conclusion of Paper I is that two \emph{diagonal strips} in the high-order PN structure of the fluxes 
(i.e., the leading logarithms at PN orders $x^{3k} \log^k(x)$ and $x^{3k+3/2} \log^k(x)$ for integers $k\ge 0$) are 
determined by the Fourier spectrum of the Newtonian quadrupole moment.  See Fig.~\ref{fig:logSequences} for a 
graphical depiction of these leading log strips in the PN structure.  The second main conclusion is that two 
additional diagonal strips, the subleading logs at PN orders $x^{3k} \log^{k-1}(x)$ and $x^{3k+3/2} \log^{k-1}(x)$ 
for integers $k\ge 1$, are also partly determined at lowest order in $\nu$ by the quadrupole spectrum, with the 
remaining eccentricity dependence having a closed form (integer order) or infinite series (half-integer order) and 
being more easily determined by BHPT.  The leading log and 3PN log sequences are represented in the 
figure as (solid and dashed) red and green lines, respectively.  The question then arises is it possible to determine 
additional entire diagonal strips in the PN structure of the fluxes with only limited additional knowledge of 
low-order source multipole moments?  As we show in this paper, the answer is yes if we focus on the 1PN corrections 
to the leading- and subleading-logarithm sequences.  

The first term in the leading-log series is the Newtonian quadrupole flux, i.e., the Peters-Mathews 
\cite{PeteMath63,Pete64} term $\mathcal{R}_{0}(e_t)$ itself.  The enhancement function in this case arises from 
simply summing the Newtonian quadrupole moment spectrum $g(n,e_t)$ over all harmonics $n$ in the eccentric motion.  
The next order term $\mathcal{R}_{1}(e_t)$ is the 1PN correction to the gravitational wave flux, which has been known 
since Wagoner and Will \cite{WagoWill76} (see also \cite{ArunETC08b,Blan14}).  In this case determining the 
enhancement function requires the Fourier spectra of the Newtonian mass octupole, the Newtonian current quadrupole, 
and the 1PN-corrected mass quadrupole moments (hereafter called the 1PN multipoles).  The $\mathcal{R}_{1}(e_t)$ 
flux is the first term in one of the two new diagonal sequences of 1PN-corrected leading logarithms, which we will
refer to as a \emph{1PN log series} (Fig.~\ref{fig:logSequences}, solid blue line).  This sequence has PN orders 
$x^{3k+1} \log^{k}(x)$ for $k\ge 0$.  The other (half-integer) 1PN log series (dashed blue line in the figure) 
begins with the 2.5PN tail at $x^{5/2}$ and has PN orders $x^{3k+5/2} \log^{k}(x)$ for $k\ge 0$.  A principal result 
of this paper is to show that it is merely the spectra of the three 1PN multipoles that are required to determine 
these two 1PN log series in their entirety to arbitrary PN order (at lowest order in $\nu$).

Calculation of the Newtonian mass octupole and current quadrupole for eccentric bound motion is 
fairly straightforward and can be found in the original, earlier papers \cite{WagoWill76,DamoDeru85}, the review by 
Blanchet \cite{Blan14}, or extrapolated from techniques reviewed in Paper I.  Calculation of the 1PN correction to 
the mass quadrupole is more involved.  At 1PN order, the determination of the mass quadrupole must account for 
relativistic orbital precession, which means that the spectrum cannot be represented as a single Fourier series but 
instead requires a double Fourier sum over harmonics of the two different frequencies, $\O_r$ (radial libration) 
and $\O_{\vp}$ \cite{Damo83,DamoDeru85,GopaIyer02,DamoGopaIyer04,ArunETC08a}.  Once these spectra are computed 
for given orbital parameters, their sums weighted by powers of $n$ over all harmonics combine to give terms in the 
1PN log series.  One key difference though between the 1PN log series and the leading logs themselves is that the 
former now have contributions beyond lowest order in the mass ratio $\nu$.  Because the multipole moment analysis in 
this paper makes no \textit{a priori} assumptions on the mass ratio, we are able to extract the likely forms for these 
$\mathcal{O}(\nu)$ corrections, though without (presently) second-order BHPT to assist in verification.  At lowest 
order in $\nu$, the analysis found in this paper provided a theoretical underpinning for several previously known 
closed-form flux terms \cite{ForsEvanHopp16,MunnEvan19a,MunnETC20}.

With the 1PN log series thus understood, we then find that the same set of 1PN multipoles again appear in the 
1PN correction to the subleading logarithms.  These sequences will be referred to as \emph{4PN log series}, since 
for a given power of $\log(x)$ each term in this series occurs at order $x^4$ relative to the corresponding leading 
log term.  In other words, the 4PN log sequences are two diagonal strips in the high PN order flux structure that 
appear at orders $x^{3k+1} \log^{k-1}(x)$ and $x^{3k+5/2} \log^{k-1}(x)$ for $k\ge 1$ (solid and dashed orange 
lines in the figure).  The first sequence begins with the 4PN non-log flux and the second with the 5.5PN non-log 
term.  In direct analogy to our findings in Paper I, the set of 1PN multipoles provides essential separable portions 
of the terms in the 4PN log series, which include all transcendental coefficients, leaving only rational series which 
at lowest order in $\nu$ can then be calculated (more) easily with BHPT.  For the first (integer-order) sequence 
(solid orange line), the remaining parts can be factored into closed forms with rational coefficients, and it is 
possible to determine their entire analytic eccentricity dependence in this manner.  For the second 
(half-integer-order) sequence (dashed orange line), the remaining eccentricity dependence is an infinite power series 
with rational coefficients, and BHPT can be used to determine coefficients to some depth in the $e^2$ expansion.  We 
illustrate this procedure in detail by obtaining the 4PN non-log energy and angular momentum fluxes at lowest order 
in the mass ratio.

The layout of this paper is as follows.  In Sec.~\ref{sec:PNexp} we review the PN expansion for radiated energy and 
angular momentum, with an illustration of the terms that will be computed in this analysis.  There we also derive 
the Fourier expansion for each of the 1PN multipole moments, and in Sec.~\ref{sec:knownLogs} we detail their 
previously known contributions to the energy and angular momentum flux expansions.  Sec.~\ref{sec:1PNLogs} 
shows how these source multipole spectra contribute to the 1PN log series, with explicit general formulae which 
generate all members of those sequences.  We proceed in Sec.~\ref{sec:4PNTail} to derive the 4PN tail flux using 
these same Fourier spectra in order to check various results and to aid our extraction of the full 4PN log series 
fluxes at lowest order in $\nu$.  Then, in Sec.~\ref{sec:full4PN} we illustrate how the various 1PN Fourier summations 
manifest specifically in the 4PN flux (and more generally in higher-order terms in the 4PN log series) and combine 
these observations with BHPT flux calculations from \cite{MunnETC20} to compute $\mathcal{R}_{4}(e_t)$ and 
$\mathcal{Z}_{4}(e_t)$ in compact form.  This result is quite timely, as it will provide a valuable check for the 
PN community as they close in on a full description of the orbital mechanics and radiative losses at 4PN.  That 
section also gives our analysis of the 5.5PN non-log energy flux, showing the procedure carries over to half-integer 
order 4PN log terms.  We conclude in Sec.~\ref{sec:conclusions} with discussion of potentially extending his process 
to the 2PN-corrected logarithm series (i.e., the 2PN log sequence and 5PN log sequence).

In this paper we use units such that $c = G = 1$.  As in Paper I, for any pair of functions with names distinguished 
by a tilde (e.g. $g(n,e_t)$ and $\tilde{g}(n,e_t)$), the ``plain'' quantity will relate to the energy flux while the 
``tilde'' version will correspond to its angular momentum counterpart (see e.g., \cite{ArunETC09a} also).

\begin{widetext}
\section{Eccentric-orbit PN flux expansion and Fourier decomposition of 1PN multipoles}
\label{sec:PNexp}

In this section we lay out the parts of the PN expansion of the orbit-averaged fluxes that are of interest in this 
paper and review the calculation of the Fourier spectra of the 1PN multipoles.  The focus is on eccentric EMRIs with 
the binary consisting of two non-spinning bodies of mass $m_1$ (primary) and mass $m_2$ (secondary).  We are 
primarily concerned with $m_2 \ll m_1$ but keep the symmetric mass ratio $\nu=m_1m_2/(m_1+m_2)^2$ as a variable.

\subsection{PN flux expansions}

In the modified harmonic gauge \cite{ArunETC08a,ArunETC08b,ArunETC09a,Blan14}, the flux expansions are parameterized 
by the aforementioned $\nu$ and compactness parameter $x$, as well as the quasi-Keplerian time eccentricity $e_t$ 
(also reviewed below).  The expansion of the energy flux at infinity has the following form 
\cite{Blan14,GoldRoss10,Fuji12a,Fuji12b,ForsEvanHopp16,MunnEvan19a,MunnETC20}:
\begin{align}
\label{eqn:energyfluxInf}
\bigg\langle \frac{dE}{dt} &\bigg\rangle = 
\frac{32}{5} \nu^2 x^5 
\biggl[\mathcal{R}_0 + x\mathcal{R}_1 + x^{3/2}\mathcal{R}_{3/2}
+x^2\mathcal{R}_2 + x^{5/2}\mathcal{R}_{5/2}
+ x^3\biggl(\mathcal{R}_3+ \log(x) \mathcal{R}_{3L} \biggr)
+ x^{7/2}\mathcal{R}_{7/2}
\notag\\&+ x^4\biggl(\mathcal{R}_4+ \log(x) \mathcal{R}_{4L} \biggr)
+x^{9/2}\Bigl(\mathcal{R}_{9/2}+\log(x)\mathcal{R}_{9/2L}\Bigr)
+x^5\Bigl(\mathcal{R}_5
+\log(x)\mathcal{R}_{5L}\Bigr)
+x^{11/2}\Bigl(\mathcal{R}_{11/2}+\log(x)\mathcal{R}_{11/2L}\Bigr)
\notag\\& 
+ x^6\Bigl(\mathcal{R}_6 + \log(x)\mathcal{R}_{6L}
+ \log^2(x)\mathcal{R}_{6L^2} \Bigr)
+x^{13/2}\Bigl(\mathcal{R}_{13/2}+\log(x)\mathcal{R}_{13/2L}\Bigr)
+\cdots
\biggr] .
\end{align}
In this expression the Newtonian circular-orbit energy flux has been factored out.  Each quantity 
$\mathcal{R}_i = \mathcal{R}_i(e_t,\nu)$ is a function of eccentricity and mass ratio that helps determine the 
flux radiated at PN order $i$.  The scripts denoting PN order track both the power of $x$ and the presence of 
powers of $\log(x)$.  The dependence of each term on $e_t$ and $\nu$ differs notationally from Paper I, where the 
flux terms were considered only at lowest order in $\nu$ and thus taken to be functions of $e_t$ alone.  
In this paper, while we retain both parameters, we will be interested occasionally in just the lowest order in 
$\nu$ limit.  In those circumstances we revert to writing explicitly $\mathcal{R}_i(e_t)$ or $\mathcal{R}_i(e_t,0)$.  
With $x$ as the compactness parameter, each flux function is known to diverge as $e_t \rightarrow 1$ (see however 
\cite{MunnETC20} for an alternative).  

In both Paper I and this paper we are concerned with diagonal strips in the high order PN structure where for 
each unit increase in power of $\log(x)$ there is an increase of three powers of $x$.  As mentioned, the first 
example of such strips were the two leading-logarithm series, with (integer) orders $x^{3k} \log^k(x)$ and 
(half-integer) orders $x^{3k+3/2} \log^k(x)$ (for $k\ge 0$), which were given by Eq.~(2.2) in Paper I.  That work 
also dealt with what were there called the subleading-logarithm sequences, which here we refer to as the 3PN log 
sequences, with (integer) orders $x^{3k} \log^{k-1}(x)$ and (half-integer) orders $x^{3k+3/2} \log^{k-1}(x)$ 
(for $k\ge 1$).  

In this paper our attention is initially on a pair of diagonal sequences that can be considered the 1PN 
correction to the two leading-log series and which form the following subset of the flux terms in 
\eqref{eqn:energyfluxInf}
\begin{align}
\label{eqn:1PNenergyfluxInf}
\left\langle \frac{dE}{dt} \right\rangle^{\text{1L}} =
\frac{32}{5} \nu^2 x^5 
\biggl[ x \mathcal{R}_1 
&+ x^{5/2}\mathcal{R}_{5/2}
+ x^4 \log(x)\mathcal{R}_{4L} +x^{11/2}\log(x)\mathcal{R}_{11/2L}  
\\
\notag 
&+ x^7\log^2(x)\mathcal{R}_{7L^2}
+x^{17/2}\log^2(x) \mathcal{R}_{17/2L^2}
+ x^{10} \log^3(x) \mathcal{R}_{10L^3}+\cdots \biggr] .
\end{align}
These 1PN log series, with integer PN order $x^{3k+1} \log^k(x)$ and half-integer PN order 
$x^{3k+5/2} \log^k(x)$ (for $k\ge 0$), are evident.  Later in this paper we focus on yet another pair of diagonal 
sequences, the 4PN log series, which make up another subset of the flux terms in \eqref{eqn:energyfluxInf}
\begin{align}
\label{eqn:4PNenergyfluxInf}
\left\langle \frac{dE}{dt} \right\rangle^{\text{4L}} =
\frac{32}{5} \nu^2 x^5 
\biggl[ x^4 \mathcal{R}_4 
&+ x^{11/2}\mathcal{R}_{11/2}
+ x^7 \log(x)\mathcal{R}_{7L} +x^{17/2}\log(x)\mathcal{R}_{17/2L}  
\\
\notag 
&+ x^{10}\log^2(x)\mathcal{R}_{10L^2}
+x^{23/2}\log^2(x) \mathcal{R}_{23/2L^2}
+ x^{13} \log^3(x) \mathcal{R}_{13L^3}+\cdots \biggr] .
\end{align}

The average loss of angular momentum is an expansion similar to \eqref{eqn:energyfluxInf} but with a different 
Newtonian circular-orbit factor and with new flux (enhancement) functions that are referred to by 
$\mathcal{Z}_i(e_t,\nu)$ instead of $\mathcal{R}_i(e_t,\nu)$.  The analogous 1PN and 4PN log series in angular 
momentum are
\begin{align}
\label{eqn:1PNangmomfluxInf}
\left\langle \frac{dL}{dt} \right\rangle^{\text{1L}} =
\frac{32}{5} \nu^2 (m_1 + m_2) x^{7/2} 
\biggl[ x \mathcal{Z}_1 
&+ x^{5/2}\mathcal{Z}_{5/2}
+ x^4 \log(x)\mathcal{Z}_{4L} +x^{11/2}\log(x)\mathcal{Z}_{11/2L}  
\\
\notag 
&+ x^7\log^2(x)\mathcal{Z}_{7L^2}
+x^{17/2}\log^2(x) \mathcal{Z}_{17/2L^2}
+ x^{10} \log^3(x) \mathcal{Z}_{10L^3}+\cdots \biggr] ,
\end{align}
and
\begin{align}
\label{eqn:4PNangmomfluxInf}
\left\langle \frac{dL}{dt} \right\rangle^{\text{4L}} =
\frac{32}{5} \nu^2 (m_1 + m_2) x^{7/2} 
\biggl[ x^4 \mathcal{Z}_4 
&+ x^{11/2}\mathcal{Z}_{11/2}
+ x^7 \log(x)\mathcal{Z}_{7L} +x^{17/2}\log(x)\mathcal{Z}_{17/2L}  
\\
\notag 
&+ x^{10}\log^2(x)\mathcal{Z}_{10L^2}
+x^{23/2}\log^2(x) \mathcal{Z}_{23/2L^2}
+ x^{13} \log^3(x) \mathcal{Z}_{13L^3}+\cdots \biggr] .
\end{align}
\end{widetext}

\subsection{The 1PN equations of motion}

The 1PN source multipoles include the 1PN correction to the mass quadrupole moment.  Its computation requires the 1PN 
correction to the equations of motion, i.e., treatment of the two-body motion as a precessing ellipse.  The other 
two 1PN multipoles, the mass octupole and current quadrupole, need only be computed to Newtonian order.  

We take the total mass to be $M = m_1 + m_2$ and assume the motion occurs in the $x,y$ plane.  Coordinates 
$r=r(t)$ and $\vp=\vp(t)$ represent the separation distance and the azimuthal angle, respectively.  We introduce 
then the well-known quasi-Keplerian parameterization of the motion \cite{DamoDeru85, DamoScha88, 
SchaWex93, MemmGopaScha04} involving three anomalies, $u(t)$, $l(t)$, $V(u)$, three eccentricities, 
$e_t$, $e_\vp$, $e_r$, the two frequencies, $\O_r$, $\O_\vp$, and the semimajor axis, $a$.  In this 
description, $u(t)$ is the eccentric anomaly, $l(t)$ is the mean anomaly, $V(u)$ is the true anomaly, $e_\vp$ is 
the azimuthal eccentricity, $e_r$ is the radial eccentricity, and $e_t$ is the aforementioned time eccentricity.  
At the 1PN level these quantities can be related by the following equations
\begin{align}
\begin{autobreak}
\MoveEqLeft
\qquad  r= a(1 - e_r \cos u),  \qquad \vp = \bigg(\frac{\O_\vp}{\O_r}\bigg) V(u) ,
l = \O_r (t - t_P) = \frac{2 \pi}{T_r} (t - t_P) = u - e_t \sin u , 
\frac{du}{dt} = \frac{\O_r}{1 - e_t \cos u}  , \qquad \beta_\vp = \frac{1 - (1 - e_\vp^2)^{1/2}}{e_\vp} ,
V(u) = u + 2 \arctan\Big(\frac{\beta_\vp \sin u}{1 - \beta_\vp \cos u} \Big) , 
\end{autobreak}
\end{align}
where $t_P$ is the time of last periastron crossing and $V(u)$ is written in a form that preserves continuity 
across $u = 2 \pi$.  A more detailed description of these equations is given in 
\cite{DamoDeru85, MemmGopaScha04, Blan14}.

Our goal is to obtain all quantities in terms of $u$, $e_t$, and $x=(M \O_\vp)^{2/3}$ prior to transformation 
to the frequency domain.  As part of this process, $e_r$ and $e_\vp$ must be expressed in terms of $e_t$ to 1PN 
order.  We find 
\begin{align}
e_r &= e_t \, \left[ 1 + (4 - \frac{3}{2} \nu) x + \cdots \right] , 
\\
e_\vp &= e_t \, \left[ 1 + (4 - \nu) x + \cdots \right] .
\end{align}
The semimajor axis can be expressed simply in terms of the (dimensionless) energy $\ve$ \cite{Blan14} and 
$\ve$ can itself be PN expanded.  Through 1PN order these are found to be
\begin{align}
a &= \frac{M}{\ve}\left(1 + \frac{\ve}{4}(-7 + \nu)\right) ,
\\
\ve &= x + \Big(\frac{3+5 e_t^2}{4 (-1+e_t^2)}-\frac{\nu }{12} \Big) x^2 ,
\end{align}
from which we obtain the 1PN expansion of $a$
\begin{equation}
a =  \frac{M}{x} \Big(1-\frac{(1-3 e_t^2)}{1-e_t^2} x +\frac{\nu}{3} x \Big).
\end{equation}
Similarly, the radial frequency $\O_r$ can be PN expanded in straightforward fashion, 
simultaneously providing the expansion for the frequency ratio $K = \O_{\vp} / \O_r$.  We obtain
\begin{gather}
\O_r = \frac{x^{3/2}}{M} \Big(1-\frac{3 x}{1-e_t^2}\Big), \qq
 K = 1 + \frac{3 x}{1-e_t^2}.
\end{gather}

The motion in the coordinate $\vp$ combines a mean advance at the rate $\O_\vp$ and a periodic motion at the 
frequency $\O_r$.  In the Fourier expansion of gravitational wave source terms this produces a biperiodic 
expansion.  Defining the 1PN difference in the mean angular advance as $k l = (K-1) l$ and starting with 
$\vp = K V$, we separate the advance of $\vp(t)$ into parts as follows 
\begin{equation}
\vp(t) = k l + l + K(V(u) - l) .
\end{equation}
With this done, all of the previous relations can be combined to give the coordinate positions and velocities in 
terms of $x$, $u$, and $e_t$ to the desired order.  Because of the particular manifestation of velocity in the 
1PN mass quadrupole, only the lowest order in $x$ is required for the coordinate velocity components.  We obtain
\begin{widetext}
\begin{align}
\frac{r}{M} &= \frac{1-e_t \cos u}{x} - \frac{1-3 e_t^2+3 e_t \cos u- e_t^3 \cos u}{1-e_t^2} 
+\frac{1}{6}  (2+7 e_t \cos u) \, \nu \, , \\
\vp &= kl + u + 2 \arctan\left(\frac{e_t \sin u}{1+\sqrt{1-e_t^2}-e_t \cos u}\right)
+\Bigg[\frac{4 e_t \sin u}{\sqrt{1-e_t^2} (1-e_t \cos u)} +\frac{3 e_t \sin u}{1-e_t^2} 
\notag 
\\
&\qq \qq 
+\left(\frac{6}{1-e_t^2}\right)\arctan\left(\frac{e_t \sin u}{1+\sqrt{1-e_t^2}-e_t \cos u}\right)\Bigg]x 
-\frac{e_t \sin u }{\sqrt{1-e_t^2} (1-e_t \cos u)} x\nu \, ,  \\
\frac{dr}{dt} &= \frac{e_t \sin u}{1-e_t \cos u}  \, x^{1/2} , \qq
M \frac{d\vp}{dt} =\frac{\sqrt{1-e_t^2} }{(1-e_t \cos u)^2} \, x^{3/2} ,  \qq
v^2 =  \Big(\frac{1 + e_t \cos u}{1 - e_t \cos u}\Big) \, x .
\end{align}
\end{widetext}

\subsection{Review of calculating the Newtonian mass octupole and current quadrupole moments}

\subsubsection{Fourier decomposition}

We review here the calculation of the Fourier series of the mass octupole and current quadrupole.  For more details 
see \cite{ArunETC08a,ArunETC08b,ArunETC09a} and the review \cite{Blan14}.  The calculation is also a 
straightforward extension of our review of the mass quadrupole Fourier calculation presented in Paper I.  

The symmetric tracefree (STF) Newtonian mass octupole tensor is defined as
\begin{align}
I_{ijk} &= Q_{ijk} - \delta_{ij} Q_{aak}/5 - \delta_{jk}Q_{ibb}/5 - \delta_{ik} Q_{cjc}/5, \notag \\
Q_{ijk} &= \sum_{\alpha} m_{\alpha} x_i^\alpha x_j^\alpha x_k^\alpha .
\end{align}
The nonzero components are given by
\begin{align}
I_{xxx} &= \frac{\mu}{20}  \sqrt{1-4 \nu } \,r^3 (3 \cos \vp+5 \cos 3 \vp), \notag \\ 
I_{xxy} &= I_{xyx} = I_{yxx} = \frac{\mu}{20} \sqrt{1-4 \nu } \,r^3 ( \sin \vp+5 \sin 3 \vp), \notag \\ 
I_{xyy} &= I_{yxy} = I_{yyx} = \frac{\mu}{20}   \sqrt{1-4 \nu } \,r^3 ( \cos \vp - 5 \cos 3 \vp), \notag \\ 
I_{yyy} &= \frac{\mu}{20} \sqrt{1-4 \nu } \,r^3 (3 \sin \vp - 5 \sin 3 \vp), \notag \\ 
I_{xzz} &= I_{zxz} = I_{zzx} = -\frac{\mu}{5} \sqrt{1-4 \nu } \,r^3 \cos \vp, \notag \\ 
I_{yzz} &= I_{zyz} = I_{zzy} = -\frac{\mu}{5} \sqrt{1-4 \nu } \,r^3 \sin \vp ,
\end{align} 
where $\mu$ is the reduced mass.

Similarly, the STF form of the current quadrupole is given by
\begin{align}
J_{ij} &= \frac{1}{2} \sum_{\alpha} m_{\alpha} 
\left[ x_i (\vec{x} \times \vec{v})_j + x_j (\vec{x} \times \vec{v})_i \right] , 
\notag 
\\
J_{xz} &= J_{zx} = \frac{1}{2} \mu \sqrt{1- 4 \nu} \cos(\vp) r^3 \frac{d\vp}{dt} , \notag \\
J_{yz} &= J_{zy} = \frac{1}{2} \mu \sqrt{1- 4 \nu} \sin(\vp) r^3 \frac{d\vp}{dt} .
\end{align}

The transformation is now made from $(r,\vp)$ to variables $(x, \nu, e_t, u)$ using the relations of the previous 
section.  Without listing every component, we find for example
\begin{align}
I_{xxx} &= \frac{M^3 \mu\sqrt{1-4 \nu} }{x^3} \Big[(e_t-\cos u)^3 \notag \\ & \qq \qq \q
 - \frac{3}{5}(e_t - \cos u) (1 - e_t \cos u)^2\Big], \\
J_{xz} &= - \frac{M^2 \mu \sqrt{1-4 \nu } }{x^{3/2}}  \Big(\frac{\sqrt{1-e_t^2}}{2} \Big)  (e_t - \cos u) ,
\end{align}
with obvious extension to the other tensor components.  There is no difference between the time eccentricity 
and the Keplerian eccentricity at Newtonian order, but we use the notation $e_t$ uniformly to prepare for more 
general expansions.  This also allows us to reserve the symbol $e$ for the relativistic Darwin eccentricity.  

In each multipole component, the scale and dimension are carried by the initial prefactor.  Since we are concerned 
with the dimensionless eccentricity enhancement functions that will appear in the fluxes, we remove these factors 
now and define
\be
\label{eqn:dimMOCQ}
\hat{I}_{ijk} = \frac{x^3}{M^3 \mu \sqrt{1-4 \nu }} I_{ijk}, \q
\hat{J}_{ij} = \frac{x^{3/2}}{M^2 \mu \sqrt{1-4 \nu }} J_{ij}. 
\ee
It is then these scaled multipole moment tensors that we represent with Fourier series
\begin{align}
\label{eqn:fourier}
\hat{I}_{ijk} &= \sum_{n=-\infty}^{n=\infty} \,\, \underset{\!\!\!\!\!\!(n)}{\hat{I}_{ijk}} e^{i n l}, 
\q \, \underset{\!\!\!\!\!\!(n)}{\hat{I}_{ijk}} = \frac{1}{2\pi} \int_{0}^{2 \pi} \hat{I}_{ijk} e^{-i n l} dl, 
\end{align}
with a similar expression for $\hat{J}_{ij}$.  As mentioned in Paper I, the Fourier components are most easily 
evaluated as integrals over $u$.  For instance,
\begin{align}
\underset{\!\!\!\!\!\!(n)}{\hat{I}_{ijk}} 
= \frac{1}{2\pi} \int_{0}^{2\pi}\hat{I}_{ijk} \, e^{-in(u-e_t \sin u)}(1-e_t \cos u ) \, du . \notag
\end{align}
Then, a closed-form expression can be obtained through multiple applications of the Bessel integral formula
\begin{equation}
J_p (x) = \frac{1}{2\pi} \int_{0}^{2\pi} e^{-ipu + x \sin u } \, du  .
\end{equation}
We find the following expressions for the mass octupole moment components
\begin{widetext}
\begin{align}
\underset{\!\!\!\!\!\!(n)}{\hat{I}_{xxx}}  
& =  -\frac{3 (3 - 4 e_t^2 + e_t^4)}{e_t^3 n^2} J_n(n e_t)
+\frac{3 (10 - 6 e_t^2 + 5(1-e_t^2)^2 n^2)}{5 e_t^2 n^3} J_n'(n e_t) , 
\notag 
\\
\underset{\!\!\!\!\!\!(n)}{\hat{I}_{xxy}} 
&= -\frac{3 i \sqrt{1-e_t^2} (2 (5 - e_t^2) + 5 (1 - e_t^2)^2 n^2)}{5 e_t^3 n^3} 
J_n(n e_t)  +  \frac{i \sqrt{1-e_t^2} (9 - 5 e_t^2) }{e_t^2 n^2}  J_n'(n e_t) ,
\notag 
\\
\underset{\!\!\!\!\!\!(n)}{\hat{I}_{xyy}}  
&=  \frac{9-13 e_t^2+4 e_t^4 }{e_t^3 n^2} J_n(n e_t)
 - \frac{3 (10 - 8 e_t^2 + (1-e_t^2)^2 n^2)}{5 e_t^2 n^3} J_n'(n e_t) , 
\notag 
\\
\underset{\!\!\!\!\!\!(n)}{\hat{I}_{yyy}}  
&= \frac{3 i \sqrt{1-e_t^2} (10 - 4 e_t^2 + 5 (1 - e_t^2)^2 n^2)}{5 e_t^3 n^3}  J_n(n e_t) 
- \frac{3 i \sqrt{1-e_t^2} (3 - 2 e_t^2)}{e_t^2 n^2} J_n'(n e_t) , 
\notag 
\\
\underset{\!\!\!\!\!\!(n)}{\hat{I}_{xzz}}  
&= \frac{1 - e_t^2}{e_t n^2} J_n(n e_t) - \frac{6}{5 n^3}  J_n'(n e_t), \qq \qq
\underset{\!\!\!\!\!\!(n)}{\hat{I}_{yzz}} 
= \frac{6i \sqrt{1 - e_t^2}}{5 e_t n^3} J_n(n e_t) - \frac{i \sqrt{1 - e_t^2}}{n^2}  J_n'(n e_t) ,
\end{align}
and the following for the current quadrupole moment components
\begin{align}
\underset{\!\!\!\!(n)}{\hat{J}_{xz}}  = -\frac{1}{2 n} \sqrt{1-e_t^2} J_n'(n e_t) , 
\qq \qq
\underset{\!\!\!\!(n)}{\hat{J}_{yz}}  = - \frac{i}{2 e_t n}  (1 - e_t^2) J_n(n e_t) .
\end{align}
\end{widetext}
We successively applied the well known Bessel function identities
\begin{gather}
J_{n+1}(n e_t)=-\frac{n J_n'(n e_t)}{n}+\frac{J_n(n e_t)}{e_t}, \notag \\
\qquad J_{n-1}(n e_t)=-J_{n+1}(n e_t)+\frac{2 J_n(n e_t)}{e_t}, 
\end{gather}
in order to simplify the above expressions for the components of the multipoles (see also 
\cite{LoutYune17,MunnEvan19a}).

\subsubsection{Partial flux functions}
\label{sec:MOCQamp}

To derive the 1PN log series, the Fourier amplitudes of the two multipoles given above are not used directly but 
rather go into forming a pair of (flux) spectral functions.  This is similar to the derivation of the Newtonian 
(Peters-Mathews \cite{PeteMath63,Pete64}) energy flux term $f(e_t)$, which was called $\mathcal{R}_0(e_t)$ in 
Paper I.  In that case a quadrupole Fourier spectrum $g(n,e_t) = (1/16) n^6 |{}_{(n)}\hat{I}_{ij} |^2$ is derived 
from the complex square of the Newtonian quadrupole Fourier amplitudes.  The function $g(n,e_t)$ was derived 
by Peters and Mathews \cite{PeteMath63} (with a correction to their printed expression pointed out by 
\cite{BlanScha93}).  The power spectrum then produces $\mathcal{R}_0(e_t)$ as the direct sum over the harmonics 
\be 
\mathcal{R}_0(e_t) = \sum_{n=0}^{\infty} g(n,e_t) 
= \frac{1}{(1-e_t^2)^{7/2}}\bigg(1+\frac{73}{24}e_t^2 + \frac{37}{96}e_t^4\bigg) .
\ee
In Paper I we showed that $g(n,e_t)$ (and its angular momentum counterpart $\tilde{g}(n,e_t)$) could generate the 
entire leading log series through sums of $g(n,e_t)$ over different powers of $n$.  Here we show that spectral 
functions similar to $g(n,e_t)$ are formed from complex squares of the mass octupole (MO) and current quadrupole 
(CQ) Fourier amplitudes.  Then, later in the paper, these spectral functions are shown to generate part of, but not 
all of, the various 1PN log series terms. 

The Fourier amplitudes of $\hat{I}_{ijk}$ and $\hat{J}_{ij}$ each contribute to both the energy flux and the 
angular momentum flux.  Calculation of all four pieces follows in close analogy to that of the mass quadrupole 
as reviewed in Paper I.  The corresponding lowest order energy and angular momentum fluxes are written as 
\cite{Papa62,CoopBoot69,Papa71,Blan14}
\begin{align}
\left<\frac{dE}{dt}\right>_1^{\text{MO}} &= \frac{1}{189}\left<\ddddot{I}_{ijk}\ddddot{I}_{ijk}\right>, \\
\left<\frac{dL}{dt}\right>_1^{\text{MO}} &=\frac{1}{63}\e_{ijl} \hat{L}_i \left<\dddot{I}_{jab}\ddddot{I}_{lab}\right>,  \\
\left<\frac{dE}{dt}\right>_1^{\text{CQ}} &= \frac{16}{45}\left<\dddot{J}_{ij}\dddot{J}_{ij}\right>,  \\
\left<\frac{dL}{dt}\right>_1^{\text{CQ}} &=\frac{32}{45}\e_{ijl} \hat{L}_i \left<\ddot{J}_{ja}\dddot{J}_{la}\right>,
\end{align} 
where angled brackets denote the time average over an orbital period, the subscript $1$ indicates these are 
contributions to the 1PN fluxes, and $\hat{L}_i$ is the unit vector in the direction of the angular momentum 
vector (which we take to be in the $z$ direction).  To compute these 1PN fluxes the mass octupole and current 
quadrupole moments need only be calculated at Newtonian order.

Inserting the Fourier expansions, integrating, and pulling out the Newtonian circular orbit limit and added 
power of $x$ for a 1PN term (see \eqref{eqn:energyfluxInf}), we obtain the following functions as 
analogs of $g(n,e_t)$ and $\tilde{g}(n,e_t)$:
\begin{widetext}
\begin{align} 
\begin{autobreak}
\MoveEqLeft
h(n,e_t) = \frac{5}{3024} n^8 | \underset{\!\!\!\!\!\!(n)}{\hat{I}_{ijk}}  |^2  = \frac{n^2 (1-e_t^2)}{504 e_t^6} 
\bigg[12 (10-5 e_t^2+e_t^4)+5 n^2 (78-153 e_t^2+91 e_t^4-16 e_t^6) 
+30 n^4 (1-e_t^2)^4 \bigg] J_n(n e_t)^2 
+ \frac{n^2}{504 e_t^4} \bigg[12 (10-15 e_t^2+6 e_t^4)
+5 (78-183 e_t^2+142 e_t^4-37 e_t^6)
 n^2+30 (1-e_t^2)^4 n^4\bigg] J_n'(n e_t)^2 
  -  \frac{5 n^3}{14 e_t^5} \bigg[(2-3 e_t^2+e_t^4) 
(2-e_t^2+(1-e_t^2)^2 n^2) \bigg]J_n(n e_t) J_n'(n e_t)  ,
 \end{autobreak} \\
\begin{autobreak}
\MoveEqLeft
\tilde{h}(n,e_t) = -\frac{5i}{1008} n^7 \epsilon_{ijl} \hat{L}_i \underset{\!\!\!\!\!\!(n)}{\hat{I}_{jab}}
 \underset{\!\!\!\!\!\!(n)}{\hat{I}_{lab}^*} =
 \frac{5 (1-e_t^2)^{3/2} n^2}{168 e_t^6} 
 \bigg[15 e_t^6 n^2-36 (2+n^2)-2 e_t^4 (4+33 n^2)
 +e_t^2 (48+87 n^2)\bigg] J_n(n e_t)^2
+\frac{5 \sqrt{1-e_t^2} n^2}{168 e_t^4}\bigg[-8 (3-2 e_t^2)^2
+3 (1-e_t^2)^2 (-12+7 e_t^2) n^2 \bigg]  J_n'(n e_t)^2
+\frac{n \sqrt{1-e_t^2} }{126 e_t^5} \bigg[36 (5-5 e_t^2+e_t^4) + 5 (1-e_t^2) (-3+2 e_t^2) (-39+19 e_t^2)
n^2+45 (1-e_t^2)^4 n^4\bigg]
 J_n(n e_t) J_n'(n e_t),
\end{autobreak} \\
\begin{autobreak}
\MoveEqLeft
k(n,e_t) =  \frac{1}{9} n^6 | \underset{\!\!\!\!(n)}{\hat{J}_{ij}} |^2 =
 \frac{1}{18 e_t^2}(1-e_t^2)^2 n^4 J_n(n e_t)^2 +
\frac{1}{18} (1-e_t^2) n^4 J_n'(n e_t)^2,
\end{autobreak} \\
\begin{autobreak}
\MoveEqLeft
\tilde{k}(n,e_t) = -\frac{2i}{9} n^5 \epsilon_{ijl} \hat{L}_i  \underset{\!\!\!\!(n)}{\hat{J}_{ja}} 
 \underset{\!\!\!\!(n)}{\hat{J}_{la}^*} =
 \frac{(1-e_t^2)^{3/2}}{9 e_t} n^3 J_n(n e_t) J_n'(n e_t).
\end{autobreak}
\end{align}
\end{widetext}
Contributions can then be found to the full 1PN energy and angular momentum fluxes, for example, by summing each
of these expressions over $n$.  To focus on one particular example, the mass octupole contribution to the energy 
flux is found by calculating
\begin{align}
\label{eqn:R1MO}
&\mathcal{R}_1^{\text{MO}}(e_t) = (1 - 4\nu) \sum_{n=0}^{\infty} h(n, e_t)   \\ &
= \frac{1 - 4\nu}{(1 - e_t^2)^{9/2}}\left(\frac{1367}{1008}+\frac{18509 e_t^2}{2016}+\frac{2395 e_t^4}{384}
+\frac{1697 e_t^6}{5376} \right) .  \notag
\end{align}
Additional explicit expansions for component sums like this one are given in Appendix \ref{sec:compSums}.  We note 
that this term in the flux became a simple closed form expression once the specific eccentricity singular factor 
was pulled out.  This particular eccentricity singular factor bears an extra power of $(1-e_t^2)^{-1}$ over that 
found in $\mathcal{R}_0$.  Clearly, this mass octupole contribution to the energy flux is not the entirety of the 1PN 
flux, as can be seen by examining equation (356b) of \cite{Blan14}.

\subsection{The 1PN mass quadrupole}
\label{sec:1PNmassquad}

The next step is to find the 1PN correction to the mass quadrupole.  Fourier decomposition at 1PN order presents a 
considerable increase in difficulty.  The motion no longer closes, which implies that the simple Fourier series, as 
found in the expansion of the mass octupole and current quadrupole, must be replaced by a double Fourier sum over 
harmonics of the two frequencies, $\O_r$ and $\O_\vp$.  This Fourier structure, first identified by 
\cite{Damo83, GopaIyer02, DamoGopaIyer04}, was laid out for use with hereditary contributions to the flux by 
Arun \textit{et al.} in \cite{ArunETC08a}.  

We follow some of the procedure and notation found in Loutrel and Yunes \cite{LoutYune17}, who provided a detailed 
derivation of the 1PN expansion as part of their work.  The expression for the components of the mass quadrupole 
tensor at 1PN order is
\begin{align}
I_{ij} &= \mu \bigg[ \Big(1+v^2 \Big(\frac{29}{42}-\frac{29 \nu }{14}\Big) - \frac{M}{r} 
\Big(\frac{5}{7} - \frac{8 \nu }{7}\Big)\Big)  x_{<i}x_{j>}   \notag \\ &
+ \Big(\frac{11}{21}-\frac{11 \nu }{7}\Big) r^2 v_{<i}v_{j>} - 
\Big(\frac{4}{7} - \frac{12 \nu }{7}\Big) r r' x_{<i}v_{j>} \bigg],
\end{align}
where bracketed indices denote STF projection \cite{ArunETC08a}.  Given the 1PN equations of motion, this tensor 
is converted from polar coordinates to the parameters $x$, $\nu$, $e_t$, $u$, and $k$ through 1PN order.  At the 
same time a factor $\mu M^2/x^2$ is pulled out of $I_{ij}$ to provide a dimensionless quadrupole moment tensor
\be
\hat{I}_{ij} = \frac{x^2}{\mu M^2} I_{ij} ,
\ee 
similar to what we did with $I_{ijk}$ and $J_{ij}$.

To obtain the Fourier expansion, the $u$ dependence of $\hat{I}_{ij}$ is expressed in terms of complex exponentials 
and the result is collected over powers of $e^{ikl}$.  The coefficient of each power of $e^{ikl}$ is singly periodic 
in $t$, meaning that each can themselves be expressed as a simple Fourier series.  The entire tensor can then be 
written as
\be
\label{eqn:biFour}
\hat{I}_{ij}(t) = \sum_{n=-\infty}^{\infty} \sum_{p=-2,0,2} \underset{(n,p)}{\hat{I}_{ij}}  e^{i (n + pk) l},
\ee
where the $k$-dependence has introduced a magnetic-type separation of components due to 1PN differences in 
$\O_\vp$ and $\O_r$.  The goal is then to determine the Fourier coefficients ${}_{(n,p)}\hat{I}_{ij}$. 

Proceeding further, we find that the (magnetic) term for each $p$ can be written as the product of a single 
function (e.g., one of the components) with a constant matrix.  Explicitly,
\be
\underset{(p)}{\hat{I}_{ij}} = \underset{(p)}{\hat{I}_{xx}} \underset{(p)}{M_{ij}} ,
\ee
where
\begin{widetext}
\begin{align}
	\renewcommand{\arraystretch}{1.3}
\underset{(2)}{M_{ij}} = \begin{bmatrix} 1 & -i & 0  \\ -i & -1 & 0 \\ 0 & 0 & 0 \end{bmatrix}, \qq
	\renewcommand{\arraystretch}{1.1}
\underset{(0)}{M_{ij}} = \begin{bmatrix} 1 & 0 & 0 \\ 0 & 1 & 0 \\ 0 & 0 & -2 \end{bmatrix}, \qq
\underset{(-2)}{M_{ij}} = \begin{bmatrix} 1 & i & 0 \\ i & -1 & 0 \\ 0 & 0 & 0 \end{bmatrix}.
\end{align}

It is most convenient to separate each ${}_{(n,p)}\hat{I}_{ij}$ on powers of $x$ and $\nu$ before making the 
Fourier transformation.  To facilitate the process, we introduce a superscript notation, $\hat{I}_{ij}^{ab}$, where 
$a$ represents the order in $x$ (0 or 1) and $b$ the order in $\nu$ (also 0 or 1).  Then each Fourier coefficient 
will formally separate into
\be
\underset{(n,p)}{\hat{I}_{ij}}  
= \underset{(n,p)}{\hat{I}_{ij}^{00}} 
+ x \Big( \underset{(n,p)}{\hat{I}_{ij}^{10}} 
+ \nu  \underset{(n,p)}{\hat{I}_{ij}^{11}} \Big) .
\ee
Since we will just need to compute the $\hat{I}_{xx}^{ab}$ functions, as the full tensors will be determined from 
these functions via multiplication by ${}_{(p)}M_{ij}$, we can drop the lower $ij$ indices, leaving $\hat{I}^{ab}$, 
to simplify the notation. 

Starting at lowest order, the Fourier components are found to be
\begin{align}
\underset{(n,\pm 2)}{\hat{I}^{00}} & =\frac{1}{2 e_t^2 n^2}\Big(e_t^2-2 \pm 2 n (1-e_t^2)^{3/2}\Big) J_n(n e_t)+
\frac{\sqrt{1-e_t^2}}{e_t n^2}\Big(1 \mp n \sqrt{1-e_t^2} \Big) J_n'(n e_t), \quad 
\underset{(n,0)}{\hat{I}^{00}} =-\frac{J_n(n e_t)}{3 n^2} ,
\end{align}
which are precisely the terms needed to generate $g(n,e_t)$ and reproduce the Peters-Mathews flux.  Then we jump 
to next order in both $x$ and $\nu$, and find that these coefficients can also be expressed cleanly in terms of 
Bessel functions
\begin{align}
\underset{(n,\pm 2)}{\hat{I}^{11}} 
&=-\frac{1}{84 e_t^2 n^2} \Big(134+17 e_t^2 \mp 146 n \sqrt{1-e_t^2} 
\mp 22 n e_t^2 \sqrt{1-e_t^2} +12 (1-e_t^2)^2 n^2 \Big) J_n(n e_t) 
\notag 
\\
&+ \frac{1}{42 e_t (1-e_t^2) n^2} 
\Big(\pm(67-25 e_t^2) \sqrt{1-e_t^2} \pm 6 n^2 (1-e_t^2)^{5/2} - n (73-65 e_t^2-8 e_t^4)\Big) J_n'(n e_t), 
\notag 
\\
\underset{(n,0)}{\hat{I}^{11}} &=  \frac{17 J_n(n e_t)}{126 n^2}-\frac{e_t J_n'(n e_t)}{21 n} .
\end{align}

Finally, we arrive at the portion that is first order in $x$ and zeroth order in $\nu$.  Here some difficulty 
arises, as the integrals for ${}_{(n,\pm 2)}{\hat{I}^{10}}$ have terms that apparently cannot be expressed in 
closed form.  We find
\begin{align}
\begin{autobreak}
\label{eqn:I1PNint}
\MoveEqLeft
\underset{(n,\pm 2)}{\hat{I}^{10}} =  \frac{1}{84 e_t^2 (1-e_t^2) n^3}
\bigg[\mp 756 (-2+e_t^2) +4 n^3(1 - e_t^2)^3 - 3 n 
\Big(-74+19 e_t^4+ \sqrt{1-e_t^2} (756 - 420 e_t^2) +111 e_t^2 \Big)
\pm 2 n^2 (1 - e_t^2) \Big(378 - \sqrt{1-e_t^2} (113 - 22 e_t^2) - 378 e_t^2\Big)\bigg] J_n(n e_t)

- \frac{1}{42 e_t (1 - e_t^2) n^3}\bigg[-756 \sqrt{1-e_t^2}
\mp 3n \Big(-378+37 (1-e_t^2)^{3/2} + 273 e_t^2 \Big)
- n^2 (1 - e_t^2) \Big(-113+23 e_t^2+378 \sqrt{1-e_t^2}\Big) \mp 2 n^3 (1-e_t^2)^{5/2} \bigg] J_n'(n e_t)

- \frac{3i}{16 (1 - e_t^2) \pi} \int_0^{2 \pi} e^{-i n (u - e_t \sin u)} (1 - e_t \cos u) 
\bigg[\pm 3 e_t^2 \mp 4 e_t \cos u \mp (-2+e_t^2) \cos 2 u
\mkern270mu - 4 i \sqrt{1-e_t^2} (e_t-\cos u) \sin u \bigg]
\arctan \bigg(\frac{e_t \sin u}{1+\sqrt{1-e_t^2}-e_t \cos u}\bigg) du
\end{autobreak} , 
\notag \\
\begin{autobreak}
\MoveEqLeft
\underset{(n,0)}{\hat{I}^{10}} =  - \frac{(75-19 e_t^2) J_n(n e_t)}{42 (1 - e_t^2) n^2}+
\frac{26 e_t n J_n'(n e_t)}{21 n^2} .
\end{autobreak}
\end{align}
Note that these results directly reveal the crossing relations, ${}_{(n,p)}{I^{*}} = {}_{(-n,-p)}{I}$.

When computing eccentricity enhancement functions, these unevaluated integrals must be expanded in $e_t$ before 
proceeding.  One might presume that this precludes the possibility of eventually finding closed form expressions 
in the fluxes, but surprisingly this is not the case.  Instead, in the case of certain flux terms, once the 
appropriate eccentricity singular factor is pulled out, we find that parts of the expansion of 
${}_{(n,p)}{\hat{I}^{10}}$ and of ${}_{(n,p)}{\hat{I}^{00}}$ conspire perfectly to cancel all coefficients beyond 
certain orders in $e_t^2$ in the remaining power series.

\subsection{Discussion}
\label{sec:fourSumDisc}

The mass octupole and current quadrupole power spectra $h(n,e_t), \tilde{h}(n,e_t), k(n,e_t), \tilde{k}(n,e_t)$,
along with the Fourier decomposition of the 1PN-corrected mass quadrupole (MQ), will be shown to generate the
entire 1PN log series.  In order to more clearly explain the calculation of each flux term, we introduce the notation 
$\mathcal{R}_i = \mathcal{R}_i^{\rm MQ} + \mathcal{R}_i^{\rm MO} + \mathcal{R}_i^{\rm CQ}$ (with a similar form for 
$\mathcal{Z}$) to represent the contributions from the (1PN) mass quadrupole, mass octupole, and current quadrupole, 
respectively.  For the latter two (Newtonian) multipole moments, this categorization will be sufficient, 
as the spectral functions presented in Sec.~\ref{sec:MOCQamp} compactly express the entirety of
those contributions to the 1PN logarithms.  Note also that in $\mathcal{R}_i^{\rm MO}$ and $\mathcal{R}_i^{\rm CQ}$ 
the $\mathcal{O}(\nu)$ contributions will be immediately accessible through the $\nu$-dependent prefactors in 
\eqref{eqn:dimMOCQ}. 

Unfortunately, the 1PN mass quadrupole contribution is not encoded in a single spectral function and so instead we 
work directly with the Fourier components introduced in the last section.  The dependence on the two orders in 
$\nu$ is more subtle also.  In what follows we are led to separate the relative flux $\mathcal{R}_i^{\rm MQ}$ into 
five terms
\be
\label{eqn:MQ5terms}
\mathcal{R}_{i}^{\rm MQ} = \mathcal{R}_{i}^{\rm MQ0} + \nu \mathcal{R}_{i}^{\rm MQ1} 
= \mathcal{R}_{i}^{\rm MQ01} + \mathcal{R}_{i}^{\rm MQ02} +
\mathcal{R}_{i}^{\rm MQ03} + \nu \big(\mathcal{R}_{i}^{\rm MQ11} + \mathcal{R}_{i}^{\rm MQ12} \big) ,
\ee 
distinguished by the $a,b$ superscripts in $\mathcal{R}_{i}^{\text{MQ}ab}$.  The $a$ represents the relative order 
in $\nu$ and $b \in \{1,2,3\}$ represents a particular ``type'' of summation over different parts in 
the decomposition of $\hat{I}_{ij}$ (see the next section for explicit examples).  This notation for separating the 
relative flux functions carries over to the corresponding absolute flux, e.g.,
\be
\left<\frac{dE}{dt}\right>_1^{\rm MQ03}
\ee
represents the mass quadrupole contribution of the ``3-type" summation to the full 1PN (subscript) flux at lowest 
order in $\nu$ (0 superscript).

\section{Recovering the 1PN and 2.5PN relative fluxes: first elements in the 1PN log sequences}
\label{sec:knownLogs}

Using the frequency-domain tools developed above, this section demonstrates the recovery of the previously-known 
first elements in the 1PN log sequences---namely the instantaneous 1PN fluxes $\mathcal{R}_{1}(e_t)$ and 
$\mathcal{Z}_{1}(e_t)$ and the hereditary 2.5PN tail fluxes $\mathcal{R}_{5/2}(e_t)$ and $\mathcal{Z}_{5/2}(e_t)$. 

\subsection{The full mass octupole and current quadrupole relative flux contributions}

The contributions from the spectra of the two Newtonian-order moments are intuitive in form and mirror the way 
$g(n,e_t)$ contributed to the leading logarithms (see the discussion in Paper I).  We examine first the 1PN fluxes. 
These enhancement functions have been known from PN analysis for some time and, since they are entirely instantaneous 
in nature, are easily calculated through time domain methods \cite{WagoWill76,BlanScha89}.  Here we give for the 
first time (as far as we know) their calculation via frequency domain analysis.  The mass octupole and current 
quadrupole contributions to energy flux are trivial in this approach, and are simply given by sums over $h(n,e_t)$ 
and $k(n,e_t)$ 
\be 
\mathcal{R}_1^{MO} = (1 - 4 \nu) \sum_{n=1}^{\infty} h(n, e_t),  \qq  \qq
\mathcal{R}_1^{CQ} = (1 - 4 \nu) \sum_{n=1}^{\infty} k(n, e_t). 
\ee
Similarly the angular momentum terms are found by substituting the use of $\tilde{h}(n,e_t)$ and 
$\tilde{k}(n,e_t)$ 
\be 
\mathcal{Z}_1^{MO} = (1 - 4 \nu) \sum_{n=1}^{\infty} \tilde{h}(n, e_t),  \qq  \qq
\mathcal{Z}_1^{CQ} = (1 - 4 \nu) \sum_{n=1}^{\infty} \tilde{k}(n, e_t). 
\ee

The 2.5PN tail functions require a bit more work \cite{ArunETC08a, LoutYune17} as these hereditary terms do not 
lend themselves to a time domain approach.  The results, though, follow exactly what one would expect from 
Newtonian-order moments based on the analysis found in Paper I (see that paper for a review of the construction 
of the 1.5PN tail flux from an analogous sum over $g(n,e_t)$).  We find 
\begin{align}
\mathcal{R}_{5/2}^{MO} = 2\pi (1 - 4 \nu) \sum_{n=1}^{\infty} n \, h(n, e_t),  \qq  \qq
\mathcal{R}_{5/2}^{CQ} = 2\pi (1 - 4 \nu) \sum_{n=1}^{\infty} n \, k(n, e_t),  
\\
\mathcal{Z}_{5/2}^{MO} = 2\pi (1 - 4 \nu) \sum_{n=1}^{\infty} n \, \tilde{h}(n, e_t),  \qq  \qq
\mathcal{Z}_{5/2}^{CQ} = 2\pi (1 - 4 \nu) \sum_{n=1}^{\infty} n \, \tilde{k}(n, e_t). 
\end{align}

\subsection{Relative flux contributions from the mass quadrupole at lowest order in $\nu$}
\label{sec:MQManip}

As discussed in Sec.~\ref{sec:1PNmassquad}, the contribution to the fluxes from the mass quadrupole, calculated 
through 1PN order, is more involved and is best split into parts.  A significant part of the split involves 
considering the two orders in $\nu$ separately.  This subsection focuses only on the flux terms at lowest order 
in $\nu$, which in the previously defined notation means
\be 
\mathcal{R}_{i}^{\rm MQ0} = \mathcal{R}_{i}^{\rm MQ01} + \mathcal{R}_{i}^{\rm MQ02} +
\mathcal{R}_{i}^{\rm MQ03}, 
\qq \qq  
\mathcal{Z}_{i}^{\rm MQ0} = \mathcal{Z}_{i}^{\rm MQ01} 
+ \mathcal{Z}_{i}^{\rm MQ02} +  \mathcal{Z}_{i}^{\rm MQ03} .    
\ee  
(The next subsection will handle the next order in $\nu$ terms.)  These expressions add up the three summation types 
and $i$ refers to $1$ or $5/2$ order.  Subject to this split, we show in this subsection the contributions from the 
1PN-corrected mass quadrupole to all of the 1PN and 2.5PN relative flux terms.

\subsubsection{Contributions to the 1PN relative energy flux from the mass quadrupole at lowest order in $\nu$}

The 1PN mass quadrupole energy flux follows from retaining 1PN corrections to the well-known quadrupole formula
\begin{align}
\Big<\frac{dE}{dt}\Big>^{\rm MQ} &= \frac{1}{5}\Big<\dddot{I}_{ij}\dddot{I}_{ij}\Big>_1
 = \frac{1}{5} \bigg\langle \sum_{n,m=-\infty}^{\infty} \sum_{p,s=-2}^{2} (\O_r)^6
 \, \left(i(n+pk) \right)^3 \, \left(i(m+sk) \right)^3 
\underset{(n,p)}{I_{ij}}  \underset{(m,s)}{I_{ij}} e^{i(n+m+(p+s)k)l}\bigg\rangle_1 ,
\end{align}
which is here converted in the second equality from the time domain to the frequency domain.  All of the terms on 
the right hand side must be expanded and retained through 1PN order including the quadrupole moment $I_{ij}$, the 
frequency, the polynomial terms, and the exponential factor.  We recall the 1PN expansions for $\O_r$ and $k$, 
given by
\begin{align}
k= \frac{3x}{1-e_t^2}+\mathcal{O}(x^2) \qquad \qquad
 \O_r=\frac{x^{3/2}}{M}\Big(1-\frac{3x}{1-e_t^2}\Big)+\mathcal{O}(x^{7/2}) 
= \O_{\varphi} \Big(1-\frac{3x}{1-e_t^2}\Big)+\mathcal{O}(x^{7/2}) . 
\end{align}
Expanding some of the terms and retaining factors linear in $k$ and $x$, we can write this summation as
\begin{align}
\label{eqn:MQfullsum}
\Big<\frac{dE}{dt}\Big>^{\rm MQ} = - \frac{1}{5}  \sum_{n,m=-\infty}^{\infty} \sum_{p,s=-2}^{2} (\O_r)^6
 \, \left(m^3 n^3 + 3m^2 n^2 (n s + m p) k \right)
\underset{(n,p)}{I_{ij}}  \underset{(m,s)}{I_{ij}} \bigg\langle e^{i(n+m+(p+s)k)l}\bigg\rangle.
\end{align}

Next we expand the time average.  The deficit in the frequency ratio, $k$, is a small quantity, so the integrand
in the integral for the time average can be expanded about $k=0$
\begin{align}
\bigg\langle e^{i(n+m+(p+s)k)l}\bigg\rangle = \int_{0}^{2 \pi} \frac{e^{i(n+m+(p+s)k)l}}{2 \pi} dl
\simeq \int_{0}^{2 \pi} \frac{e^{i(n+m)l}}{2 \pi} \Big(1+i(p+s)k l\Big) \, dl .
\end{align}
This leads to two cases.  If $m \neq -n$, the lowest order term ($k=0$) vanishes, leaving
\begin{align}
\int_{0}^{2 \pi} \frac{e^{i(n+m)l}}{2 \pi} \Big(i(p+s)k l\Big) dl = \frac{p+s}{n+m}k .
\end{align}
On the other hand, when $m=-n$ we find
\begin{align}
\int_{0}^{2 \pi} \frac{1+i(p+s)k l}{2 \pi} dl = 1 + i \pi (p+s) k .
\end{align}

However, it turns out that when these averages are inserted in the full sums in \eqref{eqn:MQfullsum} the linear 
in $k$ parts vanish in both cases.  To see this, consider the matrices ${}_{(p)}M_{ij}$.  Direct calculation shows 
that the sum ${}_{(p)}M_{ij} \, {}_{(s)}M_{ij}$ vanishes whenever $p+s \ne 0$.  Therefore,
\begin{align}
\label{eqn:pssum}
(p+s) \underset{(n,p)}{I_{ij}}  \underset{(m,s)}{I_{ij}} = 0 ,
\end{align}
which is precisely the form of the terms produced when the two linear-in-$k$ terms above are inserted in 
\eqref{eqn:MQfullsum}.  This identity turns out to have strong consequences on the calculation of 1PN log series 
fluxes (see Appendix \ref{sec:fourSumId} for details).  Here the result is that the 1PN time average reduces to 
the simple Kronecker delta, $\delta_{m,-n}$, leaving  
\begin{align}
\Big<\frac{dE}{dt}\Big>^{\rm MQ} = 
\frac{1}{5}  \sum_{n=-\infty}^{\infty} \sum_{p,s=-2}^{2} (\O_r)^6
 \left(n^6 + 3n^5 (p - s) k \right) \underset{(n,p)}{I_{ij}}  \underset{(-n,s)}{I_{ij}} .
\end{align}
The other consequence of \eqref{eqn:pssum} is that only elements in the double sum with $s = -p$ will survive, 
so that
\begin{align}
\label{eqn:1PNpresplit}
\Big<\frac{dE}{dt}\Big>^{\rm MQ} = 
\frac{1}{5}  \sum_{n=-\infty}^{\infty} \sum_{p=-2}^{2} (\O_r)^6
 (n^6 + 6 n^5 p k) \underset{(n,p)}{I_{ij}}  \underset{(-n,-p)}{I_{ij}} .
\end{align}

We then make a PN expansion of \eqref{eqn:1PNpresplit}, combining expansions for the moments, the frequency, and 
the polynomial factor.  Once the Newtonian order flux is discarded, the remainder is the 1PN mass quadrupole 
flux, which we split into three sums 
\begin{align}
\Big<\frac{dE}{dt}\Big>_1^{\rm MQ01} 
&= \frac{x}{5}  \sum_{n=-\infty}^{\infty} \sum_{p=-2}^{2} (\O_{\varphi})^6 n^6 
\bigg[\underset{(n,p)}{I_{ij}^{00}}  \underset{(-n,-p)}{I_{ij}^{10}} 
+ \underset{(n,p)}{I_{ij}^{10}}  \underset{(-n,-p)}{I_{ij}^{00}} \bigg], \notag \\
\Big<\frac{dE}{dt}\Big>_1^{\rm MQ02} 
 &= -\frac{1}{5}  \sum_{n=-\infty}^{\infty} \sum_{p=-2}^{2} \Big( \frac{18x}{1-e_t^2} \Big) (\O_{\varphi})^6
 (n^6) \underset{(n,p)}{I_{ij}^{00}}  \underset{(-n,-p)}{I_{ij}^{00}},  \notag \\
\Big<\frac{dE}{dt}\Big>_1^{\rm MQ03} 
&= \frac{1}{5}  \sum_{n=-\infty}^{\infty} \sum_{p=-2}^{2} (\O_{\varphi})^6
 (6n^5 p k) \underset{(n,p)}{I_{ij}^{00}}  \underset{(-n,-p)}{I_{ij}^{00}} .
\end{align}
In each case, negative $n$ terms duplicate positive $n$ terms (see Appendix \ref{sec:fourSumId}).  Applying the 
crossing relations and pulling out the Newtonian circular-orbit factor of $(32/5) \nu^2 x^5$, we arrive at the 
following relative flux contributions
\begin{align}
\mathcal{R}_{1}^{\rm MQ01} &= \frac{1}{16}  \sum_{n=1}^{\infty} n^6 
\bigg[\underset{(n)}{\hat{I}_{ij}^{00}}  \underset{(n)}{\hat{I}_{ij}^{10*}} 
+ \underset{(n)}{\hat{I}_{ij}^{10}}  \underset{(n)}{\hat{I}_{ij}^{00*}} \bigg],  \notag \\
\mathcal{R}_{1}^{\rm MQ02} &= -\frac{9}{8(1-e_t^2)} \sum_{n=1}^{\infty}n^6 | 
\underset{(n)}{\hat{I}_{ij}^{00}} |^2,   \notag  \\
\mathcal{R}_{1}^{\rm MQ03} &= \frac{9}{8(1-e_t^2)}  
\sum_{n=1}^{\infty} \sum_{p=-2}^{2}  n^5 p \, | \underset{(n,p)}{\hat{I}_{ij}^{00}} |^2,
\end{align}
where we define $\underset{(n)}{\hat{I}_{ij}^{00}}  = \underset{(n,-2)}{\hat{I}_{ij}^{00}} 
+ \underset{(n,0)}{\hat{I}_{ij}^{00}} + \underset{(n,2)}{\hat{I}_{ij}^{00}}$. 

\subsubsection{Contributions to the 1PN relative angular momentum flux from the mass quadrupole at lowest order 
in $\nu$}

Similarly, the angular momentum flux is given by the 1PN correction to the formula
\begin{align}
\Big<\frac{dL}{dt}\Big>_1^{\rm MQ} =&\frac{2}{5}\epsilon_{ijl} \hat{L}_i \Big<\ddot{I}_{ja}\dddot{I}_{la}\Big>_1
=\frac{2}{5} \epsilon_{3jl} \bigg\langle \sum_{n,m=-\infty}^{\infty} \sum_{p,s=-2}^{2} (\O_r)^5
 (i(n+pk))^2 (i(m+sk))^3 
\underset{(n,p)}{I_{ja}}  \underset{(m,s)}{I_{la}} e^{i(n+m+(p+s)k)l}\bigg\rangle_1 \hat{z}  
\label{eqn:H0},
\end{align}
where as mentioned earlier $\hat{L}_i = \hat{z}$ for Kepler motion in the $x,y$ plane.  This sum simplifies in 
almost the same manner as the energy flux.  There is a key identity involving $(p+s)$ in the angular momentum 
summations that is analogous to the one in the energy flux.  We find
\be
\sum_{p,s} (p+s) \epsilon_{3jl} \underset{(n,p)}{I_{ja}}  \underset{(m,s)}{I_{la}} = 0 .
\ee
The angular momentum also has the identity
$ \epsilon_{3jl} \underset{(n,0)}{I_{ja}}  \underset{(m,0)}{I_{la}} = 0$, so that only 
$ \epsilon_{3jl} \underset{(\pm2)}{M_{ja}}   \underset{(\mp2)}{M_{la}} = \pm 4 i$ survives.  Inserting 
$\delta_{m,-n}$ for the time average and taking $s \rightarrow -s$ as above, the expression reduces to
\begin{align}
\Big<\frac{dL}{dt}\Big>_1^{\rm MQ}
= - \frac{2 i}{5} \epsilon_{3jl} \sum_{n=-\infty}^{\infty} \sum_{p=-2,2} (\O_r)^5 
 (n^5 + 5 n^4 p k ) \underset{(n,p)}{I_{ja}} \underset{(-n,-p)}{I_{la}}  \hat{z} .
\end{align}

As expected, we are left with three sums, all similar in form to their energy counterparts.  We apply the crossing 
relation and simplify to obtain the following flux contributions
\begin{align}
\mathcal{Z}_{1}^{\rm MQ01} &= - \frac{i}{8}\epsilon_{3jl} \sum_{n=-\infty}^{\infty}
 n^5 \bigg[\underset{(n)}{\hat{I}_{ja}^{00}}  \underset{(n)}{\hat{I}_{la}^{10*}} 
 + \underset{(n)}{\hat{I}_{ja}^{10}}  \underset{(n)}{\hat{I}_{la}^{00*}} \bigg] \hat{z}, \notag \\
\mathcal{Z}_{1}^{\rm MQ02} &= \frac{15 i}{8(1-e_t^2)} \epsilon_{3jl} \sum_{n=1}^{\infty}
 n^5 \underset{(n)}{\hat{I}_{ja}^{00}} \underset{(n)}{\hat{I}_{la}^{00*}},  \notag \\
\mathcal{Z}_{1}^{\rm MQ03}
 &= -\frac{15 i}{8(1-e_t^2)} \epsilon_{3jl} \sum_{n=1}^{\infty} \sum_{p=-2,2} n^4 p
 \underset{(n,p)}{\hat{I}_{ja}^{00}} \underset{(n,p)}{\hat{I}_{la}^{00*}} \hat{z} .
\end{align} 
As stated previously, the biperiodicity of the 1PN mass quadrupole introduces three separate sums in the calculation
of the 1PN flux.  It turns out that these three sums characterize the entirety of both 1PN logarithm series at lowest 
order in $\nu$.  As we will see next, transition to the next highest 1PN logarithm flux (at 2.5PN order) will 
involve an increase in the power of $n$ in the sums, along with multiplication by a different leading coefficient.

\subsubsection{Contributions to the 2.5PN relative energy flux from the mass quadrupole at lowest order in $\nu$}

In the time domain, the mass quadrupole part of the energy tail flux \cite{ArunETC08a,LoutYune17} is given by
\begin{equation}
\mathcal{P}^{\text{MQtail}}_\infty =
\frac{4\mathcal{M}}{5}\dddot{I}_{ij}(t)\int_{0}^{\infty}I_{ij}^{(5)}(t-\tau)\bigg[\log\Big(\frac{\tau}{2 r_0}\Big)
+\frac{11}{12}\bigg] d\tau,
\end{equation}
where $\mathcal{M}$ is the ADM mass $\mathcal{M}=M(1-\nu x/2+\mathcal{O}(x^2))$.  This expression gives the 
time-dependent flux, which will subsequently be time averaged over an orbital libration.  It represents a 
nonlinear interaction between the mass quadrupole and ADM mass monopole of the system.  However, because we are 
currently working at lowest order in $\nu$, $\mathcal{M}$ can be replaced with $M$.

We insert the biperiodic Fourier expansion \eqref{eqn:biFour} for the quadrupole moment, replace time derivatives, 
and take the time average to find
\begin{align}
\Big<\frac{dE}{dt}\Big>_{5/2}^{\rm MQ} = 
\frac{4M}{5} \sum_{n=-\infty}^{\infty} \sum_{p,s}(\O_r)^8 \left(-n^8+ n^7(5s-3p)k \right)
\underset{(n,p)}{I_{ij}} \underset{(-n,s)}{I_{ij}}
 \int_{0}^{\infty} e^{i(n-sk)\O_r \tau}\bigg[\log\Big(\frac{\tau}{2 r_0}\Big)
+\frac{11}{12}\bigg] d\tau .
\end{align}
The only significant difference between this summation and that at 1PN order is the last integral term, which can 
be rewritten slightly to aid subsequent evaluation 
\begin{align}
\label{eqn:tailint}
\int_{0}^{\infty} e^{i(n-sk)\O_r \tau}\bigg[\log\Big(\frac{\tau}{2 r_0}\Big)
+\frac{11}{12}\bigg] d\tau 
= \int_{0}^{\infty} e^{i(n-sk)\O_r \tau}\log\Big(\frac{\tau}{2 r_0 e^{-(11/12)}}\Big)d\tau .
\end{align}
This expression is regularized by rotating the mean motion into the complex plane.  We refer the reader to 
\cite{ArunETC08a,LoutYune17,MarcBlanFaye16} as well as Paper I (Sec.~IV C and Appendix A) for details.  The result is
\begin{gather}
\label{eqn:lnTailInt}
-\frac{i}{(n-sk)\O_r}\left[\frac{\pi i}{2}\text{sign}(-n) + \log(2 \O_r |n - sk | r_0) 
+ \gamma_E - \frac{11}{12}\right]    \\
\approx -\frac{1}{n \O_r} \left[\frac{\pi}{2}\text{sign}(n) 
+i \left( \log(2 \O_{\vp} | n | r_0) + \gamma_E - \frac{11}{12}\right) \right] 
-\frac{sk}{n^2 \O_r} \left[\frac{\pi}{2}\text{sign}(n) 
+i \left( \log(2 \O_{\vp} | n | r_0) - \frac{n}{s} + \gamma_E - \frac{23}{12}\right) \right],  
\end{gather}
where the second line is an expansion to first order in $k$.  

Appendix \ref{sec:fourSumId} shows that the imaginary portion will identically vanish in sums over positive and 
negative $n$, thus allowing those terms to be eliminated.  Using the remaining factor, taking $s \rightarrow -s$, 
and then setting $s = p$, as in our earlier derivation, leads to
\begin{align}
\Big<\frac{dE}{dt}\Big>_{5/2}^{\rm MQ}  = \frac{4M}{5} \sum_{n=-\infty}^{\infty} \sum_{p}(\O_r)^7 
(n^7+ 8 n^6 p k) \underset{(n,p)}{I_{ij}} \underset{(-n,-p)}{I_{ij}} \left[\frac{\pi}{2}\text{sign}(n) 
- \left( \frac{p k}{n} \right) \frac{\pi}{2}\text{sign}(n)  \right] .
\end{align}
As in the 1PN case, this result splits into three well-defined sums, which can be written as
\begin{align}
\mathcal{R}_{5/2}^{\rm MQ01} &= \frac{\pi x}{8} \sum_{n=1}^{\infty} n^7
\Big[\underset{(n)}{\hat{I}_{ij}^{00}} \underset{(n)}{\hat{I}_{ij}^{10*}} 
+ \underset{(n)}{\hat{I}_{ij}^{10}} \underset{(n)}{\hat{I}_{ij}^{00*}} \Big],   \notag \\
\mathcal{R}_{5/2}^{\rm MQ02} &= - \frac{21 \pi x}{8(1-e_t^2)}  \sum_{n=1}^{\infty}
n^7 | \underset{(n)}{\hat{I}_{ij}^{00}} |^2, \notag \\
\mathcal{R}_{5/2}^{\rm MQ03} &= \frac{21 \pi x}{8(1-e_t^2)}
\sum_{n=1 }^{\infty} \sum_{p} n^6 p \, | \underset{(n,p)}{\hat{I}_{ij}^{00}} |^2 .
\end{align}
Summed together and normalized, these terms will recover the enhancement function $\alpha(e_t)$ defined by 
Arun et al.~\cite{ArunETC08a,LoutYune17}.

\subsubsection{Contributions to the 2.5PN relative angular momentum flux from the mass quadrupole at lowest order 
in $\nu$}

Similarly, the (time-dependent) angular momentum tail flux \cite{LoutYune17} is given by
\begin{align}
\mathcal{G}^{\text{MQtail}}_\infty =
\frac{4\mathcal{M}}{5}\epsilon_{3jl} \bigg\{\ddot{I}_{ja}(t)\int_{0}^{\infty}I_{la}^{(5)}(t-\tau)
\bigg[\log\Big(\frac{\tau}{2 r_0}\Big) +\frac{11}{12}\bigg] d\tau 
+\dddot{I}_{la}(t)\int_{0}^{\infty}I_{ja}^{(4)}(t-\tau)
\bigg[\log\Big(\frac{\tau}{2 r_0}\Big) +\frac{11}{12}\bigg] d\tau \bigg\} \hat{z} .
\end{align}
By inserting the Fourier series and performing the same simplifications as in the energy case, we arrive at
\begin{align}
\mathcal{Z}_{5/2}^{\rm MQ01} &=
- \frac{\pi i}{4} \epsilon_{3jl} \sum_{n=1}^{\infty}
 n^6 \bigg[\underset{(n)}{\hat{I}_{ja}^{00}}  \underset{(n)}{\hat{I}_{la}^{10*}} + \underset{(n)}{\hat{I}_{ja}^{10}}  
\underset{(n)}{\hat{I}_{la}^{00*}} \bigg] \hat{z} ,   \notag \\
 \mathcal{Z}_{5/2}^{\rm MQ02} &= \frac{9 \pi i}{2(1-e_t^2)} \epsilon_{3jl} \sum_{n=1}^{\infty}
 n^6 \underset{(n)}{\hat{I}_{ja}^{00}} \underset{(n)}{\hat{I}_{la}^{00*}}\hat{z} ,   
\notag 
\\
\mathcal{Z}_{5/2}^{\rm MQ03} &=
 -\frac{9 \pi i }{2 (1-e_t^2)} \epsilon_{3jl} \sum_{n=1}^{\infty} \sum_{p=-2,2} (n^5 p)
 \underset{(n,p)}{\hat{I}_{ja}^{00}} \underset{(n,p)}{\hat{I}_{la}^{00*}}\hat{z} .  
\end{align}
Despite the factor of $i$ that is pulled out of each sum, the complex conjugation and presence of the 
Levi Civita tensor ensure that all of these terms are real. 

\subsection{Relative flux contributions from the mass quadrupole at next order in $\nu$}

We next need to consider the linear-order-in-$\nu$ contributions to \eqref{eqn:MQ5terms}, i.e., the
$\mathcal{R}_{i}^{\rm MQ1}$ and $\mathcal{Z}_{i}^{\rm MQ1}$ terms.  Fortunately, much of the procedure is identical 
to that in the previous subsection, with only minor modifications to generate the corresponding reductions.  One 
difference lies in the fact that there can be no appearance of $\nu$ at 0PN order.  Thus, contributions from the 
radial frequency and magnetic factor $p$ (that is, of type $\mathcal{R}_{1}^{\rm MQ02}$ and 
$\mathcal{R}_{1}^{\rm MQ03}$, respectively) will not recur here.  This eliminates two potential types of terms that 
involve the Newtonian portion of the mass quadrupole.  However, a sum involving the Newtonian mass quadrupole 
(termed $\mathcal{R}_{i}^{\rm MQ12}$) will still manifest at $\mathcal{O}(\nu)$ in all 1PN logarithms except 
$\mathcal{R}_1$, first appearing in the 2.5PN tail through a factor of the ADM mass 
$\mathcal{M} = M (1 - \nu x/2 + \mathcal{O}(x^2))$.

As a result, the 1PN order $\nu$ flux terms are straightforward, containing only the $\mathcal{O}(\nu)$ 
correction induced by the corresponding portion of the quadrupole moment.  All aspects of their derivations 
are functionally identical to those of $\mathcal{R}_{1}^{\rm MQ01}$ and $\mathcal{Z}_{1}^{\rm MQ01}$ at $\nu^0$, 
with the simple substitution $I^{10} \rightarrow I^{11}$.  We find
\begin{align}
\mathcal{R}_{1}^{\rm MQ11} &= \frac{1}{16}  \sum_{n=1}^{\infty} n^6 
\bigg[\underset{(n)}{\hat{I}_{ij}^{00}}  \underset{(n)}{\hat{I}_{ij}^{11*}} 
+ \underset{(n)}{\hat{I}_{ij}^{11}}  \underset{(n)}{\hat{I}_{ij}^{00*}} \bigg] ,  
\notag 
\\
\mathcal{Z}_{1}^{\rm MQ11} &= - \frac{i}{8}\epsilon_{3jl} \sum_{n=-\infty}^{\infty}
 n^5 \bigg[\underset{(n)}{\hat{I}_{ja}^{00}}  \underset{(n)}{\hat{I}_{la}^{11*}} 
 + \underset{(n)}{\hat{I}_{ja}^{11}}  \underset{(n)}{\hat{I}_{la}^{00*}} \bigg] \hat{z} ,
\end{align}
with $\mathcal{R}_{1}^{\rm MQ12} = \mathcal{Z}_{1}^{\rm MQ12} = 0$.  The square bracket in the second sum has 
the same anti-hermetian behavior as before.

The 2.5PN terms are only slightly more involved.  The first contribution emerges from the same substitution in 
$\mathcal{R}_{5/2}^{\rm MQ01}$, with $I^{10} \rightarrow I^{11}$.  However, now there is a second term that comes 
from the 1PN correction to the ADM mass.  We find
\begin{align}
\mathcal{R}_{5/2}^{\rm MQ11} &= \frac{\pi}{8} \sum_{n=1}^{\infty} n^7
\Big[\underset{(n)}{\hat{I}_{ij}^{00}} \underset{(n)}{\hat{I}_{ij}^{11*}} 
+ \underset{(n)}{\hat{I}_{ij}^{11}} \underset{(n)}{\hat{I}_{ij}^{00*}} \Big],    \qq \qq \q \mkern47mu
\mathcal{R}_{5/2}^{\rm MQ12} = - \frac{\pi}{16} \sum_{n=1}^{\infty} n^7
| \underset{(n)}{\hat{I}_{ij}^{00}} |^2 ,    
\notag 
\\
\mathcal{Z}_{5/2}^{\rm MQ11} &=
- \frac{\pi i}{4} \epsilon_{3jl} \sum_{n=1}^{\infty}
n^6 \bigg[\underset{(n)}{\hat{I}_{ja}^{00}}  \underset{(n)}{\hat{I}_{la}^{11*}} 
+ \underset{(n)}{\hat{I}_{ja}^{11}}  \underset{(n)}{\hat{I}_{la}^{00*}} \bigg] \hat{z},    \qq \mkern40mu
\mathcal{Z}_{5/2}^{\rm MQ12} = \frac{\pi i}{8} \epsilon_{3jl} \sum_{n=1}^{\infty}
n^6 \underset{(n)}{\hat{I}_{ja}^{00}} \underset{(n)}{\hat{I}_{la}^{00*}} \hat{z} .
\end{align}
These 2.5PN contributions to the flux can be summed together and normalized to generate the enhancement functions 
$\theta(e_t)$ and $\tilde{\theta}(e_t)$ defined in \cite{ArunETC08a,ArunETC09a,LoutYune17}.

\subsection{Eccentricity singular factors and full flux functions}
\label{sec:singFacs}

The various sums over Fourier amplitude products derived above will produce, when added together, the power series 
in eccentricity for the full flux contributions at 1PN and 2.5PN.  As with the leading logarithms \cite{MunnEvan19a}, 
each such sum will have an associated eccentricity singular factor governing its divergent behavior as 
$e_t \rightarrow 1$.  For each of the separated parts (except one) its own singular behavior 
is easily determined using the asymptotic analysis developed in \cite{ForsEvanHopp16} (specific examples are given 
in Appendix \ref{sec:compSums}).  The exception is the term labeled MQ01, which involves the quadrupole components 
with unevaluated integrals \eqref{eqn:I1PNint}.  Because this part does not have a clean representation in terms of 
Bessel functions, it is not amenable to the exact same asymptotic analysis technique.  Nevertheless, its divergent 
behavior appears to adhere to the same patterns, and we have demonstrated apparent convergence through 22PN (see 
App.~\ref{sec:compSums} for more details) and verified the behavior with a new all-analytic perturbation code 
\cite{Munn20}.  The conclusion is that the terms in the various 1PN log sequences have the following singular 
behavior 
\begin{align}
\mathcal{R}_{(3k+1)L(k)} &\sim \frac{1}{(1 - e_t^2)^{k + 9/2}}, \qq 
\mathcal{R}_{(3k+5/2)L(k)} \sim \frac{1}{(1 - e_t^2)^{k + 6}}, \notag \\ 
\mathcal{Z}_{(3k+1)L(k)} &\sim \frac{1}{(1 - e_t^2)^{k + 3}}, \qq \q
\mathcal{Z}_{(3k+1)L(k)} \sim \frac{1}{(1 - e_t^2)^{k + 9/2}}. 
\end{align}
With the divergent behavior understood, the remaining eccentricity dependence is found to be closed-form 
(polynomial) expressions for the integer-order 1PN logarithms and convergent power series at the half-integer 
orders.

Putting all of these elements together involves summing the results of the previous sections and extracting the 
appropriate overall eccentricity singular factor.  Focusing on low PN order, we can re-derive the known energy and 
angular momentum flux functions.  This frequency domain approach leads to the well-known closed-form expressions 
at 1PN
\begin{align}
\mathcal{R}_1(e_t,\nu) &= \frac{1}{(1-e_t^2)^{9/2}} \left(-\frac{1247}{336}+\frac{10475 e_t^2}{672}+\frac{10043 
e_t^4}{384}+\frac{2179 e_t^6}{1792} \right) - \frac{\nu}{(1-e_t^2)^{9/2}} \left(\frac{35}{12}+\frac{1081 e_t^2}{36}
+\frac{311 e_t^4}{12}+\frac{851 e_t^6}{576}\right) ,
\notag 
\\
\mathcal{Z}_1(e_t,\nu) &= \frac{1}{(1-e_t^2)^3} \left(-\frac{1247}{336}+\frac{3019 e_t^2}{336}+\frac{8399 e_t^4}{2688}
\right) - \frac{\nu}{(1-e_t^2)^3} \left( \frac{35}{12} + \frac{335 e_t^2}{24}+\frac{275 e_t^4}{96}  \right) ,
\end{align}
which (being purely instantaneous) were previously derived through time domain analysis 
\cite{WagoWill76,BlanScha89}. 

The 2.5PN flux functions on the other hand do not have closed-form representations.  The original work in 
\cite{ArunETC08a, ArunETC09a} showed numerical results and presented expansions in eccentricity only through 
$e_t^4$.  Forseth et al.~\cite{ForsEvanHopp16} used a frequency domain procedure similar to the present one to 
generate $\mathcal{R}_{5/2}$ to $e_t^{70}$ and developed the asymptotic analysis to investigate the behavior as 
$e_t \rightarrow 1$ at lowest order in $\nu$.  Later, Loutrel and Yunes \cite{LoutYune17} also derived asymptotics of 
these functions as $e_t \rightarrow 1$ and for both orders in $\nu$.  We have now calculated the terms in the power 
series to $e_t^{120}$ using the methods described above, with the ability to push to much higher order should it 
prove necessary.  These two series have leading behavior
\begin{align}
\mathcal{R}_{5/2}(e_t, \nu) & = \frac{1}{(1-e_t^2)^6}\left(-\frac{8191}{672}+\frac{36067 e_t^2}{336}
+ \frac{19817891 e_t^4}{43008}+\frac{62900483 e_t^6}{387072}+\frac{26368199 e_t^8}{7077888}
-\frac{1052581 e_t^{10}}{34406400}+ \cdots \right)  \notag \\ &
+ \frac{\nu}{(1-e_t^2)^6}\left( -\frac{583}{24}-\frac{717733 e_t^2}{2016}-\frac{21216061 e_t^4}{32256}
-\frac{78753305 e_t^6}{387072}-\frac{208563695 e_t^8}{37158912}
+\frac{46886227 e_t^{10}}{3715891200} + \cdots \right) , 
\notag 
\\
\mathcal{Z}_{5/2}(e_t, \nu) &= \frac{1}{(1-e_t^2)^{9/2}} \left(-\frac{8191}{672}+\frac{108551 e_t^2}{1344}+\frac{5055125 e_t^4}{43008}+\frac{4125385 e_t^6}{774144}-\frac{11065099 e_t^8}{49545216}
+\frac{68397463 e_t^{10}}{2477260800} + \cdots \right) \notag \\ &
+ \frac{\nu}{(1-e_t^2)^{9/2}} \left( -\frac{583}{24}-\frac{32821 e_t^2}{168}-\frac{1566125 e_t^4}{10752}
-\frac{712219 e_t^6}{96768}+\frac{457507 e_t^8}{12386304}-\frac{792569 e_t^{10}}{309657600} + \cdots \right) .
\end{align}
As $e_t \rightarrow 1$, these series approach approximately $(722.1524014 - 1247.1117956 \nu)/(1 - e_t^2)^6$ and 
$(191.2520614 - 372.6399916 \nu)/(1 - e_t^2)^{9/2}$, respectively (see discussion in Sec.~III~C of \cite{MunnETC20} 
regarding prior tabulated numerical values \cite{ArunETC09a} of these series in the vicinity of $e_t = 1$).

\section{Higher-order elements of the 1PN log sequences}
\label{sec:1PNLogs}

With the derivations in the previous sections, plus the leading logarithm series \cite{MunnEvan19a} and numerical 
input from BHPT, we now have enough information to generalize to the form of the 1PN logarithm series for all PN 
orders.  As in Paper I, this process will involve incrementing powers of $n$ within sums over products of the 
Fourier amplitudes and determining the correct rational-number prefactor at each order.  

\subsection{Mass octupole and current quadrupole contributions to higher-order 1PN log terms}

We begin with the two 1PN source multipole moments (mass octupole and current quadrupole) that can be calculated 
(for present purposes) using Newtonian dynamics.  These moments give rise to the spectra $h(n,e_t)$ and $k(n,e_t)$.  
Sums over these multipole spectra with higher powers of $n$ lead to their contributions to the higher-order 1PN log 
fluxes, much as sums over the Newtonian mass quadrupole spectra did in contributing to the higher-order leading logs 
as shown in Paper I.  For integers $k \ge 0$, the mass octupole contributions to the 1PN log (energy) fluxes are 
given by
\begin{align}
\mathcal{R}_{(3k+1)L(k)}^{\rm MO} &= (1- 4\nu)
\bigg(-\frac{26}{21}\bigg)^{k}\bigg(\frac{1}{k!}\bigg)\sum_{n=1}^{\infty} n^{2k}h(n,e_t), 
\\
\mathcal{R}_{(3k+5/2)L(k)}^{\rm MO} &= 
(1- 4\nu) \bigg(-\frac{26}{21}\bigg)^{k}\bigg(\frac{2 \pi}{k!}\bigg)\sum_{n=1}^{\infty} n^{2k+1}h(n,e_t) . 
\end{align}
The current quadrupole series are even closer in appearance to the leading logarithms of Paper I, taking the 
following forms
\begin{align}
\mathcal{R}_{(3k+1)L(k)}^{\rm CQ} &= (1- 4\nu)
\bigg(-\frac{214}{105}\bigg)^{k}\bigg(\frac{1}{k!}\bigg)\sum_{n=1}^{\infty} n^{2k} k(n,e_t), 
\\
\mathcal{R}_{(3k+5/2)L(k)}^{\rm CQ} &= (1- 4\nu)
\bigg(-\frac{214}{105}\bigg)^{k}\bigg(\frac{2 \pi}{k!}\bigg)\sum_{n=1}^{\infty} n^{2k+1} k(n,e_t) .
\end{align}
In each case, the angular momentum analog $\mathcal{Z}_i$ is obtained by simply substituting $h \rightarrow \tilde{h}$
or $k \rightarrow \tilde{k}$, as appropriate.

\subsection{Mass quadrupole (at lowest order in $\nu$) contributions to higher-order 1PN log terms}

\subsubsection{The energy flux}

At lowest order in the mass ratio, three separate sums over Fourier amplitudes must be handled.  The simplest of 
the three to derive (though the hardest to compute explicitly), $\mathcal{R}_{(3k+1)L(k)}^{\rm MQ01}\mkern1mu$, 
comes from the correction to the mass quadrupole itself.  Careful inspection reveals that this term must be 
identical in form to the leading logarithm series, except with the Newtonian part of the mass quadrupole supplanted 
by its 1PN counterpart.  Thus, the prefactor must be the same, and we can simply adjust the result of Paper I to get 
the following energy flux contributions
\begin{align}
\mathcal{R}_{(3k+1)L(k)}^{\rm MQ01} &=  \frac{1}{16(k!)} \bigg(-\frac{214}{105}\bigg)^k
 \sum_{n=1}^{\infty} n^{2k+6}
\Big[\underset{(n)}{\hat{I}_{ij}^{00}} \underset{(n)}{\hat{I}_{ij}^{10*}} 
+ \underset{(n)}{\hat{I}_{ij}^{10}} \underset{(n)}{\hat{I}_{ij}^{00*}} \Big] , 
\\
\mathcal{R}_{(3k+5/2)L(k)}^{\rm MQ01} &=  \frac{\pi}{8(k!)} \bigg(-\frac{214}{105}\bigg)^k
 \sum_{n=1}^{\infty} n^{2k+7}
\Big[\underset{(n)}{\hat{I}_{ij}^{00}} \underset{(n)}{\hat{I}_{ij}^{10*}} 
+ \underset{(n)}{\hat{I}_{ij}^{10}} \underset{(n)}{\hat{I}_{ij}^{00*}} \Big] .
\end{align}
(We note again that in these and all sums in this section, $k$ refers to any non-negative integer, rather than the 
ratio of frequencies $k = \O_\vp/\O_r-1$.)

The next sum type, $\mathcal{R}_{i}^{\rm MQ02}$, which in our scheme involves the 1PN correction to $\O_r$, can 
be found in a similar manner.  The portion of the quadrupole moment involved is just the Newtonian part and the $k$ 
dependent coefficient follows from a binomial expansion of powers of $\O_r = \O_{\vp} (1 - 3x/(1-e_t^2))$ to 1PN 
order.  We find
\begin{align}
\mathcal{R}_{(3k+1)L(k)}^{\rm MQ02} &=  -\frac{3k+9}{8(k!)(1-e_t^2)} \bigg(-\frac{214}{105}\bigg)^k
 \sum_{n=1}^{\infty} n^{2k+6} | \underset{(n)}{\hat{I}_{ij}^{00}} |^2 ,
\\
\mathcal{R}_{(3k+5/2)L(k)}^{\rm MQ02} &=  -\frac{3 \pi (2 k + 7)}{8(k!)(1-e_t^2)} \bigg(-\frac{214}{105}\bigg)^k
 \sum_{n=1}^{\infty} n^{2k+7}  | \underset{(n)}{\hat{I}_{ij}^{00}} |^2 .
\end{align}

Finally, the third sum type is $\mathcal{R}_{k}^{\rm MQ03}$, whose definition involves the magnetic factor $p$
with $\hat{I}_{ij}^{00}$.  We find (and illustrate in the discussion below) that the $k$-dependent coefficient 
prefacing this summation is equal and opposite to that of $\mathcal{R}^{\rm MQ02}$, or
\begin{align}
\mathcal{R}_{(3k+1)L(k)}^{\rm MQ03} &= \frac{3k+9}{8(k!)(1-e_t^2)} \bigg(-\frac{214}{105}\bigg)^k
 \sum_{n=1}^{\infty}  \sum_{p= -2,2}  n^{2k+5} p \, | \underset{(n,p)}{\hat{I}_{ij}^{00}} |^2 ,
\\
\mathcal{R}_{(3k+5/2)L(k)}^{\rm MQ03} &=  \frac{3 \pi (2 k + 7)}{8(k!)(1-e_t^2)} \bigg(-\frac{214}{105}\bigg)^k
 \sum_{n=1}^{\infty} \sum_{p= -2,2}  n^{2k+6} p \, | \underset{(n,p)}{\hat{I}_{ij}^{00}} |^2 .
\end{align}

\subsubsection{The angular momentum flux}

As seen throughout Sec.~\ref{sec:knownLogs}, the contributions to the angular momentum flux are nearly identical in 
form, only requiring minor adjustments in the moments and prefactors.   The first sum mirrors that of the leading 
logarithm series, giving
\begin{align}
\mathcal{Z}_{(3k+1)L(k)}^{\rm MQ01} &= - \frac{i}{8(k!)} \bigg(-\frac{214}{105}\bigg)^k \e_{3jl}
 \sum_{n=1}^{\infty} n^{2k+5}
\Big[\underset{(n)}{\hat{I}_{ja}^{00}} \underset{(n)}{\hat{I}_{la}^{10*}} 
+ \underset{(n)}{\hat{I}_{ja}^{10}} \underset{(n)}{\hat{I}_{la}^{00*}} \Big]\hat{z}, 
\notag 
\\
\mathcal{Z}_{(3k+5/2)L(k)}^{\rm MQ01} &= - \frac{\pi i}{4(k!)} \bigg(-\frac{214}{105}\bigg)^k \e_{3jl}
 \sum_{n=1}^{\infty} n^{2k+6}
\Big[\underset{(n)}{\hat{I}_{ja}^{00}} \underset{(n)}{\hat{I}_{la}^{10*}} 
+ \underset{(n)}{\hat{I}_{ja}^{10}} \underset{(n)}{\hat{I}_{la}^{00*}} \Big]\hat{z} .
\end{align}
The second sum type, $\mathcal{Z}_{i}^{\rm MQ02}$, has one lower power of $\O_r$ than the corresponding energy 
flux term, $\mathcal{R}_{i}^{\rm MQ02}$, and is found to be
\begin{align}
\mathcal{Z}_{(3k+1)L(k)}^{\rm MQ02} &=  \frac{3 (2k+5) i}{8(k!)(1-e_t^2)} \bigg(-\frac{214}{105}\bigg)^k \e_{3jl}
 \sum_{n=1}^{\infty} n^{2k+5} \underset{(n)}{\hat{I}_{ja}^{00}} \underset{(n)}{\hat{I}_{la}^{00*}}\hat{z},   
\notag 
\\
\mathcal{Z}_{(3k+5/2)L(k)}^{\rm MQ02} 
&=  \frac{3 \pi (k + 3) i}{2(k!)(1-e_t^2)}  \bigg(-\frac{214}{105}\bigg)^k \e_{3jl}
 \sum_{n=1}^{\infty} n^{2k+6}  \underset{(n)}{\hat{I}_{ja}^{00}} \underset{(n)}{\hat{I}_{la}^{00*}} \hat{z} ,
\end{align}
with the antisymmetry and factor of $i$ guaranteeing the flux is real.  Finally, terms of the third sum type emerge 
with identical $k$-dependent factors (up to sign), and are found to be
\begin{align}
\mathcal{Z}_{(3k+1)L(k)}^{\rm MQ03} &= -\frac{3 (2k+5) i}{8(k!)(1-e_t^2)} \bigg(-\frac{214}{105}\bigg)^k \e_{3jl}
 \sum_{n=1}^{\infty}  \sum_{p= -2,2}  n^{2k+4} p \underset{(n,p)}{\hat{I}_{ja}^{00}} 
 \underset{(n,p)}{\hat{I}_{la}^{00*}} \hat{z} ,  
\notag 
\\
\mathcal{Z}_{(3k+5/2)L(k)}^{\rm MQ03} 
&=  -\frac{3 \pi (k + 3) i}{2(k!)(1-e_t^2)} \bigg(-\frac{214}{105}\bigg)^k \e_{3jl}
\sum_{n=1}^{\infty} \sum_{p= -2,2}  n^{2k+5} p  \underset{(n,p)}{\hat{I}_{ja}^{00}} 
\underset{(n,p)}{\hat{I}_{la}^{00*}}\hat{z} .
\end{align}

\subsection{Mass quadrupole (next order in $\nu$) contributions to higher-order 1PN log terms}
\label{sec:1PNlogsMQNu}

There is an expected contribution at next order in $\nu$ to the flux in each higher-order 1PN log term, just as 
there was with the base terms of these sequences: $\mathcal{R}_1$, $\mathcal{Z}_1$, $\mathcal{R}_{5/2}$, and 
$\mathcal{Z}_{5/2}$.  These contributions emerge from two summations---one involving the 1PN part of the quadrupole 
moment, $\hat{I}^{11}$, and one containing its Newtonian counterpart, $\hat{I}^{00}$.  From the earlier discussion 
of the 1PN and 2.5PN relative order fluxes, we can see that the coefficients for $\mathcal{R}^{\rm MQ11}$ in the 
1PN log sequence must exactly match those of their $\mathcal{R}^{\rm MQ01}$ counterparts in the previous subsection.  

The $k$-dependent factor preceding the sum for $\mathcal{R}^{\rm MQ12}$ is less straightforward.  This sum involves 
the Newtonian-order mass quadrupole and is of a form that did not make an appearance in $\mathcal{R}_1$.  Instead, 
it first shows up with the ADM mass in the 2.5PN tail.  The appearance of the ADM mass in the known hereditary flux 
terms is fairly regular: Each higher-order tail merely sees an increment in the power of $\mathcal{M}$ (see, for 
example, Eq.~(4.8) of \cite{MarcBlanFaye16}), making the tail portion of $\mathcal{R}^{\rm MQ12}$ calculable to 
high PN order.  Moreover, in Paper I we used a combination of BHPT and PN results to show that for leading logarithms 
(starting with $\mathcal{R}_{3L}$), all instantaneous contributions uniformly equal a factor of -2/3 of their 
hereditary counterparts.  A similar line of reasoning might be applied to 1PN log terms at $\mathcal{O}(\nu^0)$.  
However, because that argument relied upon information from BHPT, which is presently limited to first order in the 
mass ratio, it cannot be extended as written for next order in $\nu$ (i.e., $\mathcal{O}(\nu^1)$) results.  

Nevertheless, the PN regularization parameter $r_0$ \cite{Blan14}, which exists in all hereditary integrals but 
which must cancel in the overall flux and thus implies corresponding factors in the instantaneous flux, lends strong 
credence to the notion that the simple relationship also exists at $\mathcal{O}(\nu^1)$.  For the time being we 
conjecture that this is the case and present the results that follow from this assumption.  If the conjecture is 
correct, then the coefficients on the $\mathcal{R}_{i}^{\rm MQ12}$ terms become nearly identical to those of 
$\mathcal{R}_{i}^{\rm MQ11}$, except the binomial expansion of $\mathcal{M}^q = M^q (1 - \nu x/2)^q$ introduces 
a factor of $-q/2$ for the $(q+1)$th element of the 1PN log series.  We are led to the following expected forms 
of the next order in $\nu$ eccentricity-dependent flux functions
\begin{align}
\mathcal{R}_{(3k+1)L(k)}^{\rm MQ11} &=  \frac{1}{16(k!)} \bigg(-\frac{214}{105}\bigg)^k
 \sum_{n=1}^{\infty} n^{2k+6} \Big[\underset{(n)}{\hat{I}_{ij}^{00}} \underset{(n)}{\hat{I}_{ij}^{11*}} 
+ \underset{(n)}{\hat{I}_{ij}^{11}} \underset{(n)}{\hat{I}_{ij}^{00*}} \Big] ,   
\notag 
\\
\mathcal{R}_{(3k+5/2)L(k)}^{\rm MQ11} &=  \frac{\pi}{8(k!)} \bigg(-\frac{214}{105}\bigg)^k
 \sum_{n=1}^{\infty} n^{2k+7} \Big[\underset{(n)}{\hat{I}_{ij}^{00}} \underset{(n)}{\hat{I}_{ij}^{11*}} 
+ \underset{(n)}{\hat{I}_{ij}^{11}} \underset{(n)}{\hat{I}_{ij}^{00*}} \Big],    
\notag 
\\
\mathcal{R}_{(3k+1)L(k)}^{\rm MQ12} &= - \frac{1}{16(k-1)!} \bigg(-\frac{214}{105}\bigg)^k
 \sum_{n=1}^{\infty} n^{2k+6} | \underset{(n)}{\hat{I}_{ij}^{00}} |^2  ,   
\notag 
\\
\mathcal{R}_{(3k+5/2)L(k)}^{\rm MQ12} &= - \frac{\pi (2k+1)}{16(k!)} \bigg(-\frac{214}{105}\bigg)^k
 \sum_{n=1}^{\infty} n^{2k+7} | \underset{(n)}{\hat{I}_{ij}^{00}} |^2 , 
\label{eqn:1PNLogMQ1}
\end{align}
and
\begin{align}
\mathcal{Z}_{(3k+1)L(k)}^{\rm MQ11} &= - \frac{i}{8(k!)} \bigg(-\frac{214}{105}\bigg)^k \e_{3jl}
 \sum_{n=1}^{\infty} n^{2k+5}
\Big[\underset{(n)}{\hat{I}_{ja}^{00}} \underset{(n)}{\hat{I}_{la}^{11*}} 
+ \underset{(n)}{\hat{I}_{ja}^{11}} \underset{(n)}{\hat{I}_{la}^{00*}} \Big] ,   
\notag 
\\
\mathcal{Z}_{(3k+5/2)L(k)}^{\rm MQ11} &= - \frac{\pi i}{4(k!)} \bigg(-\frac{214}{105}\bigg)^k \e_{3jl}
 \sum_{n=1}^{\infty} n^{2k+6} \Big[\underset{(n)}{\hat{I}_{ja}^{00}} \underset{(n)}{\hat{I}_{la}^{11*}} 
+ \underset{(n)}{\hat{I}_{ja}^{11}} \underset{(n)}{\hat{I}_{la}^{00*}} \Big],    
\notag 
\\
\mathcal{Z}_{(3k+1)L(k)}^{\rm MQ12} &=  \frac{i}{8(k-1)!} \bigg(-\frac{214}{105}\bigg)^k \e_{3jl}
 \sum_{n=1}^{\infty} n^{2k+5}  \underset{(n)}{\hat{I}_{ja}^{00}}  \underset{(n)}{\hat{I}_{la}^{00*}}  ,   
\notag 
\\
\mathcal{Z}_{(3k+5/2)L(k)}^{\rm MQ12} &= \frac{(2k+1) \pi i}{8(k!)} \bigg(-\frac{214}{105}\bigg)^k \e_{3jl}
 \sum_{n=1}^{\infty} n^{2k+6}  \underset{(n)}{\hat{I}_{ja}^{00}}  \underset{(n)}{\hat{I}_{la}^{00*}}.
\end{align}

Unfortunately, if the above conjecture were to break down for some $k$, the representations for 
$\mathcal{R}^{\rm MQ12}$ and $\mathcal{Z}^{\rm MQ12}$ would cease to hold.  However, we would expect that 
the MQ11 summations, as well as all components of $\mathcal{R}^{\rm MQ0}$ and $\mathcal{Z}^{\rm MQ0}$, would 
continue to remain valid.

\subsection{Assembling the complete 1PN log sequences}

We now draw together all of the preceding computations into compact expressions for the terms in each 1PN 
logarithm sequence.  To make this assembly for, say, the integer-order energy flux terms involve the following sum 
of terms 
\be
\mathcal{R}_{(3k+1)L(k)}  = \mathcal{R}_{(3k+1)L(k)}^{\rm MQ01} + \mathcal{R}_{(3k+1)L(k)}^{\rm MQ02} +
\mathcal{R}_{(3k+1)L(k)}^{\rm MQ03} + \nu \big(\mathcal{R}_{(3k+1)L(k)}^{\rm MQ11} 
+ \mathcal{R}_{(3k+1)L(k)}^{\rm MQ12} \big)
+ \mathcal{R}_{(3k+1)L(k)}^{\rm MO} + \mathcal{R}_{(3k+1)L(k)}^{\rm CQ} .
\ee
The full expressions for the integer-order and half-integer-order energy fluxes are given by
\begin{align}
\label{eqn:1PNRk}
\mathcal{R}_{(3k+1)L(k)} & = \frac{1}{16 (k!)} \left(-\frac{214}{105}\right)^k \sum_{n=1}^{\infty} 
\bigg[  \Big( \frac{6k+18}{1-e_t^2} \Big) \Big( \sum_{p= -2,2} n^{2k+5} p \, 
| \underset{(n,p)}{\hat{I}_{ij}^{00}} |^2\Big) +
n^{2k+6} \Big(\underset{(n)}{\hat{I}_{ij}^{00}} \underset{(n)}{\hat{I}_{ij}^{10*}} 
+ \underset{(n)}{\hat{I}_{ij}^{10}} \underset{(n)}{\hat{I}_{ij}^{00*}} \Big) 
\notag 
\\
&- \Big( \frac{6k+18}{1-e_t^2} \Big)  \Big( n^{2k+6}  | \underset{(n)}{\hat{I}_{ij}^{00}} |^2 \Big)
- \nu \Big( n^{2k+6} (k) | \underset{(n)}{\hat{I}_{ij}^{00}} |^2 \Big) + \nu \, n^{2k+6}
\Big(\underset{(n)}{\hat{I}_{ij}^{00}} \underset{(n)}{\hat{I}_{ij}^{11*}} 
+ \underset{(n)}{\hat{I}_{ij}^{11}} \underset{(n)}{\hat{I}_{ij}^{00*}} \Big) \bigg] 
\notag 
\\ 
&+ \, (1- 4\nu) \bigg(-\frac{26}{21}\bigg)^{k}\bigg(\frac{1}{k!}\bigg)\sum_{n=1}^{\infty} n^{2k}h(n,e_t)
+ (1 - 4\nu) \bigg(-\frac{214}{105}\bigg)^{k}\bigg(\frac{1}{k!}\bigg)\sum_{n=1}^{\infty} n^{2k}k(n,e_t),
\end{align}
and
\begin{align}
\mathcal{R}_{(3k+5/2)L(k)} & = \frac{\pi }{8 (k!)} \bigg(-\frac{214}{105}\bigg)^k \sum_{n=1}^{\infty} 
\bigg[ \Big( \frac{6k+21}{1-e_t^2}\Big) 
\Big( \sum_{p= -2,2} n^{2k+6} p \, | \underset{(n,p)}{\hat{I}_{ij}^{00}} |^2\Big) +
n^{2k+7} \Big(\underset{(n)}{\hat{I}_{ij}^{00}} \underset{(n)}{\hat{I}_{ij}^{10*}} 
+ \underset{(n)}{\hat{I}_{ij}^{10}} \underset{(n)}{\hat{I}_{ij}^{00*}} \Big) 
\notag 
\\
&-  \Big( \frac{6k+21}{1-e_t^2}\Big) \Big( n^{2k+7}  | \underset{(n)}{\hat{I}_{ij}^{00}} |^2 \Big)
- \frac{\nu}{2} \Big( n^{2k+7} (2 k + 1) | \underset{(n)}{\hat{I}_{ij}^{00}} |^2 \Big) + \nu \, n^{2k+7}
\Big(\underset{(n)}{\hat{I}_{ij}^{00}} \underset{(n)}{\hat{I}_{ij}^{11*}} 
+ \underset{(n)}{\hat{I}_{ij}^{11}} \underset{(n)}{\hat{I}_{ij}^{00*}} \Big) \bigg] 
\notag 
\\ 
&+ \, (1- 4\nu) \bigg(-\frac{26}{21}\bigg)^{k}\bigg(\frac{2 \pi}{k!}\bigg)\sum_{n=1}^{\infty} n^{2k+1}h(n,e_t)
+ (1 - 4\nu) \bigg(-\frac{214}{105}\bigg)^{k}\bigg(\frac{2 \pi}{k!}\bigg)\sum_{n=1}^{\infty} n^{2k+1}k(n,e_t) .
\end{align}
In these expressions (and in the angular momentum analogs that will follow), we emphasize once again that the 
validity of the portion from MQ12, which determines in part the linear-in-$\nu$ piece of the flux, depends on the 
conjecture made in the previous subsection.  If that supposition were to fail at some PN order, these expressions 
would not be accurate at 1st order in $\nu$ but would, of course, continue to be valid for the $\mathcal{O}(\nu^0)$ 
portion.  

The last essential consideration when using these expressions to generate high-order eccentricity functions or power 
series is that of their eccentricity singular behavior.  As mentioned in Sec.~\ref{sec:singFacs}, past work 
\cite{ForsEvanHopp16,MunnEvan19a,MunnETC20,LoutYune17,Munn20} shows that each 1PN logarithm will be characterized 
by a divergence as $e_t \rightarrow 1$ in the form of an eccentricity singular factor.  For PN order $r$, that singular 
factor will have the form $(1-e_t^2)^{-(r+7/2)}$.  In fact, once we account for the presence of a singular factor 
$(1-e_t^2)^{-(3k+9/2)}$, we find closed-form expressions for the integer-order terms $\mathcal{R}_{(3k+1)L(k)}$.  
The half-integer sequence $\mathcal{R}_{(3k+5/2)L(k)}$ almost surely admits no closed representations.  However, 
here too the removal of the singular factor $(1-e_t^2)^{-(3k+6)}$ is beneficial, and leads to a remaining power 
series that is convergent as $e_t \rightarrow 1$.  We have demonstrated convergence in these terms to 22PN
through direct eccentricity expansion to high order.

Returning to the assembly of the entire flux terms, the terms in the angular momentum 1PN log sequences are given by
\be
\mathcal{Z}_i = \mathcal{Z}_{i}^{\rm MQ01} + \mathcal{Z}_{i}^{\rm MQ02} +
\mathcal{Z}_{i}^{\rm MQ03} + \nu \big(\mathcal{Z}_{i}^{\rm MQ11} + \mathcal{Z}_{i}^{\rm MQ12} \big)
+ \mathcal{Z}_{i}^{\rm MO} + \mathcal{Z}_{i}^{\rm CQ},
\ee
which for the integer-order sequence can be shown to be
\begin{align}
\label{eqn:1PNZk}
\mathcal{Z}_{(3k+1)L(k)} & = \frac{-i}{8 (k!)} \bigg(-\frac{214}{105}\bigg)^k \e_{3jl} \sum_{n=1}^{\infty} 
\bigg[  \Big( \frac{6k+15}{1-e_t^2}\Big) 
\Big( \sum_{p= -2,2}  n^{2k+4} p \, \underset{(n,p)}{\hat{I}_{ja}^{00}} \underset{(n,p)}{\hat{I}_{la}^{00*}} \Big) +
n^{2k+5} \Big(\underset{(n)}{\hat{I}_{ja}^{00}} \underset{(n)}{\hat{I}_{la}^{10*}} 
+ \underset{(n)}{\hat{I}_{ja}^{10}} \underset{(n)}{\hat{I}_{la}^{00*}} \Big) 
\notag 
\\
& -  \Big( \frac{6k+15}{1-e_t^2}\Big) \Big( n^{2k+5} \underset{(n)}{\hat{I}_{ja}^{00}} \underset{(n)}{\hat{I}_{la}^{00*}} \Big)
- \nu \Big( n^{2k+5} (k) \underset{(n)}{\hat{I}_{ja}^{00}} \underset{(n)}{\hat{I}_{la}^{00*}}  \Big) + \nu \, n^{2k+5}
\Big(\underset{(n)}{\hat{I}_{ja}^{00}} \underset{(n)}{\hat{I}_{la}^{11*}} 
+ \underset{(n)}{\hat{I}_{ja}^{11}} \underset{(n)}{\hat{I}_{la}^{00*}} \Big) \bigg] 
\notag 
\\ 
&+ \, (1- 4\nu) \bigg(-\frac{26}{21}\bigg)^{k}\bigg(\frac{1}{k!}\bigg)\sum_{n=1}^{\infty} n^{2k} \tilde{h}(n,e_t)
+ (1 - 4\nu) \bigg(-\frac{214}{105}\bigg)^{k}\bigg(\frac{1}{k!}\bigg)\sum_{n=1}^{\infty} n^{2k}\tilde{k}(n,e_t),
\end{align}
and for the half-integer-order sequence becomes
\begin{align}
\mathcal{Z}&_{(3k+5/2)L(k)}  = \frac{- \pi i}{4 (k!)} \bigg(-\frac{214}{105}\bigg)^k \e_{3jl} \sum_{n=1}^{\infty} 
\bigg[ \Big( \frac{6k+18}{1-e_t^2}\Big)
\Big( \sum_{p= -2,2}  n^{2k+5} p \, \underset{(n,p)}{\hat{I}_{ja}^{00}} \underset{(n,p)}{\hat{I}_{la}^{00*}} \Big) +
n^{2k+6} \Big(\underset{(n)}{\hat{I}_{ja}^{00}} \underset{(n)}{\hat{I}_{la}^{10*}} 
+ \underset{(n)}{\hat{I}_{ja}^{10}} \underset{(n)}{\hat{I}_{la}^{00*}} \Big) 
\notag 
\\
&-\Big( \frac{6k+18}{1-e_t^2}\Big) \Big( n^{2k+6} \underset{(n)}{\hat{I}_{ja}^{00}} \underset{(n)}{\hat{I}_{la}^{00*}}  \Big)
- \frac{\nu}{2} \Big( n^{2k+6} (2 k + 1)  \underset{(n)}{\hat{I}_{ja}^{00}} \underset{(n)}{\hat{I}_{la}^{00*}} \Big) 
+ \nu \, n^{2k+6}
\Big(\underset{(n)}{\hat{I}_{ja}^{00}} \underset{(n)}{\hat{I}_{la}^{11*}} 
+ \underset{(n)}{\hat{I}_{ja}^{11}} \underset{(n)}{\hat{I}_{la}^{00*}} \Big) \bigg] 
\notag 
\\ 
&+ \, (1- 4\nu) \bigg(-\frac{26}{21}\bigg)^{k}\bigg(\frac{2 \pi}{k!}\bigg)\sum_{n=1}^{\infty} n^{2k+1}\tilde{h}(n,e_t)
+ (1 - 4\nu) \bigg(-\frac{214}{105}\bigg)^{k}\bigg(\frac{2 \pi}{k!}\bigg)\sum_{n=1}^{\infty} n^{2k+1}\tilde{k}(n,e_t).
\end{align}
To reduce further, the relevant singular factors, which are respectively $(1-e_t^2)^{-(3k+3)}$ and 
$(1-e_t^2)^{-(3k+9/2)}$, would be pulled out.  While it is difficult to see until after that step and after the 
source multipoles are inserted and expanded, the integer-order flux terms all produce residual polynomials in 
$e_t^2$ while the half-integer-order terms have residual convergent power series.

\subsection{Some explicit results from the 1PN log sequences}
\label{sec:explicit1PNlogs}

These formulas can now be utilized to generate explicit eccentricity functions or power series for higher-order 
members of the 1PN log sequences.  In fact, each term from 4PN to 8.5PN at lowest order in $\nu$ has already been 
calculated to high order in Darwin $e$ in a companion paper \cite{MunnETC20} to this one and Paper I.  Those results 
were obtained by combining BHPT numerical calculations with the PSLQ integer-relation algorithm on a $lmn$ mode 
basis to extract the coefficients in analytic form.  The eccentricity functions in that paper (upon conversion from 
$e$ to $e_t$) provide a valuable check on our results.  Unfortunately, the portions at next order in $\nu$ cannot 
be similarly validated by BHPT yet and thus remain a conjecture as discussed in the previous two subsections.  

We consider first the pair of fluxes at 4PN log order, $\mathcal{R}_{4L}$ and $\mathcal{Z}_{4L}$, which are the second 
elements in the integer-order 1PN log sequences.  With the appropriate eccentricity singular function removed, we find 
that each provides a closed-form expression
\begin{align}
\label{eqn:R4L}
\mathcal{R}_{4L}(e_t,\nu) &= \frac{1}{(1-e_t^2)^{15/2}} \left(\frac{232597}{8820}-\frac{1020553 e_t^2}{5880}
-\frac{85136197 e_t^4}{35280}-\frac{194295169 e_t^6}{70560}-\frac{570319469 e_t^8}{1128960}
-\frac{1677429 e_t^{10}}{250880} \right)  
\notag 
\\ 
& \qq + \frac{\nu}{(1-e_t^2)^{15/2}} \left(\frac{34889}{735}+\frac{7327462 e_t^2}{6615}+\frac{1218464 e_t^4}{315}
+\frac{10895825 e_t^6}{3528}+\frac{89571401  e_t^8}{169344}+\frac{284187 e_t^{10}}{31360}\right) ,  
\end{align}
\begin{align}
\label{eqn:Z4L}
\mathcal{Z}_{4L}(e_t,\nu) &= \frac{1}{(1-e_t^2)^6} \left(\frac{232597}{8820}-\frac{1761619 e_t^2}{8820}
-\frac{6412241 e_t^4}{7056}-\frac{22800487 e_t^6}{70560}-\frac{803961 e_t^8}{125440} \right)  
\notag 
\\ 
& + \frac{\nu}{(1-e_t^2)^6} \left( \frac{34889}{735}+\frac{2961167 e_t^2}{4410}+\frac{4210447 e_t^4}{3528}
+\frac{12389779 e_t^6}{35280}+\frac{413339 e_t^8}{47040} \right) .
\end{align}
The order $\nu^0$ part of $\mathcal{R}_{4L}(e_t)$ was previously discovered and described in \cite{ForsEvanHopp16} 
(actually as a closed-form function $\mathcal{L}_{4L}(e)$ in $e$ which is easily converted from $e$ to $e_t$ to 
compare to $\mathcal{R}_{4L}(e_t)$).  The order $\nu^0$ part of $\mathcal{Z}_{4L}(e_t)$ was also effectively 
previously found \cite{Fors16} (again as a closed-form function $\mathcal{J}_{4L}(e)$ in $e$, convertible to 
$\mathcal{Z}_{4L}(e_t)$). 

Turning next to the 5.5PN log fluxes, which are the second elements in the half-integer-order 1PN log sequences, 
we find a pair of convergent infinite series that begin with
\begin{align}
\mathcal{R}_{11/2L}(e_t, \nu) & = \frac{1}{(1-e_t^2)^9}\bigg(\frac{177293}{2352}-\frac{33467083 e_t^2}{35280}
-\frac{55188741467 e_t^4}{2257920}-\frac{41837812001 e_t^6}{677376}-\frac{90806230499893 e_t^8}{2601123840} 
\notag 
\\
& -\frac{52599122781397 e_t^{10}}{13005619200}-\frac{743899262983 e_t^{12}}{18874368000}
-\frac{1035559450883 e_t^{14}}{32774160384000} + \cdots \bigg)  
\notag 
\\ 
& + \frac{\nu}{(1-e_t^2)^9}\bigg(\frac{2880637}{8820}+\frac{168033239 e_t^2}{15120}
+\frac{107321219093 e_t^4}{1693440}+\frac{132390912895 e_t^6}{1354752}
+\frac{86038451049547 e_t^8}{1950842880}  
\notag 
\\
& +\frac{956641092501793 e_t^{10}}{195084288000}+\frac{223041145823171 e_t^{12}}{3745618329600}
+\frac{493441625881693 e_t^{14}}{3670705963008000} + \cdots \bigg),    
\notag 
\\
\mathcal{Z}_{11/2L}(e_t, \nu) &= \frac{1}{(1-e_t^2)^{15/2}} \bigg(\frac{177293}{2352}-\frac{31011679 e_t^2}{23520}
-\frac{8782000063 e_t^4}{752640}-\frac{11351372383  e_t^6}{903168}-\frac{661373524417 e_t^8}{289013760}
\notag 
\\
& -\frac{102517313999 e_t^{10}}{3715891200}-\frac{8739859767749 e_t^{12}}{8323596288000}
+\frac{70533378963781 e_t^{14}}{349591044096000} + \cdots \bigg) 
\notag 
\\ 
& + \frac{\nu}{(1-e_t^2)^{15/2}} \bigg(\frac{2880637}{8820}+\frac{349633 e_t^2}{49}+\frac{127083053 e_t^4}{5376}
+\frac{9184388503 e_t^6}{508032}+\frac{647130113243 e_t^8}{216760320}  
\notag 
\\
& +\frac{199406895011 e_t^{10}}{4064256000}+\frac{214645510549 e_t^{12}}{2080899072000}
-\frac{3051343498747 e_t^{14}}{152946081792000} + \cdots \bigg).
\end{align}

The third elements in the integer-order 1PN log sequences are the 7PN $\log^2(x)$ fluxes.  These flux contributions 
also have closed-form expressions, as anticipated
\begin{align}
\mathcal{R}_{7L2}(e_t,\nu) &= \frac{1}{(1-e_t^2)^{21/2}} \bigg(-\frac{52525903}{617400}
-\frac{3278576291 e_t^2}{3704400}+\frac{74771270471 e_t^4}{2116800}
+\frac{6205145514833 e_t^6}{29635200}+\frac{1870078245281 e_t^8}{6773760}  
\notag 
\\
& +\frac{6721297402199 e_t^{10}}{67737600}+\frac{728355846523 e_t^{12}}{90316800}
+\frac{976909819 e_t^{14}}{16859136}\bigg)   
\notag 
\\
& + \frac{\nu}{(1-e_t^2)^{21/2}} \bigg(\frac{146507731}{463050}+\frac{11066346794 e_t^2}{694575}
+\frac{32900565808 e_t^4}{231525}   +  \frac{8296487145703 e_t^6}{22226400}
+\frac{401207614757 e_t^8}{1209600}   
\notag 
\\
& +\frac{164075647087 e_t^{10}}{1693440}
+\frac{30213176177 e_t^{12}}{4064256}+\frac{812183599 e_t^{14}}{12644352}\bigg),   
\notag 
\\
\mathcal{Z}_{7L2}(e_t,\nu) &= \frac{1}{(1-e_t^2)^{9}} \bigg(-\frac{52525903}{617400}
+\frac{1285805299 e_t^2}{1852200}+\frac{346615894427 e_t^4}{14817600}
+\frac{449325541709  e_t^6}{7408800}+\frac{327681660679 e_t^8}{9483264}  
\notag 
\\
& +\frac{189470991041 e_t^{10}}{47416320}+\frac{1602170813 e_t^{12}}{42147840}\bigg)  
+ \frac{\nu}{(1-e_t^2)^{9}} \bigg(-\frac{146507731}{463050}-\frac{9851216819 e_t^2}{926100}   
\notag 
\\ 
& -\frac{220777622729 e_t^4}{3704400}-\frac{332158696507  e_t^6}{3704400}
-\frac{463335196631 e_t^8}{11854080}  
-\frac{97794785317 e_t^{10}}{23708160}-\frac{161911009 e_t^{12}}{3512320}  \bigg).
\end{align}

At order $\nu^0$ these functions and power series show complete agreement with those found using BHPT fitting.  
The convergent power series for $\mathcal{R}_{11/2L}$ and $\mathcal{Z}_{11/2L}$ were verified to $e_t^{30}$ in 
the power series expansion and those for $\mathcal{R}_{17/2L2}$ and $\mathcal{Z}_{17/2L2}$ were checked and verified 
to order $e_t^{20}$.  Additionally, we extended the validation to 22PN at the level 
$e_t^{10}$ by combining BHPT results with Johnson-McDaniel's $S_{lmn}$ factorization \cite{JohnMcDa14} (see 
Sec.~IV D of Paper I), again at $\mathcal{O}(\nu^0)$.  As we will explain in the next subsection, we now have the 
means to compute these and all other members of the 1PN log series to at least $e_t^{120}$ with manageable 
computational cost. 
\end{widetext}

\subsection{Discussion}

To summarize, despite an increase in calculational complexity, the pair of 1PN log sequences (shown in blue in 
Fig.~\ref{fig:logSequences}) are determined in their entirety by a few low-order source multipoles---namely, the 
Newtonian mass octupole and current quadrupole moments and the 1PN-order mass quadrupole moment.  This behavior 
is exactly analogous to, if more complicated than, the way the Newtonian quadrupole moment provided all the 
information necessary to derive all elements of the leading-log sequences (as shown in Paper I).  The Fourier 
amplitudes of these moments appear in sums as complex products weighted by successively higher powers of $n$, the 
harmonics of orbital frequency that are present in eccentric motion.  As such, these terms represent in the time 
domain higher and higher order time derivatives of the low-order source multipole moments. 

The greater complexity is due in part to the fact that the 1PN quadrupole moment gives rise to five different 
sums over squares of Fourier amplitudes.  In compensation, however, simplifying patterns emerge amongst these sums.  
For example, we found an exact correspondence between the higher-order quadrupole sums MQ01 and MQ11 and the sums 
over the Newtonian-order quadrupole moment in the leading-logarithm sequence.  Specifically, the substitution
$I^{00} \rightarrow I^{10}$ or $I^{00} \rightarrow I^{11}$ in terms where the former appears, along with changes 
in the normalization, leads to parts of the flux at 1PN order higher.  Secondly, a relationship exists between the 
sums we denoted by MQ02 and MQ03, which are related to the 1PN correction in the frequency $\O_r$ and the ``magnetic'' 
harmonics, $p$, respectively.  The $k$-dependent prefactors on these sums turn out to be the additive inverse of 
each other.  The reason for this symmetry is that the harmonics (as defined and manipulated in 
Sec.~\ref{sec:MQManip}) ultimately satisfy $m = -n$ and $s = -p$, given orthogonality, and so $\O_r$ and 
$p$ only appear in the combination $\pm \O_r (n + p k)$.  Through 1PN order this can be rewritten as 
$\pm \O_{\vp} [n + (p-n) k]$, which means that a 1PN contribution will emerge with $p - n$ times the rest of the 
quadrupole factors.  We had simply split this into two separate sums originally, with otherwise identical forms.

The open question concerns the sum that we labeled MQ12, which involves the appearance of the $I^{00}$ 
(Newtonian quadrupole) at next order in the mass ratio and which first arises with the ADM mass at 2.5PN order.  
As we mentioned in Sec.~\ref{sec:1PNlogsMQNu}, in PN theory it is expected that progressively higher powers of 
the ADM mass will appear in progressively higher corrections to the tail.  Thus, we expect that this will lead to a 
simple factor from the relevant binomial expansion of $(1 - \nu x/2)^q$.  However, it is not clear how else the 
Newtonian quadrupole might manifest at this order in $\nu$.  If, for instance, the ADM mass in the tail were the sole 
appearance of this type of sum, then the partial cancellation between instantaneous and hereditary contributions 
discussed in \cite{MunnEvan19a} would not occur, enhancing any orders with both types of flux by a factor of 3.  
According to \cite{ArunETC08b}, this would include all orders 3PN and above.  However, this would leave an 
unphysical normalization constant $r_0$ in the full flux (see, for example, Sec.~\ref{sec:fullTail}), which cannot 
exist.  Therefore, the likeliest possibility is that a corresponding summation exists on the instantaneous side and 
the cancellation seen at $\mathcal{O}(\nu^0)$ does continue here, leading to the result above.  Regardless, further 
developments in full PN theory or second order BHPT should soon be able to resolve this question definitively.
At that point, even if our conjecture of Sec.~\ref{sec:1PNlogsMQNu} fails to hold, the Fourier infrastructure 
presented here should be able to provide accurate $\mathcal{O}(\nu)$ expansions in eccentricity for all elements of 
the 1PN log sequences once the correct prefactor is supplied by other means.

Equally important to the generation of high-order expansions is the question of computational implementation
and cost.  The procedures we describe in this paper turn out to be quite manageable computationally, though the 
calculation of complete flux terms tends to be more than an order of magnitude more time-consuming than the leading 
logarithm calculations of Paper I.  Of the seven required sums, three (MQ02, MQ12, CQ) are roughly equal in expense 
to the corresponding leading logarithms.  Three (MQ03, MQ11, MO) are 1.5-4 times more expensive to compute, owing 
to their lengthier Bessel function representations.  In any event, calculation of all of these terms only amounts to 
a matter of at most minutes for computation to hundreds of orders in $e_t$ on an average laptop in 
\textsc{Mathematica}.

However, the remaining summation MQ01, with the 1PN amplitudes ${}_{(n)}{I_{ij}^{10}}$, is the ultimate bottleneck.  
As noted in Sec.~\ref{sec:1PNmassquad}, these Fourier coefficients cannot be expressed cleanly in terms of Bessel 
functions, and the unevaluated integral in \eqref{eqn:I1PNint} is cumbersome to handle.  We had partial success
in handling it by expanding the integrand in $e_t$ directly before integrating.  However, the arctangent function with 
its complicated argument remained a prime source of difficulty, leading to a series of integrals that can require hours 
to expand, as well as require large quantities of memory, on cluster computers that support \textsc{Mathematica}.  
We found that a convenient way to proceed was to precompute the expansion of this arctangent function on the UNC 
cluster KillDevil to $e_t^{120}$, a task which required about 1.5 hours and 20 GB of RAM.  Once this expansion was 
calculated, the rest of the process became much more manageable.  Indeed, with the arctangent series in hand, we 
are now able to expand any element in the 1PN log sequences to $e_t^{120}$ via laptop in only 
a few minutes.  This process was used in particular to expand $\psi(e_t)$ and $\tilde{\psi}(e_t)$ to $e_t^{120}$, 
enhancement functions which are discussed in \cite{MunnETC20}.  Another difficult function, $\mathcal{R}_4^{\chi}$ 
(described below), can also be obtained to $e_t^{120}$ in this manner.  

\section{Deriving an essential part of the 4PN tail}
\label{sec:4PNTail}

Up to now we have focused on the 1PN log sequences of gravitational wave fluxes (depicted by the blue lines in 
Fig.~\ref{fig:logSequences}).  Drawing upon the frequency domain multipole analysis in Sec.~\ref{sec:PNexp}, we 
re-derived the known 1PN and 2.5PN relative fluxes in Sec.~\ref{sec:knownLogs}.  We then used that frequency domain 
approach in Sec.~\ref{sec:1PNLogs} to detail the analytic dependence of elements in those sequences to all higher 
PN orders.  What remains, for this section and Sec.~\ref{sec:full4PN}, is to apply a similar approach to the 4PN log 
sequences (i.e., the orange lines in Fig.~\ref{fig:logSequences}).
 
Like the subleading log sequences of Paper I (what we call here the 3PN logs), the derivation of the form of the 
4PN logs requires an assist from BHPT.  As Paper I showed, it is possible to find a theoretical explanation for 
part of each subleading log term (even absent a full PN calculation) that is based merely on knowledge of the Newtonian 
quadrupole moment.  The remaining part of each subleading log term can then in principle be determined, at lowest 
order in $\nu$, by BHPT.  A similar useful split carries over to the elements in the 4PN log sequences, though it 
requires the 1PN source multipoles.  

Because the process is somewhat involved, we focus primarily on illustrating how it is applied to the 4PN non-log 
fluxes, $\mathcal{R}_{4}(e_t)$ and $\mathcal{Z}_{4}(e_t)$, the first elements in the integer-order 4PN log sequences.  
(Sec.~\ref{sec:full4PN} also briefly touches on the 5.5PN non-log term, which is the first element in the 
half-integer-order 4PN log sequence.)  We find that an essential tail portion of these 4PN terms is 
theoretically determined by the same 1PN source multipoles that were discussed in Sec.~\ref{sec:PNexp}.  Deriving 
that tail portion is the subject of this section.  Once this essential 4PN tail portion is known, we combine it with 
knowledge of the 4PN log flux from Sec.~\ref{sec:1PNLogs} and results \cite{MunnETC20} from BHPT to determine the 
entire analytic form of the 4PN non-log fluxes $\mathcal{R}_{4}(e_t)$ and $\mathcal{Z}_{4}(e_t)$ to high order in 
an expansion in eccentricity.  This result is timely, as it will provide a valuable check for those working to 
extend PN theory to a full description of the orbital mechanics and radiative losses at 4PN.  

The portion of the 4PN tail to be addressed provides the 1PN correction to the 3PN enhancement function $\chi(e_t)$ 
\cite{ArunETC08a}.  This portion of the full tail is provided by the sum of the tail$^2$ and tail-of-tails 
corrections to the flux, and is determined by the 1PN source multipoles.  The mass octupole and current 
quadrupole orbital computations will remain at Newtonian order, mirroring the derivation of $\chi(e_t)$ itself in 
\cite{ArunETC08a}.  However, as usual, the mass quadrupole part requires extension to 1PN, as discussed in 
Sec.~\ref{sec:PNexp}. 

\begin{widetext}
\subsection{Mass octupole}

For the mass octupole the quadratic in $\mathcal{M}$ portions of the energy flux tail have the following time domain 
expressions 
\begin{gather}
\mathcal{P}^{MO (\text{tail})^2}_\infty =
\frac{4\mathcal{M}^2}{189}\bigg\{\int_{0}^{\infty}I_{ijk}^{(6)}(t-\tau)\bigg[\log\Big(\frac{\tau}{2 r_0}\Big)
+ \frac{97}{60}\bigg] d\tau\bigg\}^2 , 
\\
\mathcal{P}^{MO (\text{tail-of-tails})}_\infty =
\frac{4\mathcal{M}^2}{189}I^{(4)}_{ijk}(t)\int_{0}^{\infty}I_{ijk}^{(7)}(t-\tau)\bigg[\log\Big(\frac{\tau}{2 r_0}\Big)^2
+ \frac{183}{70} \log\Big(\frac{\tau}{2 r_0}\Big)+ \frac{13283}{8820} \bigg] d\tau ,
\end{gather}
where the tail-of-tail coefficients were taken from equation (4.9a) in \cite{MarcBlanFaye16} with constant $b$ set 
to $r_0$.  Note that a factor of 2 is pulled from their equation, with another factor of 2 coming from the polynomial 
product $U_L U_L$.

The MO part of the tail$^2$ term can be evaluated using the integral identity \eqref{eqn:lnTailInt}.  Then, because 
$k=\O_\vp/\O_r - 1 = 0$ and $\mathcal{M} = M$ for a Newtonian orbit, the time average of the MO tail$^2$ term can 
be simplified to
\begin{align}
\label{eqn:MOtailsq}
\bigg<\mathcal{P}^{\text{MO(tail)}^2}_\infty\bigg> 
= \frac{8 M^2}{189} \sum_{n=1}^{\infty} (\O_{\vp})^{10} n^{10} 
| \underset{(n)}{I_{ijk}^{00}} |^2
 \bigg[\frac{\pi^2}{4} + \Big(\log(2 \O_r |n| r_0) + \gamma_E-\frac{97}{60}\Big)^2\bigg] .
\end{align}

The tail-of-tails term requires a bit more work.  First, the $\log^2$ piece must be handled using the following
integral identity \cite{ArunETC08a,LoutYune17,MunnEvan19a}:
\begin{gather}
\label{eqn:lnSqTailInt}
 \int_{0}^{\infty} e^{i(n-sk)\O_r \tau} \log\Big(\frac{\tau}{2 r_0}\Big)^2 d\tau 
= \frac{i}{(n-sk)\O_r}\bigg[\frac{\pi^2}{6} 
+ \Big(\frac{\pi i}{2}\text{sign}(-n) + \log(2 \O_r |n-sk| r_0) + \gamma_E \Big)^2 \bigg]  \\
\approx \frac{i}{n\O_r}  \bigg[-\frac{\pi^2}{12} + \Big(\log(2 \O_{\vp} |n| r_0) + \gamma_E \Big)^2
+ \pi i \text{sign}(-n) \Big( \log(2 \O_{\vp} |n| r_0) + \gamma_E \Big)  \bigg],
\label{eqn:lnSq0PN}
\end{gather}
where in the second line we set $k=0$ and made a lowest-order PN expansion.  When the various factors of $i$ and 
$n$ are considered, it becomes clear that the last term in \eqref{eqn:lnSq0PN} cancels in a sum over positive and 
negative $n$.  Once combined with the rest of the integral, the total tail-of-tails contribution has the following 
time average
\begin{align}
\bigg<\mathcal{P}^{\text{MO(tail-of-tails)}}_\infty\bigg> 
= \frac{8 M^2}{189} \sum_{n=1}^{\infty} (\O_{\vp})^{10} n^{10}
| \underset{(n)}{I_{ijk}^{00}} |^2 \bigg[\frac{\pi^2}{12} -  \Big( \log(2 \O_{\vp} |n| r_0) + \gamma_E \Big)^2 
+ \frac{183}{70} \Big( \log(2 \O_{\vp} |n| r_0) + \gamma_E \Big) - \frac{13283}{8820}\bigg].
\label{eqn:MOtailoftails}
\end{align}
Then, \eqref{eqn:MOtailsq} and \eqref{eqn:MOtailoftails} are summed to yield the complete mass octupole flux 
contribution
\begin{align}
\bigg<\mathcal{P}^{\text{MO(tail)}^2+\text{(tail-of-tails)}}_\infty\bigg>
= \frac{8 M^2}{189} (\O_{\vp})^{10} \sum_{n=1}^{\infty}  n^{10}
| \underset{(n)}{I_{ijk}^{00}} |^2 \bigg[\frac{\pi^2}{3} - \frac{13}{21}\Big(\log(2 \O_\vp |n| r_0) + \gamma_E\Big)
 + \frac{21709}{19600}\bigg] .
\end{align}

Likewise, the angular momentum expressions have the following time dependent forms:
\begin{align}
\mathcal{G}^{\text{MO(tail)}^2}_\infty =
\frac{4\mathcal{M}^2}{63}\epsilon_{3jl} \bigg\{\int_{0}^{\infty}I_{jab}^{(5)}(t-\tau)
\bigg[\log\Big(\frac{\tau}{2 r_0}\Big) +\frac{97}{60}\bigg] d\tau \bigg\} \times \bigg\{ 
\int_{0}^{\infty}I_{lab}^{(6)}(t-\tau)
\bigg[\log\Big(\frac{\tau}{2 r_0}\Big) +\frac{97}{60}\bigg] d\tau \bigg\} \hat{z} ,
\\
\mathcal{G}^{\text{MO(tail-of-tails)}}_\infty =
\frac{2\mathcal{M}^2}{63} \e_{3jl} \bigg\{I_{jab}^{(3)} \int_{0}^{\infty}I_{lab}^{(7)}(t-\tau)
\bigg[\log\Big(\frac{\tau}{2 r_0}\Big)^2 +
\frac{183}{70} \log\Big(\frac{\tau}{2 r_0}\Big)+ \frac{13283}{8820} \bigg] d\tau  \qq \qq \qq 
\notag 
\\
+ I_{lab}^{(4)} \int_{0}^{\infty}I_{jab}^{(6)}(t-\tau)
\bigg[\log\Big(\frac{\tau}{2 r_0}\Big)^2 +
\frac{183}{70} \log\Big(\frac{\tau}{2 r_0}\Big)+ \frac{13283}{8820} \bigg]  d\tau \bigg\} \hat{z} .
\end{align}
These merge together in the same way to generate the complete mass octupole flux contribution
\begin{align}
\bigg<\mathcal{G}^{\text{MO(tail)}^2+\text{(tail-of-tails)}}_\infty\bigg> =
- \frac{8 M^2 i}{63} (\O_{\vp})^{9} \e_{3jl} \sum_{n=1}^{\infty}  n^{9} \underset{(n)}{I_{jab}^{00}} 
\underset{(n)}{I_{lab}^{00*}} \bigg[\frac{\pi^2}{3} - \frac{13}{21}\Big(\log(2 \O_\vp |n| r_0) + \gamma_E\Big)
+ \frac{21709}{19600}\bigg] \hat{z} .
\end{align}

\subsection{Current quadrupole}

The next component of the quadratic-in-$\mathcal{M}$ 4PN tail stems from the Newtonian current quadrupole.  The
energy and angular momentum time domain representations are
\begin{gather}
\mathcal{P}^{CQ (\text{tail})^2}_\infty =
\frac{64\mathcal{M}^2}{45}\bigg\{\int_{0}^{\infty}J_{ij}^{(5)}(t-\tau)\bigg[\log\Big(\frac{\tau}{2 r_0}\Big)
+ \frac{7}{6}\bigg] d\tau\bigg\}^2, \notag \\
\mathcal{P}^{CQ (\text{tail-of-tails})}_\infty =
\frac{64\mathcal{M}^2}{45}J^{(3)}_{ij}(t)\int_{0}^{\infty}J_{ij}^{(6)}(t-\tau)\bigg[\log\Big(\frac{\tau}{2 r_0}\Big)^2
+\frac{46}{35} \log\Big(\frac{\tau}{2 r_0}\Big) - \frac{26254}{22050} \bigg] d\tau ,
\end{gather}
and
\begin{align}
\mathcal{G}^{\text{CQ(tail)}^2}_\infty =
\frac{128\mathcal{M}^2}{45}\epsilon_{3jl} \bigg\{\int_{0}^{\infty}J_{ja}^{(4)}(t-\tau)
\bigg[\log\Big(\frac{\tau}{2 r_0}\Big) + \frac{7}{6}\bigg] d\tau \bigg\} \times \bigg\{ 
\int_{0}^{\infty}J_{la}^{(5)}(t-\tau)
\bigg[\log\Big(\frac{\tau}{2 r_0}\Big) +  \frac{7}{6}\bigg] d\tau \bigg\} \hat{z} , \\
\mathcal{G}^{\text{CQ(tail-of-tails)}}_\infty =
\frac{64\mathcal{M}^2}{45}  \e_{3jl} \bigg\{J_{ja}^{(2)} \int_{0}^{\infty}J_{la}^{(6)}(t-\tau)
\bigg[\log\Big(\frac{\tau}{2 r_0}\Big)^2
+\frac{46}{35} \log\Big(\frac{\tau}{2 r_0}\Big) - \frac{26254}{22050} \bigg]  d\tau  \qq \qq \qq \notag \\
+ J_{la}^{(3)} \int_{0}^{\infty}I_{ja}^{(5)}(t-\tau)
\bigg[\log\Big(\frac{\tau}{2 r_0}\Big)^2
+\frac{46}{35} \log\Big(\frac{\tau}{2 r_0}\Big) - \frac{26254}{22050} \bigg]  d\tau \bigg\} \hat{z} ,
\end{align}
respectively.  Again, the particular forms the two tails-of-tails were adapted from \cite{MarcBlanFaye16}.  The 
time averaged fluxes are then found to be
\begin{gather}
\bigg<\mathcal{P}^{\text{CQ(tail)}^2+\text{(tail-of-tails)}}_\infty\bigg>
= \frac{128 M^2}{45} \sum_{n=1}^{\infty} (\O_{\vp})^{8} n^{8}
| \underset{(n)}{I_{ij}^{00}} |^2 \bigg[\frac{\pi^2}{3} 
- \frac{107}{105}\Big(\log(2 \O_\vp |n| r_0) + \gamma_E \Big) + \frac{7517}{44100}\bigg] , 
\\
\bigg<\mathcal{G}^{\text{CQ(tail)}^2+\text{(tail-of-tails)}}_\infty\bigg>
=- \frac{256 M^2}{45} i \e_{3jl} \sum_{n=1}^{\infty} (\O_{\vp})^{7} n^{7}
\underset{(n)}{I_{ja}^{00}} \underset{(n)}{I_{la}^{00*}}  \bigg[\frac{\pi^2}{3} 
- \frac{107}{105}\Big(\log(2 \O_\vp |n| r_0) + \gamma_E \Big) + \frac{7517}{44100}\bigg] .
\end{gather}

\subsection{Mass quadrupole, lowest order in $\nu$}

The remaining order $\mathcal{M}^2$ part of the 1PN correction to the tail$^2$ and the tail-of-tails terms comes 
from the 1PN correction to the mass quadrupole moment.

\subsubsection{The energy flux tail$^2$}

In the time domain, the mass quadrupole part of the tail$^2$ is given by \cite{ArunETC08a}
\begin{equation}
\mathcal{P}^{\text{MQ(tail)}^2}_\infty =
\frac{4\mathcal{M}^2}{5}\bigg\{\int_{0}^{\infty}I_{ij}^{(5)}(t-\tau)\bigg[\log\Big(\frac{\tau}{2 r_0}\Big)
+\frac{11}{12}\bigg] d\tau\bigg\}^2 .
\end{equation}
When the quadrupole moment is taken to leading (Newtonian) order, this term contributes to the 3PN hereditary flux.  
By taking the calculation to one PN order higher approximation, we can obtain its contribution to the 4PN flux.  To 
do so, we plug in the biperiodic Fourier expansion for the quadrupole moment along with the expansion for the ADM 
mass, replace the time derivatives with powers of the frequency, and take the time average.  An intermediate step in 
the calculation is
\begin{gather}
\mathcal{P}^{\text{MQ(tail)}^2}_\infty =
\frac{4M^2}{5} \sum_{n=-\infty}^{\infty} \sum_{p=-2}^{2} (\O_r)^{10} (n^{10} + 10 n^9 p k) 
\underset{(n,p)}{I_{ij}} \underset{(-n,-p)}{I_{ij}}
 \bigg\{ \int_{0}^{\infty} e^{i(n+pk)\O_r \tau}\bigg[\log\Big(\frac{\tau}{2 r_0}\Big)
+\frac{11}{12}\bigg] d\tau \bigg\} \times \notag \\
\bigg\{ \int_{0}^{\infty} e^{-i(n+pk)\O_r \tau}\bigg[\log\Big(\frac{\tau}{2 r_0}\Big)
+\frac{11}{12}\bigg] d\tau \bigg\} .
\end{gather}
The product of integrals can be simplified through a double application of \eqref{eqn:lnTailInt}.  Collecting the 
results of an expansion through first order reduces the product of integrals to
\begin{align}
\frac{1}{n^2 \O_r^2} \left(\frac{\pi^2}{4} + \beta_0^2 \right) 
-\frac{1}{2 n^2 \O_r^2}\left[4 \beta_0 + \frac{p}{n} (\pi^2 + 4 \beta_0^2 - 4 \beta_0) \right] k ,
\end{align}
where we define $\beta_0 \equiv \log(2 \O_{\vp} |n| r_0) + \gamma_E - 11/12$.  The final result reduces 
to three compact sums:
\begin{align}
\bigg<\mathcal{P}^{\text{MQ01(tail)}^2}_\infty\bigg>
 &= \frac{2 M^2 (\O_{\vp})^8}{5} \sum_{n=1}^{\infty} (n^8)
\Big(\underset{(n)}{I_{ij}^{10}} \underset{(n)}{I_{ij}^{00*}} 
+ \underset{(n)}{I_{ij}^{00}} \underset{(n)}{I_{ij}^{10*}} \Big) \ \left(\pi^2 + 4 \beta_0^2 \right) , 
\\
\bigg<\mathcal{P}^{\text{MQ02(tail)}^2}_\infty\bigg>
 &= -\frac{48 M^2 x (\O_{\vp})^8}{5(1-e_t^2)} \sum_{n=1}^{\infty}  n^8
| \underset{(n)}{I_{ij}^{00}} |^2  \left(\pi^2 + 4 \beta_0^2 + \beta_0 \right) , 
\\
\bigg<\mathcal{P}^{\text{MQ03(tail)}^2}_\infty\bigg>
 &= \frac{48 M^2 x (\O_{\vp})^8}{5(1-e_t^2)} \sum_{n=1}^{\infty} \sum_{p} n^7 p \,
| \underset{(n,p)}{I_{ij}^{00}} |^2  \left(\pi^2 + 4 \beta_0^2 + \beta_0 \right) .
\end{align}

\subsubsection{The energy flux tail-of-tails}

The mass quadrupole part of the tail-of-tails time-dependent flux is given by \cite{ArunETC08a}
\begin{equation}
\mathcal{P}^{MQ (\text{tail-of-tails})}_\infty =
\frac{4\mathcal{M}^2}{5} I_{ij}^{(3)} \int_{0}^{\infty}I_{ij}^{(6)}(t-\tau)
\bigg[\log\Big(\frac{\tau}{2 r_0}\Big)^2 + \frac{57}{70} \log\Big(\frac{\tau}{2 r_0}\Big) 
+ \frac{124627}{44100}\bigg] d\tau .
\end{equation}
When the quadrupole moment is calculated to Newtonian order, this gives a hereditary contribution to the 3PN flux.  
The 4PN contribution we seek comes from considering 1PN orbital dynamics and the mass quadrupole through 1PN order.
The usual Fourier simplifications lead to
\begin{align}
\label{eqn:MQtotFlux}
\frac{4M^2}{5} \sum_{n,p} (\O_r)^{9} i (n^9 + 9 n^8 p)  \underset{(n,p)}{I_{ij}} \underset{(-n,-p)}{I_{ij}}
 \int_{0}^{\infty} e^{i(n + p k)\O_r \tau}
 \bigg[\log\Big(\frac{\tau}{2 r_0}\Big)^2 + \frac{57}{70} \log\Big(\frac{\tau}{2 r_0}\Big) 
+ \frac{124627}{44100}\bigg] d\tau .
\end{align} 
To handle the $\log^2$ term, we expand the integral identity \eqref{eqn:lnSqTailInt} to first order in $k$, giving
\begin{gather}
 \int_{0}^{\infty} e^{i(n + p k)\O_r \tau} \log\Big(\frac{\tau}{2 r_0}\Big)^2 d\tau 
= \frac{i}{(n + p k)\O_r}\bigg[\frac{\pi^2}{6} 
+ \Big(\frac{\pi i}{2}\text{sign}(-n) + \log(2 \O_r |n + pk| r_0) + \gamma_E \Big)^2 \bigg]  \\
\approx \frac{i}{n\O_r}  \bigg[\alpha_0^2 - \frac{\pi^2}{12} - \pi i \text{sign}(n) \alpha_0  \bigg]
+ \frac{i}{n^2 \O_r} \left[\left(\frac{\pi^2}{12} -\alpha_0^2 + 2 \alpha_0  - (1 - \alpha_0) \pi i \text{sign}(n)
\right) p - 2 n \alpha_0 + n \pi i \text{sign}(n) \right]k,  
\label{eqn:lnSq1PN}
\end{gather}
where $\alpha_0 \equiv \log(2 \O_{\vp} |n| r_0) + \gamma_E$.  The rest of the integral can be found using 
\eqref{eqn:lnTailInt}.  In all cases the terms with sign$(\pm n)$ will vanish in sums over positive and negative 
$n$, so those are dropped in what follows.  We combine what is left with the other terms in the integrand of
\eqref{eqn:MQtotFlux} to get the total contribution from that integral
\begin{align}
\frac{i}{n \O_r} \left(\alpha_0^2 - \frac{57}{70} 
\alpha_0 - \frac{\pi^2}{12} + \frac{124627}{44100}\right)
+ \frac{i}{n \O_r} \left[-2 \alpha_0+ \frac{57}{70}  - 
\frac{p}{n}\left( \alpha_0^2 - \frac{197}{70} \alpha_0 - \frac{\pi^2}{12} + \frac{160537}{44100} \right) \right] k .
\end{align}
With this factor reinserted in the expression for the flux, the tail-of-tails can be separated at 1PN order into the 
now-familiar three sums 
\begin{align}
\bigg<\mathcal{P}^{\text{MQ01(tail-of-tails)}}_\infty\bigg> 
&= \frac{8M^2 (\O_{\vp})^8}{5} \sum_{n=1}^{\infty} n^8
\Big(\underset{(n)}{I_{ij}^{10}} \underset{(n)}{I_{ij}^{00*}} 
+ \underset{(n)}{I_{ij}^{00}} \underset{(n)}{I_{ij}^{10*}} \Big) \bigg[\frac{\pi^2}{12} - \alpha_0^2 + 
\frac{57}{70}\alpha_0 - \frac{124627}{44100}\bigg],     \\
\bigg<\mathcal{P}^{\text{MQ02(tail-of-tails)}}_\infty\bigg>
 &= - \frac{24 M^2 x}{5(1-e_t^2)} \sum_{n=1}^{\infty} (\O_{\vp})^8 n^8
| \underset{(n)}{I_{ij}^{00}} |^2 \bigg[\frac{2 \pi^2}{3} - 8 \alpha_0^2 + \frac{158}{35} \alpha_0
 - \frac{480553}{22050}\bigg],     \\
\bigg<\mathcal{P}^{\text{MQ03(tail-of-tails)}}_\infty\bigg>
 &=  \frac{24 M^2 x (\O_{\vp})^8}{5(1-e_t^2)} \sum_{n=1}^{\infty} \sum_{p} n^7 p \,
| \underset{(n,p)}{I_{ij}^{00}} |^2 \bigg[\frac{2 \pi^2}{3} - 8 \alpha_0^2 + 
\frac{158}{35}\alpha_0 - \frac{480553}{22050}\bigg] .
\end{align}

\subsubsection{Summing the tail$^2$ and tail-of-tails}

We can now combine the sums of corresponding type from the tail-of-tail and tail$^2$ parts into one set of 1PN 
mass quadrupole contributions.  We find that upon fusing the tail pieces all of the $\log^2$ terms (i.e., 
$\alpha_0^2$ terms) vanish.  The result is
\begin{align}
\label{eqn:MQ01}
\bigg<\mathcal{P}^{\text{MQ01(tail)}^2+\text{(tail-of-tails)}}_\infty\bigg>
 = M^2 (\O_{\vp})^8 \sum_{n=1}^{\infty} n^8
\Big(\underset{(n)}{I_{ij}^{10}} \underset{(n)}{I_{ij}^{00*}} 
+ \underset{(n)}{I_{ij}^{00}} \underset{(n)}{I_{ij}^{10*}} \Big) 
\bigg[\frac{8\pi^2}{15} - \frac{856}{525}\Big(\log(2 \O_{\vp} |n| r_0) + \gamma_E \Big) 
- \frac{116761}{36750}\bigg] ,
\end{align}
\begin{align}
\label{eqn:MQ02}
\bigg<\mathcal{P}^{\text{MQ02(tail)}^2+\text{(tail-of-tails)}}_\infty\bigg>
 = -\frac{M^2 x (\O_{\vp})^8 }{(1-e_t^2)} \sum_{n=1}^{\infty} n^8
| \underset{(n)}{I_{ij}^{00}} |^2 \bigg[\frac{64\pi^2}{5} 
- \frac{6848}{175}\Big(\log(2 \O_{\vp} |n| r_0) + \gamma_E \Big) -\frac{497004}{6125}\bigg] ,
\end{align}
\begin{align}
\label{eqn:MQ03}
\bigg<\mathcal{P}^{\text{MQ03(tail)}^2+\text{(tail-of-tails)}}_\infty\bigg>
 = \frac{M^2 x (\O_{\vp})^8}{(1-e_t^2)} \sum_{n=1}^{\infty} \sum_{p} n^7 p \,
| \underset{(n,p)}{I_{ij}^{00}} |^2 \bigg[\frac{64\pi^2}{5} -  
\frac{6848}{175}\Big(\log(2 \O_{\vp} |n| r_0) + \gamma_E \Big) - \frac{497004}{6125}\bigg] .
\end{align}

\subsubsection{The angular momentum tail flux}

On the angular momentum side, the time-dependent tail$^2$ and tail-of-tails fluxes take the following forms
\begin{align}
\mathcal{G}^{\text{MQ(tail)}^2}_\infty =
\frac{8\mathcal{M}}{5}\epsilon_{3jl} \bigg\{\int_{0}^{\infty}I_{ja}^{(4)}(t-\tau)
\bigg[\log\Big(\frac{\tau}{2 r_0}\Big) +\frac{11}{12}\bigg] d\tau \bigg\} \times \bigg\{ 
\int_{0}^{\infty}I_{la}^{(5)}(t-\tau)
\bigg[\log\Big(\frac{\tau}{2 r_0}\Big) +\frac{11}{12}\bigg] d\tau \bigg\} \hat{z} , 
\\
\mathcal{G}^{\text{MQ(tail-of-tails)}}_\infty =
\frac{4\mathcal{M}^2}{5} \e_{3jl} \bigg\{I_{ja}^{(2)} \int_{0}^{\infty}I_{la}^{(6)}(t-\tau)
\bigg[\log\Big(\frac{\tau}{2 r_0}\Big)^2 + \frac{57}{70} \log\Big(\frac{\tau}{2 r_0}\Big) 
+ \frac{124627}{44100}\bigg] d\tau  \qq \qq \qq 
\notag 
\\
+ I_{la}^{(3)} \int_{0}^{\infty}I_{ja}^{(5)}(t-\tau)
\bigg[\log\Big(\frac{\tau}{2 r_0}\Big)^2 + \frac{57}{70} \log\Big(\frac{\tau}{2 r_0}\Big) 
+ \frac{124627}{44100}\bigg] d\tau \bigg\} \hat{z} .
\end{align}

The simplification procedure is nearly identical to that in the energy case, so we jump straight to the three 
sums that give this essential part of the tail flux
\begin{align}
\bigg<\mathcal{G}^{\text{MQ01(tail)}^2+\text{(tail-of-tails)}}_\infty\bigg>
 = - M^2 (\O_{\vp})^7 i \e_{3jl} \sum_{n=1}^{\infty} n^7
\Big(\underset{(n)}{I_{ja}^{10}}& \underset{(n)}{I_{la}^{00*}} + \underset{(n)}{I_{ja}^{00}} 
\underset{(n)}{I_{la}^{10*}} \Big) \times \notag \\
&\bigg[\frac{16\pi^2}{15} - \frac{1712}{525}\Big(\log(2 \O_{\vp} |n| r_0) + \gamma_E \Big) 
- \frac{116761}{18375}\bigg] ,
\end{align}
\begin{align}
\bigg<\mathcal{G}^{\text{MQ02(tail)}^2+\text{(tail-of-tails)}}_\infty\bigg>
= \frac{M^2 x (\O_{\vp})^7 i}{(1-e_t^2)} \e_{3jl} \sum_{n=1}^{\infty} n^7
\underset{(n)}{I_{ja}^{00}} \underset{(n)}{I_{la}^{00*}} 
\bigg[\frac{112\pi^2}{5} - \frac{1712}{25}\Big(\log(2 \O_{\vp} |n| r_0) + \gamma_E \Big) 
- \frac{17903}{125}\bigg] ,
\end{align}
\begin{align}
\bigg<\mathcal{G}^{\text{MQ03(tail)}^2+\text{(tail-of-tails)}}_\infty\bigg>
 = - \frac{M^2 x (\O_{\vp})^7 i}{(1-e_t^2)} \e_{3jl} \sum_{n=1}^{\infty} \sum_{p} n^6 p 
\underset{(n,p)}{I_{ja}^{00}} \underset{(n,p)}{I_{la}^{00*}} \bigg[\frac{112\pi^2}{5} -  
\frac{1712}{25}\Big(\log(2 \O_{\vp} |n| r_0) + \gamma_E \Big) - \frac{17903}{125}\bigg] .
\end{align}

\subsection{Mass quadrupole, next order in $\nu$}

As noted in an earlier section of the paper, only minor adjustments to the above results are required 
to obtain these parts of the 4PN tail at $\mathcal{O}(\nu)$.  The $\nu$-correction to the quadrupole moment itself 
can again be found by simple substitution, and the $\nu$-correction to the ADM mass simply provides a factor of (-1).  
Thus, these essential parts of the 4PN tail at order $\nu$ become
\begin{align}
\label{eqn:MQ11}
\bigg<\mathcal{P}^{\text{MQ11(tail)}^2+\text{(tail-of-tails)}}_\infty\bigg>
 = \nu x M^2 (\O_{\vp})^8 \sum_{n=1}^{\infty} (n^8)
\Big(\underset{(n)}{I_{ij}^{11}} \underset{(n)}{I_{ij}^{00*}} 
+ \underset{(n)}{I_{ij}^{00}} \underset{(n)}{I_{ij}^{11*}} \Big) 
\bigg[\frac{8\pi^2}{15} - \frac{856}{525}\Big(\log(2 \O_{\vp} |n| r_0) + \gamma_E \Big) 
- \frac{116761}{36750}\bigg] ,
\end{align}
\begin{align}
\label{eqn:MQ12}
\bigg<\mathcal{P}^{\text{MQ12(tail)}^2+\text{(tail-of-tails)}}_\infty\bigg>
 = -\nu x M^2 (\O_{\vp})^8 \sum_{n=1}^{\infty} (n^8) | \underset{(n)}{I_{ij}^{00}} |^2 
\bigg[\frac{8\pi^2}{15} - \frac{856}{525}\Big(\log(2 \O_{\vp} |n| r_0) + \gamma_E \Big) 
- \frac{116761}{36750}\bigg] ,
\end{align}
for the energy flux and
\begin{align}
\bigg<\mathcal{G}^{\text{MQ11(tail)}^2+\text{(tail-of-tails)}}_\infty\bigg>
 = - \nu x M^2 (\O_{\vp})^7 i \e_{3jl} \sum_{n=1}^{\infty} (n^7)
\Big(& \underset{(n)}{I_{ja}^{11}} \underset{(n)}{I_{la}^{00*}} 
+ \underset{(n)}{I_{ja}^{00}} \underset{(n)}{I_{la}^{11*}} \Big)  \times 
\notag 
\\
&\bigg[\frac{16\pi^2}{15} - \frac{1712}{525}\Big(\log(2 \O_{\vp} |n| r_0) + \gamma_E \Big) 
- \frac{116761}{18375}\bigg] ,
\end{align}
\begin{align}
\bigg<\mathcal{G}^{\text{MQ12(tail)}^2+\text{(tail-of-tails)}}_\infty\bigg>
 = \nu x M^2 (\O_{\vp})^7 i \e_{3jl} \sum_{n=1}^{\infty} (n^7) \underset{(n)}{I_{ja}^{00}} 
  \underset{(n)}{I_{la}^{00}}  
\bigg[\frac{16\pi^2}{15} - \frac{1712}{525}\Big(\log(2 \O_{\vp} |n| r_0) + \gamma_E \Big) 
- \frac{116761}{18375}\bigg] ,
\end{align}
for the angular momentum flux.

\subsection{Putting the essential part of the 4PN tail together}
\label{sec:fullTail}

We are now in a position to assemble the entire order-$\mathcal{M}^2$ part of the 4PN tail.  We will focus on the 
energy flux case first.  This net tail flux comes from summing together \eqref{eqn:MQ01}, \eqref{eqn:MQ02}, 
\eqref{eqn:MQ03}, \eqref{eqn:MQ11}, and \eqref{eqn:MQ12}.  Since this is a 4PN energy flux, we pull out the 
circular-orbit limit and an extra factor of $x^4$ to define a tail enhancement function 
$\mathcal{R}_4^{\rm tail}(e_t,\nu)$:
\begin{equation}
\left\langle \frac{dE}{dt} \right\rangle_{\rm 4L}^{\rm tail} = \frac{32}{5} \nu^2 x^9 \,
\mathcal{R}_4^{\rm tail}(e_t,\nu) . 
\end{equation}
With $\mathcal{R}_4^{\rm tail}(e_t,\nu)$ defined, we then make a new separation of this function by grouping on 
common factors like $\pi^2$, rational numbers, a variant of the eulerlog function \cite{MunnEvan19a,MunnETC20}, and 
$\log\left(\frac{n}{2}\right)$, all of which appear in \eqref{eqn:MQ01}, \eqref{eqn:MQ02}, \eqref{eqn:MQ03}, 
\eqref{eqn:MQ11}, and \eqref{eqn:MQ12}.  Then these separate groupings are each expanded in power series in $e^2$.  

We draw attention first to the grouping on the $\log\left(\frac{n}{2}\right)$ term within the sums, which defines 
a new function that we call $\mathcal{R}_4^{\chi}(e_t,\nu)$.  This function is reminiscent of the 3PN function 
$\chi(e_t)$ \cite{ArunETC08a,ForsEvanHopp16,MunnEvan19a} that leads to the related relative flux function 
$\mathcal{R}_3^{\chi}(e_t)$, 
\be
\label{eqn:R3chi}
\mathcal{R}_3^{\chi}(e_t) = -\frac{1712}{105} \chi(e_t) 
= -\frac{1712}{105} \sum_{n=1}^{\infty} \frac{n^2}{4} \log\Big(\frac{n}{2}\Big) \, g(n,e_t) .
\ee
In turn, $\chi(e_t)$ is related to an infinite sequence of functions $\Lambda_k(e_t)$ that we defined in Sec.~IV 
of Paper I.  With these connections in mind, the definition for $\mathcal{R}_4^{\chi}(e_t,\nu)$, along with its power 
series expansion, is found to be
\begin{align}
\label{eqn:chi4PN}
\mathcal{R}_4^{\chi}&(e_t,\nu) 
= \bigg(\frac{107}{420}\bigg) \sum_{n=1}^{\infty} \log\left(\frac{n}{2}\right)
\bigg[  \Big( \frac{24}{1-e_t^2} \Big) \Big(n^8  | \underset{(n)}{\hat{I}_{ij}^{00}} |^2 - \sum_{p= -2,2} n^7 p \, 
| \underset{(n,p)}{\hat{I}_{ij}^{00}} |^2\Big) -
n^8 \Big(\underset{(n)}{\hat{I}_{ij}^{00}} \underset{(n)}{\hat{I}_{ij}^{10*}} 
+ \underset{(n)}{\hat{I}_{ij}^{10}} \underset{(n)}{\hat{I}_{ij}^{00*}} \Big)  
\notag 
\\
& + \nu \, n^8 | \underset{(n)}{\hat{I}_{ij}^{00}} |^2 
- \nu \, n^8 \Big(\underset{(n)}{\hat{I}_{ij}^{00}} \underset{(n)}{\hat{I}_{ij}^{11*}} 
+ \underset{(n)}{\hat{I}_{ij}^{11}} 
\underset{(n)}{\hat{I}_{ij}^{00*}} \Big) \bigg]
- (1- 4\nu) \sum_{n=1}^{\infty} n^2 \log\left(\frac{n}{2}\right) \left[ \left(\frac{52}{21}\right) h(n,e_t)
+ \left(\frac{428}{105}\right) k(n,e_t) \right]  
\notag 
\\ 
&= -3 \, \mathcal{R}_{4L} \log(1 - e_t^2) + \frac{1}{(1-e_t^2)^{15/2}}\biggl[ \frac{133771 \log (2)}{4410} 
-\frac{47385 \log (3)}{1568}+ \biggl(-\frac{232597}{2940}-\frac{19405829 \log (2)}{8820}  
\notag 
\\
&+\frac{15792327 \log (3)}{15680}\biggr) e_t^2 + \left(\frac{1414531}{2940}+\frac{365627093 \log (2)}{17640}
+\frac{459923913 \log (3)}{71680}-\frac{15869140625 \log (5)}{903168}\right) e_t^4 
\notag 
\\
&+\left(\frac{13183159}{1764} +\frac{30959613721 \log (2)}{317520}-\frac{276844091571 \log (3)}{1003520}
+\frac{2189439453125 \log (5)}{16257024}\right) e_t^6 + \cdots \biggr]  
\notag 
\\
&+ \frac{\nu}{(1-e_t^2)^{15/2}}\biggl[ \left(-\frac{267542 \log (2)}{2205}+\frac{47385 \log (3)}{392}\right)
+\biggl(-\frac{34889}{245} 
+ \frac{7321852 \log (2)}{945} -\frac{11424159 \log (3)}{3920}\biggr) e_t^2    
\notag 
\\ 
&+\biggl(-\frac{2993785}{882}-\frac{241129100 \log (2)}{1323}+\frac{3196747377 \log (3)}{125440} 
+\frac{15869140625 \log (5)}{225792}\biggr) e_t^4     
\notag 
\\ 
&+\left(-\frac{29356142}{2205}+\frac{52459170329 \log (2)}{26460}
+\frac{3605227461 \log (3)}{7168}-\frac{4796728515625 \log (5)}{4064256}\right) e_t^6 + \cdots \biggr] .
\end{align}
While this function has no overall closed form, it \emph{does} have an isolated closed-form part that involves the 
1PN-log-sequence function $\mathcal{R}_{4L}(e_t,\nu)$ \eqref{eqn:R4L}.  The reappearance of this 1PN log function 
within a 4PN log function is exactly analogous to the way a leading log function, $F(e_t)$, reappears in the 3PN 
function $\chi(e_t)$ \cite{ForsEvanHopp16} (see also Paper I, Sec.~IV A).  Its appearance aids in isolating the 
singular behavior (as $e_t \rightarrow 1$) of $\mathcal{R}_{4}^{\chi}$ into two parts---one with algebraic divergence 
and one with a dual logarithmic/algebraic divergence.  

The remaining groupings on the other factors ($\pi^2$, rational numbers, and a variant of the eulerlog function) lead 
to the remarkable behavior that all of the rest of $\mathcal{R}_{4}^{\rm tail}$ has a closed-form appearance.  We find
\begin{align}
\mathcal{R}_4^{\rm tail} &= \frac{1}{(1 - e_t^2)^{15/2}} \biggl[\biggl(\frac{5887504939}{22226400}
+\frac{105800809423 e_t^2}{44452800}-\frac{12538208629 e_t^4}{29635200}
-\frac{778013619983 e_t^6}{177811200}-\frac{2645724108523 e_t^8}{2844979200}  
\notag 
\\ 
&-\frac{1498169789 e_t^{10}}{210739200} \biggr) + \nu \biggl(-\frac{1488040411}{5556600}
-\frac{2854515929 e_t^2}{555660}-\frac{1030726283 e_t^4}{66150}-\frac{103160580401 e_t^6}{8890560}
-\frac{289778969059 e_t^8}{142248960}   
\notag 
\\
&-\frac{1066256753 e_t^{10}}{26342400} \biggr) \biggr] - \frac{1}{(1-e_t^2)^{6}} \left(\frac{1712}{35}
+\frac{24503 e_t^2}{70} +\frac{34989 e_t^4}{140}+\frac{7383 e_t^6}{560}  \right) + \frac{\pi^2}{(1-e_t^2)^{15/2}} 
\biggl[\biggl(-\frac{1369}{126} +\frac{62107 e_t^2}{252}  
\notag 
\\
&+\frac{1011881 e_t^4}{504}+\frac{715759 e_t^6}{336} +\frac{693593 e_t^8}{1792} 
+\frac{19389 e_t^{10}}{3584} \biggr) + \nu \biggl(-\frac{3566}{63}-\frac{237442 e_t^2}{189}-\frac{38219 e_t^4}{9}
-\frac{280195 e_t^6}{84}  
\notag 
\\
&-\frac{6922325  e_t^8}{12096}-\frac{141 e_t^{10}}{14}  \biggr) \biggr]
+ 2 \left(\gamma_E + 2 \log 2 + \frac{3}{2} \log x \right) \mathcal{R}_{4L}(e_t,\nu) +\mathcal{R}_4^{\chi}(e_t,\nu) .
\end{align}
Note that the eulerlog function becomes a coefficient on another appearance of $\mathcal{R}_{4L}(e_t,\nu)$, with
a form that exactly matches the predictions laid out in Sec.~IV E of Paper I.   (This is only part of the appearance of 
$\log x$ at 4PN order; the remainder arises in the instantaneous 4PN term, which is not calculated here.)  All the 
other terms involve polynomials once the relevant eccentricity singular factors are removed.

In turning to the case of the angular momentum flux, all of the steps made for energy flux carry over almost 
identically.  At the end of the process we find that the order-$\mathcal{M}^2$ part of the 4PN tail in angular 
momentum flux is
\begin{align}
\label{eqn:4PNZtail}
\mathcal{Z}_4^{\rm tail} &= \frac{1}{(1 - e_t^2)^{6}} \biggl[\biggl(\frac{5887504939}{22226400}
+\frac{7325116643 e_t^2}{7408800}-\frac{27567910067 e_t^4}{29635200}
-\frac{98001030431 e_t^6}{177811200} - \frac{125213141 e_t^8}{21073920}  \biggr) 
\\
&+ \biggl( -\frac{1488040411}{5556600}-\frac{2260694969 e_t^2}{740880}-\frac{71048901401 e_t^4}{14817600}
-\frac{125477205683 e_t^6}{88905600}-\frac{36791617 e_t^8}{878080} \biggr) \biggr]
\notag 
\\ 
&-  \frac{1}{(1-e_t^2)^{9/2}} \biggl(\frac{1712}{35}+\frac{10379 e_t^2}{70}+\frac{749 e_t^4}{40}\biggr) 
+ \frac{\pi^2}{(1-e_t^2)^{6}} \biggl[\biggl(-\frac{1369}{126}+\frac{26519 e_t^2}{126}+\frac{366503 e_t^4}{504} 
+\frac{84109 e_t^6}{336} 
+\frac{9417 e_t^8}{1792} \biggr) 
\notag 
\\
\notag 
& + \nu \biggl(-\frac{3566}{63}-\frac{47785 e_t^2}{63} 
-\frac{330481 e_t^4}{252}-\frac{194269 e_t^6}{504}-\frac{3359 e_t^8}{336} \biggr) \biggr]  
+ 2 \left(\gamma_E + 2 \log 2 
+ \frac{3}{2} \log x \right) \mathcal{Z}_{4L}(e_t,\nu) + \mathcal{Z}_4^{\chi}(e_t,\nu) .
\end{align}
Here, $\mathcal{Z}_{4L}(e_t,\nu)$ is the second element in the integer-order 1PN log sequence defined in 
\eqref{eqn:Z4L}.  The remaining part of the above expression is a new function defined by
\begin{align}
\label{eqn:chiTild4PN}
\mathcal{Z}_{4}^{\chi}&(e_t,\nu) 
= \bigg(\frac{107i}{210}\bigg) \e_{3jl} \sum_{n=1}^{\infty} \log\left(\frac{n}{2}\right)
\bigg[ \Big( \frac{-21}{1-e_t^2}\Big) 
\Big(n^7 \underset{(n)}{\hat{I}_{ja}^{00}} \underset{(n)}{\hat{I}_{la}^{00*}}  
- \sum_{p= -2,2}  n^6 p \, \underset{(n,p)}{\hat{I}_{ja}^{00}} \underset{(n,p)}{\hat{I}_{la}^{00*}} \Big)
+ n^7 \Big(\underset{(n)}{\hat{I}_{ja}^{00}} \underset{(n)}{\hat{I}_{la}^{10*}} 
+ \underset{(n)}{\hat{I}_{ja}^{10}} \underset{(n)}{\hat{I}_{la}^{00*}} \Big) \notag \\
& \mkern-3mu - \nu \Big( n^7 \underset{(n)}{\hat{I}_{ja}^{00}} \underset{(n)}{\hat{I}_{la}^{00*}}  \Big) + \nu \, n^7
\Big(\underset{(n)}{\hat{I}_{ja}^{00}} \underset{(n)}{\hat{I}_{la}^{11*}} 
+ \underset{(n)}{\hat{I}_{ja}^{11}} \underset{(n)}{\hat{I}_{la}^{00*}} \Big) \bigg]
- (1- 4\nu) \sum_{n=1}^{\infty} n^2 \log\left(\frac{n}{2}\right) \left[\left(\frac{52}{21}\right) \tilde{h}(n,e_t)
+ \bigg(\frac{428}{105}\bigg) \tilde{k}(n,e_t)\right]  \notag \\ &
= -3 \, \mathcal{Z}_{4L} \log(1 - e_t^2) + \frac{1}{(1-e_t^2)^{6}}\biggl[ \frac{133771 \log (2)}{4410} 
-\frac{47385 \log (3)}{1568}+e_t^2 \biggl(-\frac{232597}{2940}-\frac{6267421 \log (2)}{4410} \notag \\&
+\frac{426465 \log (3)}{784}\biggr) +\left(\frac{3290641}{5880}+\frac{54495991
\log (2)}{5880}+\frac{3298439961 \log (3)}{501760}-\frac{3173828125 \log (5)}{301056}\right) e_t^4  \notag \\& 
+\left(\frac{15117563}{5040}+\frac{19854591727 \log (2)}{317520}-\frac{78260015511 \log (3)}{501760}
+\frac{88952734375 \log (5)}{1161216}\right) e_t^6 + \cdots \biggr]  \notag \\&
+ \frac{\nu}{(1-e_t^2)^{6}}\biggl[ \left(-\frac{267542 \log (2)}{2205}+\frac{47385 \log (3)}{392}\right)
+\left(-\frac{34889}{245}+\frac{1944359 \log (2)}{315}-\frac{2496339 \log (3)}{980}\right) e_t^2   \notag \\ &   
+\left(-\frac{1532917}{735}-\frac{10104547 \log (2)}{84}+\frac{2481686343 \log (3)}{125440}
+\frac{3173828125 \log (5)}{75264}\right)e_t^4     \notag \\ &
+\left(-\frac{3893383}{840}+\frac{180364779509 \log (2)}{158760}+\frac{5720993955 \log (3)}{25088}
-\frac{1302119140625 \log (5)}{2032128}\right)e_t^6 + \cdots \biggr] ,
\end{align}
where the second equality provides its power series expansion.   The full order-$\mathcal{M}^2$ tail functions, 
$\mathcal{R}_4^{\rm tail}(e_t,\nu)$ and $\mathcal{Z}_4^{\rm tail}(e_t,\nu)$, can now be used with an assist from 
BHPT to determine the flux terms $\mathcal{R}_4$ and $\mathcal{Z}_4$, at lowest order in $\nu$.

\section{The complete 4PN fluxes at lowest order in $\nu$}
\label{sec:full4PN}

\subsection{1PN correction to $\chi(e_t)$ and compact expressions for $\mathcal{L}_4(e)$ and $\mathcal{J}_4(e)$}

We demonstrated in Paper I how the threefold combination of (i) knowledge of the leading logarithm sequence, (ii) 
theoretical understanding of the role of the $\Lambda_k(e_t)$ sequence of functions (analogs of the function 
$\chi(e_t)$), and (iii) use of BHPT and fitting to finite-order expansions in eccentricity was sufficient to 
determine completely the integer-order 3PN log sequence at lowest order in the mass ratio.  This procedure involved, 
first, converting a given leading-log term and its associated $\Lambda_k(e_t)$ function from expressions and 
expansions in $e_t$ into expansions in Darwin $e$, the natural eccentricity for BHPT calculations.  Then, these known 
functions were incorporated into a model for the eccentricity power series dependence at the given PN order.  Thirdly, 
high accuracy BHPT numerical results, or a fully analytic BHPT calculation, were used to determine the remaining, 
most-often rational, coefficients in the model.  Finally, the result was then transformed back from $e$ to $e_t$.  In 
this way, the leading-log (0PN log) sequence was used to assist in finding terms in the 3PN log sequence at 
corresponding PN order.  This is a connection between the red and the green lines in Fig.~\ref{fig:logSequences}.  
We used this process to determine the $\mathcal{R}_{6L}(e_t)$ and $\mathcal{Z}_{6L}(e_t)$ terms in their entirety, 
aided by knowledge of the leading logs $\mathcal{R}_{6L2}(e_t)$ and $\mathcal{Z}_{6L2}(e_t)$.

A similar process appears to hold in being able to use terms in the 1PN log sequence to aid in determining the form 
of the corresponding term in the 4PN log sequence (i.e., a connection between the blue and orange lines in the 
figure), which we demonstrate with the first element in the 4PN log sequence---the 4PN non-log flux itself.  The 
derivations in Sec.~\ref{sec:1PNLogs} provided one key component of this process---the closed-form expressions for 
the second elements in the integer-order 1PN log sequences, $\mathcal{R}_{4L}(e_t)$ and $\mathcal{Z}_{4L}(e_t)$.  
Then the analysis in Sec.~\ref{sec:4PNTail} provided a second key component---the analytic form (including one infinite 
series) for the energy and angular momentum $\chi$-like tail fluxes, $\mathcal{R}_4^{\chi}(e_t)$ and 
$\mathcal{Z}_4^{\chi}(e_t)$, the analogs at 4PN of the 3PN tail functions, $\chi(e_t)$ and $\tilde{\chi}(e_t)$.  
Knowing how these functions make an appearance in the full 4PN non-log fluxes was sufficient to allow BHPT fitting 
to determine closed-form dependence for the rest of the 4PN non-log fluxes at lowest order in $\nu$.  

Beginning with the energy flux, we require first a high-order eccentricity expansion for $\mathcal{L}_4^{\chi}(e)$, 
which like $\mathcal{R}_4^{\chi}(e_t)$ will be an infinite series.  The process to obtain $\mathcal{L}_4^{\chi}(e)$ 
is straightforward.  We start with $\mathcal{R}_{4}^{\chi}(e_t,\nu=0)$, which can be isolated from 
\eqref{eqn:chi4PN}.  This function is expanded in $e_t$ to $e_t^{30}$.  Then, $\mathcal{R}_{4}^{\chi}(e_t)$ must be 
converted to $\mathcal{L}_4^{\chi}(e)$, that is, from a function of time eccentricity $e_t$ to one of Darwin 
eccentricity $e$.  This is achieved by expressing $e_t$ in terms of $e$, to sufficient approximation, as 
$e_t = e (1 - 3 x + \mathcal{O}(x^2))$, substituting into the full energy flux expansion, and letting the 
post-Newtonian difference between $e$ and $e_t$ ripple through the flux expressions.  Then we collect all relevant 
results at 4PN order.  The post-Newtonian corrections not only come from switching from $e_t$ to $e$ in 
$\mathcal{R}_{4}^{\chi}$ but also from a correction to $\mathcal{R}_3^{\chi}$ \eqref{eqn:R3chi}.  The result is 
that $\mathcal{L}_{4}^{\chi}(e)$ is calculated by taking
\be
\label{eqn:4PNLchi}
x^4 \mathcal{L}_{4}^{\chi}(e) = \bigg(-\frac{1712}{105} x^3 \chi(e - 3 x e) 
+ x^4 \mathcal{R}_4^{\chi}(e) \bigg)^{\rm 4PN} ,
\ee
where the superscript ``4PN'' on the right side means expand out and then collect and retain the $\mathcal{O}(x^4)$ 
terms.

We could perform a similar procedure to generate $\mathcal{L}_{4L}(e)$ from $\mathcal{R}_{3L}(e_t)$ and 
$\mathcal{R}_{4L}(e_t)$, but there is no need since we can simply use the expression already found in 
\cite{ForsEvanHopp16,Fors16,MunnETC20} via fitting.  The closed-form expression is
\be
\mathcal{L}_{4L}(e) = \frac{1}{(1-e^2)^{15/2}}
 \bigg(\frac{232597}{8820}+\frac{4923511 e^2}{5880}+\frac{142278179 e^4}{35280}
+\frac{318425291 e^6}{70560}+\frac{1256401651 e^8}{1128960}+\frac{7220691 e^{10}}{250880}\bigg) .
\ee
With those two functions, $\mathcal{L}_{4L}(e)$ and $\mathcal{L}_{4}^{\chi}(e)$, determined, the procedure now 
closely follows that of Paper I.  The tail part $\mathcal{L}_{4}^{\chi}(e)$ is expected to appear directly as a 
term in $\mathcal{L}_4(e)$, while the function $\mathcal{L}_{4L}(e)$ appears also but only after having been 
multiplied by a particular function containing $\gamma_E$ and a log term.  The sum of these two terms is expanded in 
a power series to $e^{30}$.  The model for the entire behavior of $\mathcal{L}_4(e)$, similar to one assumed in 
Paper I for $\mathcal{L}(e)$, includes these two parts as well as a power series in $e^2$ with rational coefficients 
and a second power series in $e^2$ with rational coefficients that is multiplied by $\pi^2$.  The starting point 
for these two power series is actually three closed-form expressions with relevant eccentricity singular factors.  
We subtract the known part in this model due to $\mathcal{L}_{4L}(e)$ and $\mathcal{L}_{4}^{\chi}(e)$ from the 
numerical 4PN non-log flux data provided by BHPT.  The modified numerical data should be represented by the remaining 
two rational-coefficient power series in this model.  We then progressively solve for the remaining unknown (rational) 
coefficients.  This process is successful, meaning the model was a correct ansatz, and yields
\begin{align}
\mathcal{L}_4(e) &= 
\frac{1}{(1-e^2)^{15/2}}\biggl[
\frac{18510752431}{44144100}-\frac{40934075709731 e^2}{6356750400}
-\frac{131458534402891 e^4}{2542700160}-\frac{3215698875850801e^6}{50854003200}  \notag \\
&-\frac{586522182193681 e^8}{31294771200} 
-\frac{3028139270269 e^{10}}{45203558400}+
 \frac{670101511e^{12}}{14057472}
+ \sqrt{1-e^2}\bigg(-\frac{1654225499}{3175200}+\frac{2426725501 e^2}{3175200} \notag \\
&+\frac{186636561079 e^4}{12700800}+\frac{72788261801 e^6}{8467200}
-\frac{16274063783 e^8}{7526400}-\frac{4982855 e^{10}}{18432}\bigg)
\biggr] - \frac{1369 \pi^2}{126 (1 - e^2)^{15/2}} \bigg(1+\frac{104549 e^2}{2738}  \notag \\ &
+\frac{1113487 e^4}{5476}
+\frac{2644503 e^6}{10952}+\frac{10829823 e^8}{175232}
+\frac{573939 e^{10}}{350464}\bigg)
+ 2 \left[\gamma_\text{E} + \log \left(\frac{8(1-e^2)}{1+\sqrt{1-e^2}}\right)\right] \mathcal{L}_{4L}(e)
+\mathcal{L}_{4}^{\chi}(e) .
\label{eqn:4PNL}
\end{align}
The result matches the expansion for $\mathcal{L}_4(e)$ found to $e^{30}$ in \cite{MunnETC20,BHPTK18}. 

The 4PN non-log angular momentum flux follows precisely the same procedure, and yields
\begin{align}
\mathcal{J}_4(e) &= 
\frac{1}{(1-e^2)^{6}}\biggl[\frac{139774944409}{1059458400}-\frac{35619868663789 e^2}{12713500800}
-\frac{284430057678037 e^4}{25427001600}-\frac{353931345220951 e^6}{50854003200} \notag \\ &
-\frac{66321815297809 e^8}{67805337600}
-\frac{3159752887 e^{10}}{147603456}
+\sqrt{1-e^2}\bigg(-\frac{370844347}{1587600}+\frac{2259332951 e^2}{3175200}
+\frac{2594164919 e^4}{846720} \notag \\ &
+\frac{921635651 e^6}{2822400}-\frac{572575 e^8}{8064}\bigg)
\biggr] 
- \frac{1369 \pi^2}{126 (1-e^2)^{6}} \bigg(1+\frac{22495 e^2}{1369}+\frac{259969 e^4}{5476}
+\frac{268179 e^6}{10952}+\frac{193455 e^8}{175232}\bigg) \notag \\ &
+
2 \left[\gamma_\text{E}+ 3 \log(2) + \log \left(\frac{1-e^2}{1+\sqrt{1-e^2}}\right)\right] \mathcal{J}_{4L}(e)
+\mathcal{J}_{4}^{\chi}(e),
\label{eqn:4PNJ}
\end{align}
where the second element in the angular momentum (integer-order) 1PN log sequence is
\be
\mathcal{J}_{4L}(e) = \frac{1}{(1-e^2)^6}\bigg(\frac{232597}{8820}+\frac{3482879 e^2}{8820}
+\frac{34971299 e^4}{35280}+\frac{6578731 e^6}{14112}+\frac{2503623 e^8}{125440}\bigg) .
\ee
The result in \eqref{eqn:4PNJ} also matches the expansion found by fitting given in \cite{MunnETC20,BHPTK18} but 
provides a deeper, though partial, theoretical explanation.

\subsection{Transforming from $\mathcal{L}_4(e)$ and $\mathcal{J}_4(e)$ to $\mathcal{R}_4(e_t)$ and 
$\mathcal{Z}_4(e_t)$}

In order to convert these flux terms to functions in terms of $e_t$ (i.e., $\mathcal{R}_4(e_t)$ and 
$\mathcal{Z}_4(e_t)$) in the modified harmonic gauge, we require the relationship between $e$ and $e_t$ to 4PN 
order at lowest order in the mass ratio.  With that restriction, the expansions relating $e$ and $e_t$ can be 
calculated to \emph{any} PN order by analyzing geodesic motion on a Schwarzschild background \cite{Munn20}.  We 
quote the result through the necessary order
\begin{align}
\label{eqn:eOveretExp}
\frac{e^2}{e_t^2} &= 1+6 x + \frac{\left(17 - 21 e_t^2 + 15 \sqrt{1-e_t^2}\right) x^2}{1 - e_t^2}+\frac{\left(26+
54 e_t^4+150 \sqrt{1-e_t^2}-e_t^2 \left(107+90 \sqrt{1-e_t^2}\right)\right)x^3}{\left(1 - e_t^2\right)^2} \,\, 
\\
\notag 
& -\frac{\left(880 e_t^6-10 e_t^4 \left(367+240 \sqrt{1-e_t^2}\right) -2 \left(865+3167 \sqrt{1-e_t^2}\right)
+e_t^2 \left(6120+6265 \sqrt{1-e_t^2}\right)\right) x^4}{8 \left(1 - e_t^2\right)^3}+ \mathcal{O}(\nu,x^5) .
\end{align}

We then construct the net flux by combining $\mathcal{L}_0(e)$, $\mathcal{L}_1(e)$, $\mathcal{L}_2(e)$, 
$\mathcal{L}_3(e)$, and $\mathcal{L}_4(e)$ and replacing $e$ with its relationship to $e_t$ given in 
\eqref{eqn:eOveretExp} along the lines done in \eqref{eqn:4PNLchi}.  The result is expanded, allowing the PN 
corrections to ripple through to the 4PN term, giving at last
\begin{align}
\mathcal{R}_4(e_t) =& 
\frac{1}{(1-e_t^2)^{15/2}}\biggl[
\frac{20670029551}{44144100}+\frac{90592819680523 e_t^2}{6356750400}+\frac{45374652958109
   e_t^4}{1589187600}+\frac{215773793118089 e_t^6}{50854003200}  
\notag 
\\
& -\frac{139754682191 e_t^8}{2844979200}+\frac{4853373238601 e_t^{10}}{45203558400}
-\frac{12776867 e_t^{12}}{14057472}
+ \sqrt{1-e_t^2}\bigg(-\frac{1809538139}{3175200}  
\notag 
\\ 
& -\frac{30429943463 e_t^2}{3175200}-\frac{103455982193 e_t^4}{12700800}+\frac{67397848199
   e_t^6}{8467200}+\frac{863341943 e_t^8}{501760}-\frac{308515 e_t^{10}}{64512}\bigg)  \biggr]  
\notag 
\\ 
& - \frac{1369 \pi^2}{126 (1 - e_t^2)^{15/2}} \bigg(1-\frac{62107 e_t^2}{2738}-\frac{1011881 e_t^4}{5476}
-\frac{2147277 e_t^6}{10952}-\frac{6242337 e_t^8}{175232}
-\frac{174501 e_t^{10}}{350464}\bigg)   
\notag 
\\ 
& + 2 \left[\gamma_\text{E} + 3 \log(2) + \log \left(\frac{(1-e_t^2)}{1+\sqrt{1-e_t^2}}\right)\right] 
\mathcal{R}_{4L}(e_t) + \mathcal{R}_{4}^{\chi}(e_t) .
\label{eqn:4PNR}
\end{align}
We follow precisely the same procedure in combining $\mathcal{J}_0(e)$, $\mathcal{J}_1(e)$, $\mathcal{J}_2(e)$, 
$\mathcal{J}_3(e)$, and $\mathcal{J}_4(e)$ to obtain
\begin{align}
\mathcal{Z}_4(e_t) =& 
\frac{1}{(1-e_t^2)^{6}}\biggl[\frac{191597595289}{1059458400}+\frac{99527954953927 e_t^2}{12713500800} + 
\frac{191377070535107 e_t^4}{25427001600}-\frac{101432063662609 e_t^6}{50854003200} 
\notag 
\\ 
& -\frac{13890223720171 e_t^8}{67805337600}+\frac{9732011 e_t^{10}}{21086208}
+\sqrt{1-e_t^2}\bigg(-\frac{448500667}{1587600}-\frac{14027009779 e_t^2}{3175200}-\frac{1764770893 e_t^4}{846720}  
\notag 
\\ 
& +\frac{235209407 e_t^6}{352800}-\frac{81965 e_t^8}{16128}  \bigg) \biggr] 
- \frac{1369 \pi^2}{126 (1-e_t^2)^{6}} \bigg(1-\frac{26519 e_t^2}{1369}-\frac{366503 e_t^4}{5476}
-\frac{252327 e_t^6}{10952}-\frac{84753 e_t^8}{175232}\bigg) 
\notag 
\\ 
& +2 \left[\gamma_\text{E}+ 3 \log(2) + \log \left(\frac{1-e_t^2}{1+\sqrt{1-e_t^2}}\right)\right] 
\mathcal{Z}_{4L}(e_t) + \mathcal{J}_{4}^{\chi}(e_t) .
\label{eqn:4PNZ}
\end{align}
Note that the polynomial attached to $\pi^2$ in each of these expressions now perfectly matches the corresponding 
result obtained through analysis of the 4PN tail in Sec.~\ref{sec:fullTail}.
\clearpage

\end{widetext}
\subsection{General structure in the 4PN log sequences and a simplified form for $\mathcal{L}_{11/2}(e)$}

The first part of this section has developed compact expressions for the first elements in the two integer-order 
4PN log sequences, namely the terms $\mathcal{L}_4(e)$ and $\mathcal{J}_4(e)$, by combining 1PN source multipoles 
in formulae for tail fluxes and using perturbation theory to find rational number coefficients in the remaining 
(closed-form) functions of $e$.  This is simply an extrapolation to the 4PN log sequences (solid orange line in 
Fig.~\ref{fig:logSequences}) of the procedure used in Paper I (Sec.~IV) to find comparable expressions for the 3PN 
log sequences (solid green line).  

We went about this by deriving the form of $\mathcal{R}_4^{\chi}(e_t)$ and $\mathcal{Z}_4^{\chi}(e_t)$ directly 
using the 4PN hereditary contributions.  However, strictly speaking this approach was not necessary.  At lowest order 
in the mass ratio, the analysis of Sec.~IV E in Paper I still holds, meaning that the form of the $\chi$-like 
contribution to \emph{any} 4PN logarithm can be ascertained \emph{a priori}.  Using the results of that section in 
Paper I, we see that the expressions for $\mathcal{R}_{(3k+1)L(k)}^{\chi}(e_t)$ and 
$\mathcal{Z}_{(3k+1)L(k)}^{\chi}(e_t)$ can be derived from the summations for $\mathcal{R}_{(3k+1)L(k)}(e_t)$ and 
$\mathcal{Z}_{(3k+1)L(k)}(e_t)$, which appear in \eqref{eqn:1PNRk} and \eqref{eqn:1PNZk}, respectively, by 
including in each a factor of $(2k)\log(n/2)$.  In this way we can not only reproduce the results for the 4PN non-log 
flux but also generalize to arbitrarily higher order terms in the integer-order 4PN log sequence.  The more 
general $\chi$-like functions will have dual logarithmic and algebraic divergent parts, with the former attached 
to the corresponding 1PN log term.  With these higher PN order $\chi$-like functions determined, we might then rely 
upon BHPT to determine the remaining functional dependence in these higher (integer) order 4PN log terms.  
The next of these, $\mathcal{L}_{7L}(e)$, would have a model with a set of closed-form functions with unknown 
rational coefficients.  Those functions would need to be expanded in a power series to $e^{34}$ and then BHPT would 
be used to fit for the rational coefficients to that order.

Unfortunately, the terms in the half-integer 4PN log sequences cannot be manipulated into expressions that are 
quite as compact, with closed-form parts.  However, each term in these sequences can still be reduced to a remaining 
infinite series with rational coefficients, once the roles of the mass quadrupole, mass octupole, and current 
quadrupole 1PN moments are understood.  As an example, take the first half-integer (energy) 4PN logarithm, 
$\mathcal{R}_{11/2}(e_t)$.  By applying the procedure above, the counterpart function in $e$, 
$\mathcal{L}_{11/2}(e)$, can be given the following form
\begin{widetext}
\begin{align}
\mathcal{L}&_{11/2}(e) = \frac{\pi}{(1-e_t^2)^{9}} \biggl(\frac{8399309750401}{101708006400}
-\frac{6431125434321667 e^2}{203416012800}-\frac{347369943176265227 e^4}{813664051200}
-\frac{15186120717515117243 e^6}{11716762337280}  
\notag 
\\ 
& -\frac{18230625005177349698411 e^8}{14997455791718400}   
-\frac{66989953560049801996499 e^{10}}{249957596528640000} 
-\frac{997758112480120369559 e^{12}}{3856488632156160000}    
\\
& -\frac{18489702206162114169107 e^{14}}{1476312054497280000}  
-\frac{58073989289629682554885336291 e^{16}}{5418088860997889556480000}  
-\frac{7460925685163379777975573791651 e^{18}}{877730395481658108149760000}   
\notag 
\\
\notag 
& -\frac{46468844430394780206366046707113 e^{20}}{6751772272935831601152000000} + \cdots \biggr)  
+ 2 \left[\gamma_E + 3 \log 2 + \log\left(\frac{1-e^2}{1+\sqrt{1-e^2}}\right) \right] 
\mathcal{L}_{11/2L}(e) + \mathcal{L}_{11/2}^{\chi}(e) .
\end{align}
\end{widetext}
As mentioned, the initial part of this expression is an infinite series with rational coefficients.  We have 
calculated the series to $e^{30}$ \cite{MunnETC20,BHPTK18} but omitted here the last few coefficients for brevity.  
The remainder of the expression involves two functions that together capture all of the transcendental and 
logarithmic constants.  The first of these is the 1PN log function itself at 5.5PN order, $\mathcal{L}_{11/2L}(e)$, 
which can be derived from the expression for $\mathcal{R}_{11/2L}(e_t)$ given earlier in this paper or found by 
consulting \cite{MunnETC20}.  The second is the 5.5PN $\chi$-like function, $\mathcal{L}_{11/2}^{\chi}(e)$, which 
can be found by the process described above.

\section{Conclusions and outlook}
\label{sec:conclusions}

This paper extended an approach found in Paper I for determining the eccentricity dependence of the leading log 
sequences of flux terms (depicted as solid and dashed red lines in Fig.~\ref{fig:logSequences}) and the 3PN log 
sequences (green lines) to additional strips in the higher-order PN structure.  The new strips considered here are 
the 1PN log sequences (blue lines) and 4PN log sequences (orange lines).  In the earlier paper we developed a 
complete understanding of the terms in the leading log sequences in terms of the Newtonian quadrupole moment Fourier 
spectrum $g(n,e_t)$.  The integer-order leading logs were found to have closed-form expressions and the 
half-integer-order leading logs were shown to be infinite series in $e^2$ with calculable rational coefficients.  
The 3PN log terms, at lowest order in the mass ratio, were shown to have part of their functional dependence given 
by the quadrupole spectrum, with the rest involving series with rational coefficients that could be determined 
with the assistance of BHPT fitting.  

In this paper we showed that a mirror image of those procedures could be found which would allow us to calculate the 
1PN and 4PN log sequences, provided we use PN theory to calculate additional, 1PN multipole moment spectra (i.e., 
the mass octupole, current quadrupole, and 1PN mass quadrupole moments) along with somewhat higher order in $e^2$ 
BHPT fitting.  In the case of the 1PN log sequences, the PN calculation provides as a bonus next-order-in-$\nu$ parts 
of the fluxes, with only some remaining uncertainty whether the $\mathcal{O}(\nu)$ part of the MQ12 terms 
(see Eq.~\eqref{eqn:1PNLogMQ1}) is complete.  Without a full PN theory calculation, the conjecture that the order 
$\nu$ part is complete can only be verified by an (as yet unavailable) second-order BHPT comparison.  We used the 
procedure to detail explicitly the 4PN log, 5.5PN log, and 7PN $\log^2$ terms.  However, our computational 
infrastructure allows us to compute any integer 1PN logarithm as a closed-form expression and permits the rapid 
expansion of all half-integer (non-closed) 1PN logs to at least $e_t^{120}$.

In addition to the 1PN logarithms, our approach allowed for the computation of the 1PN correction to the 
$\Lambda_k(e_t)$ and $\Xi_k(e_t)$ set of functions of Paper I.  The specific 1PN correction to 
$\Lambda_1(e_t) = \chi(e_t)$ allowed for the extraction of the full 4PN non-log fluxes at lowest order in $\nu$, 
as well as the isolation of all transcendental contributions in the 5.5PN non-log term, $\mathcal{R}_{11/2}(e_t)$.

To extend the procedures of Paper I and this paper further, we would need to calculate the Fourier spectra of the 
2PN source multipoles and use even higher-order BHPT fitting.  The algorithmic complexity and cost would increase, 
and there may be additional hereditary-term integrals that are more difficult to compute.  This 2PN extension, to 
the 2PN log sequences and the 5PN log sequences, may be the subject of future work.

We conclude by presenting an update of a table found in Paper I that summarizes the state of knowledge of the 
eccentricity dependence of high PN order (lowest order in $\nu$) flux terms.  

\acknowledgments

We thank Nathan Johnson-McDaniel and Luc Blanchet for helpful discussions.  This work was supported in part by 
NSF grants PHY-1506182 and PHY-1806447, and by the North Carolina Space Grant.  C.R.E.~acknowledges support from 
the Bahnson Fund at the University of North Carolina-Chapel Hill. 

\begin{widetext}
\begin{center}
\begin{table}[t]
\label{tab:PNterms}
\parbox{17.9cm}{
\caption{State of knowledge of eccentricity dependence of high PN order flux terms.  The second column indicates 
whether a closed form exists or to what order in $e$ the power series expansion is known.  The closed-form result for 
$\mathcal{L}_{4L}$ was previously found in Forseth et al. \cite{ForsEvanHopp16}.  All other results come from this 
paper and its companions, Paper I \cite{MunnEvan19a} and Munna et al.~\cite{MunnETC20}.  Flux terms labeled as 
``all orders" are infinite series in $e^2$ but with coefficients that can now be analytically calculated to 
arbitrary order.  Other terms are only known in analytic form up to order $e^{30}$ (or in a few 
cases less).  The fourth column gives the number of PN corrections to the leading-logs which must be calculated 
to derive the term fully.  The fifth column indicates the number of leading log (and $\Lambda(e_t)/\Xi(e_t)$) 
corrections which must be calculated to extract the term to all orders in $e$ in the manner of 
Sec.~\ref{sec:full4PN}.  A superset of these terms allow for the separation of transcendental contributions 
in the same way, as shown in column six.  Above 5PN it is more difficult to apply these methods (labeled by 
asterisk).  The last two rows represent all further leading and 1PN logarithms.\\}}
\begin{tabular}{ || c | c | c | c | c | c ||}
\hline\hline
Term & Known order in $e$ & Original source  & 
\begin{tabular}{@{}c@{}} PN Order \vspace{-1em} \\ beyond LL \end{tabular}  & 
\begin{tabular}{@{}c@{}} Order for fitting \vspace{-1em} \\ extraction\end{tabular}  & 
\begin{tabular}{@{}c@{}}  Order to find  \vspace{-1em} \\ transcendental part \end{tabular} \\
\hline
${L}_{7/2}$ &\cellcolor{Yellow} Fitted to $e^{30}$ & Munna et al. & 2PN & --- & ---  \\
\hline
$\mathcal{L}_{4}$ &\cellcolor{GreenYellow} All orders &  This paper  & 4PN & 1PN & 1PN \\
\hline
$\mathcal{L}_{4L}$ &\cellcolor{lime} Closed Form & Forseth et al.  & 1PN & --- & --- \\
\hline
$\mathcal{L}_{9/2}$ &\cellcolor{Yellow} Fitted to $e^{30}$ &  Munna et al.  & 3PN & --- & 0PN \\
\hline
$\mathcal{L}_{9/2L}$ &\cellcolor{GreenYellow} All Orders &  Paper I  & --- & --- & --- \\
\hline
$\mathcal{L}_{5}$ &\cellcolor{Yellow} Fitted to $e^{30}$ &  Munna et al.  & 5PN & 2PN & 2PN \\
\hline
$\mathcal{L}_{5L}$ &\cellcolor{lime} Closed Form &  Munna et al.  & 2PN & --- & --- \\
\hline
$\mathcal{L}_{11/2}$ &\cellcolor{Yellow} Fitted to $e^{30}$ &  Munna et al.  & 4PN & --- & 1PN \\
\hline
$\mathcal{L}_{11/2L}$ &\cellcolor{GreenYellow} All orders &  This paper  & 1PN & --- & --- \\
\hline
$\mathcal{L}_{6}$ &\cellcolor{YellowOrange} Fitted to $e^{20}$ &  Munna et al.  & 6PN & 3PN* & 3PN* \\
\hline
$\mathcal{L}_{6L}$ &\cellcolor{GreenYellow} All Orders &  Paper I   & 3PN & 0PN & 0PN \\
\hline
$\mathcal{L}_{6L2}$ &\cellcolor{lime} Closed Form &  Paper I  & --- & --- & --- \\
\hline
$\mathcal{L}_{13/2}$ &\cellcolor{Yellow} Fitted to $e^{30}$ &  Munna et al.  & 5PN & --- & 2PN \\
\hline
$\mathcal{L}_{13/2L}$ &\cellcolor{Yellow} Fitted to $e^{30}$ &  Munna et al.  & 2PN & --- & --- \\
\hline
$\mathcal{L}_{7}$ &\cellcolor{Orange} Fitted to $e^{12}$  &  Munna et al. & 7PN & 4PN* & 4PN* \\
\hline
$\mathcal{L}_{7L}$ &\cellcolor{YellowOrange} Fitted to $e^{26}$  &  Munna et al. & 4PN & 1PN & 1PN \\
\hline
$\mathcal{L}_{7L2}$ &\cellcolor{lime} Closed Form &  This paper  & 1PN & --- & --- \\
\hline
$\mathcal{L}_{15/2}$ &\cellcolor{Orange} Fitted to $e^{12}$ &  Munna et al.  & 6PN & --- & 3PN* \\
\hline
$\mathcal{L}_{15/2L}$ &\cellcolor{YellowOrange} Fitted to $e^{26}$ &  Munna et al.  & 3PN & --- & 0PN \\
\hline
$\mathcal{L}_{15/2L2}$ &\cellcolor{GreenYellow} All Orders &  Paper I  & --- & --- & --- \\
\hline
$\mathcal{L}_{(3k)L(k)}$ &\cellcolor{lime} Closed Form &  Paper I  & --- & --- & --- \\
\hline
$\mathcal{L}_{(3k+3/2)L(k)}$ &\cellcolor{GreenYellow} All Orders &  Paper I  & --- & --- & --- \\
\hline
$\mathcal{L}_{(3k+1)L(k)}$ &\cellcolor{lime} Closed Form &  This paper  & 1PN & --- & --- \\
\hline
$\mathcal{L}_{(3k+5/2)L(k)}$ &\cellcolor{GreenYellow} All Orders &  This paper  & 1PN & --- & --- \\
\hline \hline
\end{tabular}
\end{table}
\end{center}

\appendix

\section{Component sums for 1PN logarithms}
\label{sec:compSums}

We provide here some low-order examples of the component sums that, once added together, produce the corresponding
1PN log sequence term.  These illustrate (i) some of the steps in the procedure, (ii) the particular eccentricity 
dependence of individual terms, (iii) how different source multipoles contribute, and (iv) the presence of 
next-order-in-$\nu$ dependence.  At 1PN order itself, we find
\begin{align}
\mathcal{R}_1^{\rm MQ01} &= \frac{1}{(1-e_t^2)^{9/2}} \left(\frac{271}{21}+\frac{1705 e_t^2}{28}
+\frac{2555 e_t^4}{96}+\frac{1189 e_t^6}{1344} \right) - \frac{1}{(1-e_t^2)^{3}} \left(18 + \frac{63 e_t^2}{4} \right),  \notag \\
\mathcal{R}_1^{\rm MQ02} &= \frac{1}{(1-e_t^2)^{9/2}} \left(-18-\frac{219 e_t^2}{4}-\frac{111 e_t^4}{16} \right),
\qq  \q \mkern8mu \mathcal{R}_1^{\rm MQ03} =  \frac{1}{(1-e_t^2)^{3}} \left(18 + \frac{63 e_t^2}{4} \right),  \notag \\
\mathcal{R}_1^{\rm MQ11} &= \frac{1}{(1-e_t^2)^{9/2}} \left(\frac{55}{21}+\frac{3907 e_t^2}{504}-\frac{e_t^4}{96}
-\frac{307 e_t^6}{2016} \right), \qq \mathcal{R}_1^{\rm CQ} = \frac{1-4\nu}{(1-e_t^2)^{9/2}} \left(\frac{1}{36}
+\frac{19 e_t^2}{72}+\frac{23 e_t^4}{96}+\frac{e_t^6}{64}\right) .
\end{align}
The mass octupole portion is given in \eqref{eqn:R1MO}.  The Newtonian moments match their expected forms, but the 
mass quadrupole functions are more interesting, with the pieces displaying somewhat distinct singular behavior 
as $e_t \rightarrow 1$.  We have confirmed that a similar pattern exists in all integer-order 1PN log terms through 
22PN, with
\begin{align}
\mathcal{R}_{(3k+1)L(k)}^{\rm MQ01} &= \frac{1}{(1-e_t^2)^{3k+9/2}} \, f^{(1)}_k(e_t) - \frac{1}{(1-e_t^2)^{3k+3}} \,
f^{(2)}_k(e_t), \qq  \mathcal{R}_{(3k+1)L(k)}^{\rm MQ02} = \frac{1}{(1-e_t^2)^{3k+9/2}} \, f^{(3)}_k(e_t),  \notag \\  
\mathcal{R}_{(3k+1)L(k)}^{\rm MQ03} &= \frac{1}{(1-e_t^2)^{3k+3}} \, f^{(2)}_k(e_t),  \qq  \qq \qq \qq \qq \qq
 \mathcal{R}_{(3k+1)L(k)}^{\rm MQ11} = \frac{1}{(1-e_t^2)^{3k+9/2}} \, f^{(4)}_k(e_t) ,
\end{align}
where $f^{(1)}_k(e_t), f^{(2)}_k(e_t), f^{(3)}_k(e_t), f^{(4)}_k(e_t)$ are polynomials in $e_t$.  It is not difficult 
to prove that the trends in singular behavior continue to all orders for MQ02, MQ03, MQ11, MO, and CQ using the 
methods of asymptotic analysis laid out in \cite{ForsEvanHopp16,LoutYune17,MunnETC20}.  Unfortunately, a similar 
proof for MQ01 has remained elusive, though there are overlapping reasons to believe that the same behavior arises 
in this term as well, including the fact that all divergences as $e_t \rightarrow 1$ must vanish in a PN expansion 
that uses $1/p$ (the semi-latus rectum) as the compactness parameter instead of $x$ (see \cite{MunnETC20}). 
Nearly identical trends exist in the integer-order 1PN angular momentum log sequence terms.

At half-integer orders, the component terms are not closed in form.  For future reference, at 2.5PN we find 
\begin{align}
\mathcal{R}_{5/2}^{\rm MQ01} &= \frac{1}{(1-e_t^2)^6} \left(\frac{1336}{21}+\frac{29083 e_t^2}{48}
+\frac{137933 e_t^4}{192}+\frac{3704005 e_t^6}{27648}+\frac{3902585 e_t^8}{1548288}
-\frac{54803587 e_t^{10}}{619315200}+\cdots \right) 
-\mathcal{R}_{5/2}^{\rm MQ03},  
\notag \\
\mathcal{R}_{5/2}^{\rm MQ02} &= \frac{1}{(1-e_t^2)^6} \left(-84-\frac{9625 e_t^2}{16}-\frac{27545 e_t^4}{64}
-\frac{70049 e_t^6}{3072}-\frac{16247 e_t^8}{73728}+\frac{1664999 e_t^{10}}{29491200}
-\frac{1280041 e_t^{12}}{353894400} + \cdots \right), 
\notag \\
\mathcal{R}_{5/2}^{\rm MQ03} &=  \frac{1}{(1-e_t^2)^{9/2}} \left(84+\frac{2037 e_t^2}{8}+\frac{1029 e_t^4}{32}
-\frac{343 e_t^6}{1536}-\frac{763 e_t^8}{12288}-\frac{17969 e_t^{10}}{4915200}+\frac{32543 e_t^{12}}{58982400} 
+ \cdots \right),  \notag \\
\mathcal{R}_{5/2}^{\rm MQ11} &= \frac{1}{(1-e_t^2)^6} \left(\frac{220}{21}+\frac{9841 e_t^2}{144}
+\frac{16891 e_t^4}{576}-\frac{216235 e_t^6}{27648}-\frac{2088109 e_t^8}{4644864}
+\frac{4380643 e_t^{10}}{371589120}-\frac{33875507 e_t^{12}}{22295347200} + \cdots \right),  \notag \\ 
\mathcal{R}_{5/2}^{\rm MQ12} &= \frac{1}{(1-e_t^2)^6} \left(-2-\frac{1183 e_t^2}{96}+\frac{1565 e_t^4}{384}+\frac{178873 e_t^6}{18432}+\frac{237847 e_t^8}{442368}+\frac{1166257
   e_t^{10}}{176947200}-\frac{3037147 e_t^{12}}{2123366400} + \cdots \right),  \notag \\
\mathcal{R}_{5/2}^{\rm MO} &= \frac{1-4\nu}{(1-e_t^2)^6} \left(\frac{16403}{2016}+\frac{34163 e_t^2}{336}
+\frac{21836233 e_t^4}{129024}+\frac{57821777 e_t^6}{1161216}+\frac{67599745e_t^8}{49545216}
+\frac{241631 e_t^{10}}{132710400} + \cdots \right), \notag \\
\mathcal{R}_{5/2}^{\rm CQ} &= \frac{1-4\nu}{(1-e_t^2)^6} \left(\frac{1}{18}+\frac{4 e_t^2}{3}+\frac{2041 e_t^4}{576}
+\frac{7991 e_t^6}{5184}+\frac{2989 e_t^8}{49152}-\frac{6307 e_t^{10}}{16588800}
+\frac{11669 e_t^{12}}{212336640} + \cdots \right).
\end{align}
In each infinite series, the coefficients drop off rapidly in magnitude with power of $e_t$, indicating likely 
convergence as $e_t \rightarrow 1$.  As with integer orders, similar singular behavior can be proven to hold to all 
orders in each type of sum except for that of MQ01.  Nevertheless, we have used high order expansions to 
demonstrate apparent convergence for the MQ01 sums (and the rest) through 20.5PN order.  There is nearly identical 
structure observed again in the angular momentum flux case.

\section{Fourier sum identities}
\label{sec:fourSumId}

In this section, we briefly provide a couple of the Fourier series identities used in the various 1PN mass quadrupole
derivations.  We start with sums of the following form:
\begin{align}
\sum_{n=-\infty}^{\infty} \sum_{p,s=-2}^{2} n^{2r} p \, \underset{(n,p)}{I_{ij}}  \underset{(-n,s)}{I_{ij}}
= \sum_{n=-\infty}^{\infty} \sum_{p=-2}^{2} n^{2r} p \, \underset{(n,p)}{I_{ij}}  \underset{(-n,-p)}{I_{ij}}, 
\end{align}
where $r$ is an integer, and where on the right hand side we noted that only terms with $s = -p$ will survive. 
Then,
\begin{gather}
 \sum_{n=-\infty}^{\infty} \sum_{p} n^{2r} p \,
\underset{(n,p)}{I_{ij}} \underset{(-n,-p)}{I_{ij}} 
= \sum_{n=-\infty}^{\infty} n^{2r}
 \bigg[2 \underset{(n,2)}{I_{ij}} \underset{(-n,-2)}{I_{ij}} - 2 \underset{(n,-2)}{I_{ij}} \underset{(-n,2)}{I_{ij}} \bigg]
\notag \\
= \sum_{n=1}^{\infty} n^{2r}
 \bigg[2 \underset{(n,2)}{I_{ij}} \underset{(-n,-2)}{I_{ij}} - 2 \underset{(n,-2)}{I_{ij}} \underset{(-n,2)}{I_{ij}} \bigg] +
 \sum_{n=-1}^{-\infty} n^{2r}
 \bigg[2 \underset{(n,2)}{I_{ij}} \underset{(-n,-2)}{I_{ij}} - 2 \underset{(n,-2)}{I_{ij}} \underset{(-n,2)}{I_{ij}} \bigg]
 \notag \\
= \sum_{n=1}^{\infty} n^{2r}
 \bigg[2 \underset{(n,2)}{I_{ij}} \underset{(-n,-2)}{I_{ij}} - 2 \underset{(n,-2)}{I_{ij}} \underset{(-n,2)}{I_{ij}} \bigg] +
 \sum_{n=1}^{\infty} n^{2r}
 \bigg[2 \underset{(-n,2)}{I_{ij}} \underset{(n,-2)}{I_{ij}} - 2 \underset{(-n,-2)}{I_{ij}} \underset{(n,2)}{I_{ij}} \bigg]
 = 0.
\end{gather}
In the same way, we can prove that 
\begin{gather}
\sum_{n=-\infty}^{\infty} \sum_{p,s=-2}^{2} n^{2r+1} \text{sign}(n) \, p \, 
\underset{(n,p)}{I_{ij}}  \underset{(-n,s)}{I_{ij}},    \notag \\
\e_{3jl} \sum_{n=-\infty}^{\infty} \sum_{p,s=-2}^{2} n^{2r+1}  p \, 
\underset{(n,p)}{I_{ja}}  \underset{(-n,s)}{I_{la}},  \notag \\
\e_{3jl} \sum_{n=-\infty}^{\infty} \sum_{p,s=-2}^{2} n^{2r} \text{sign}(n) \, p \, 
\underset{(n,p)}{I_{ja}}  \underset{(-n,s)}{I_{la}},   \notag
\end{gather}
all vanish and
\begin{gather}
\sum_{n=-\infty}^{\infty} \sum_{p,s=-2}^{2} n^{2r+1} p \, \underset{(n,p)}{I_{ij}}  \underset{(-n,s)}{I_{ij}}, \notag \\
\sum_{n=-\infty}^{\infty} \sum_{p,s=-2}^{2} n^{2r} \text{sign}(n) \, p \, 
\underset{(n,p)}{I_{ij}}  \underset{(-n,s)}{I_{ij}},    \notag \\
\e_{3jl} \sum_{n=-\infty}^{\infty} \sum_{p,s=-2}^{2} n^{2r}  p \, 
\underset{(n,p)}{I_{ja}}  \underset{(-n,s)}{I_{la}},  \notag \\
\e_{3jl} \sum_{n=-\infty}^{\infty} \sum_{p,s=-2}^{2} n^{2r+1} \text{sign}(n) \, p \, 
\underset{(n,p)}{I_{ja}}  \underset{(-n,s)}{I_{la}},   \notag
\end{gather}
all gain a factor of 2 when expressed in terms of positive-$n$ sums.

\end{widetext}

\clearpage

\bibliography{1PNlogs}

\end{document}